\documentclass[reqno,11pt]{book}
\usepackage{amsmath,german,umlaut} 
\usepackage{amsfonts} 
\usepackage{amssymb}
\usepackage{graphicx,epic,eepic}
\usepackage[dvips]{color}
\usepackage[bf]{caption2}

\newcommand{\bra}[1]{\langle #1\vert}
\newcommand{\ket}[1]{\arrowvert #1 \rangle}

\newcommand{\kin}{-\frac{\hbar^2}{2m}  \nabla^2}
\newcommand{\oppdag}[2]{\hat{#1}^\dagger(t,\vec{#2})}
\newcommand{\opp}[2]{\hat{#1}(t,\vec{#2})}
\newcommand{\bog}{\hat{\Psi}(t,\vec{x})=\Phi(t,\vec{x}) \, + \, \varepsilon \hat{\psi}(t,\vec{x})\, + \, \ldots}
\newcommand{\zeit}{i\hbar \frac{\partial}{\partial t}}

\begin{document}
\ifx\href\undefined\else\hypersetup{linktocpage=true}\fi

\bibliographystyle{plain}

\begin{titlepage}

\center


{\Huge \bf Simulation}\\[0.3cm]
{\Huge \bf von}\\[0.3cm]
{\Huge \bf Gravitationsobjekten}\\[0.3cm]
{\Huge \bf im}\\[0.3cm]
{\Huge \bf Bose-Einstein-Kondensat}\\[3cm]
{\sf \large
Diplomarbeit\\[1mm]

von\\[1mm]

\Large Silke E. Ch. Weinfurtner\\[1cm]

Betreuer der Diplomarbeit:
Prof. Dr. I. Cirac\\[2mm]

\large Max-Planck-Institut für Quantenoptik\\[2mm]

Technische Universität München\\
Fakulät für Physik\\[0.5cm]
}
\begin{minipage}[t][4cm][c]{1cm}
\includegraphics[height=3cm]{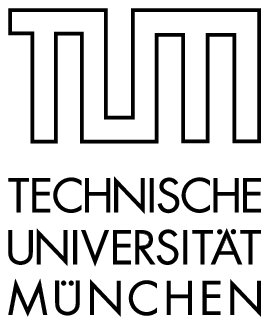}
\end{minipage}
\hfill
\begin{minipage}[t][4cm][c]{2cm}
\center
\includegraphics{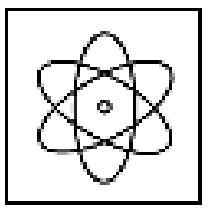}\\
{\small \sf PHYSIK\\DEPARTMENT}\end{minipage}
\hfill
\begin{minipage}[t][4cm][c]{1cm}
\includegraphics[height=3cm]{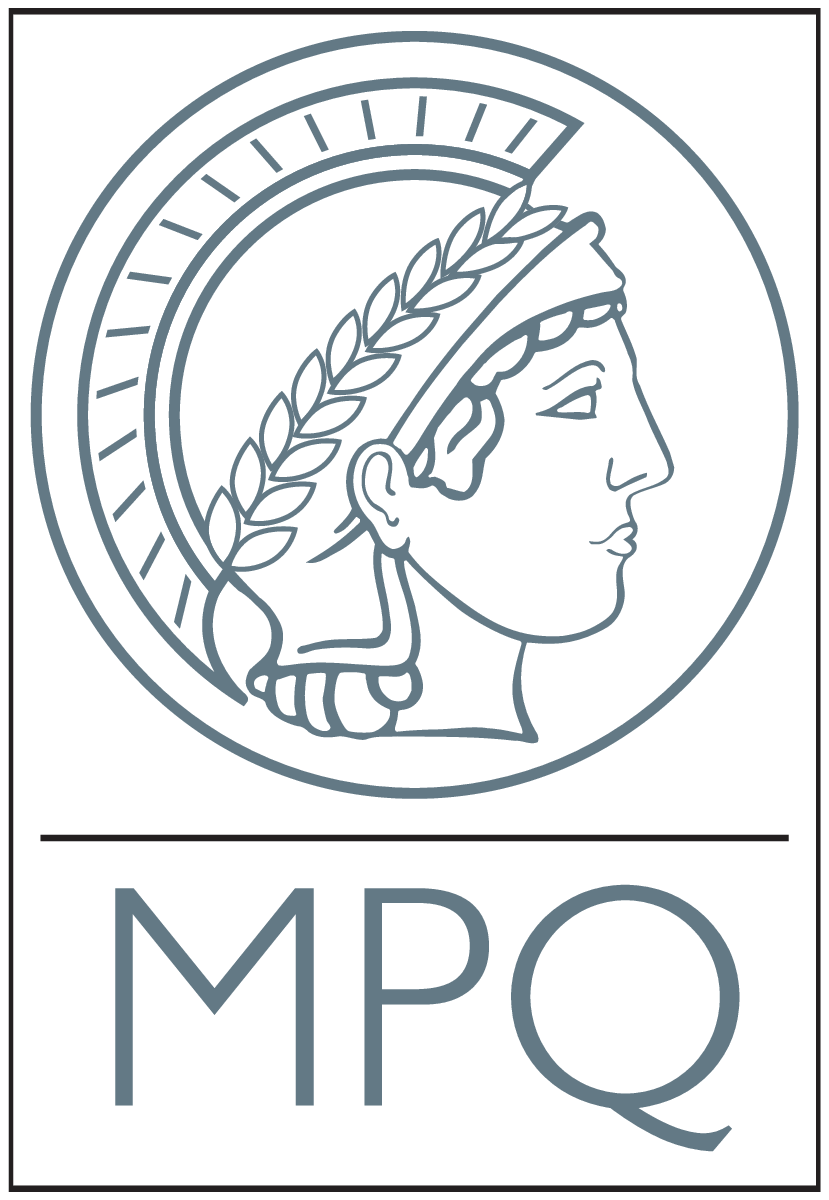}
\end{minipage}
\\

\end{titlepage}

          \tableofcontents

          \frontmatter

                              \chapter{Einleitung}

In den letzten Jahren hatte sich herausgestellt, dass zwischen den
Bose-Einstein-Kondensaten (BEC) und Objekten aus der Allgemeinen Relativitätstheorie (ART)
eine Analogie besteht \cite{cirac1}.
Die Bose-Einstein-Kondensation konnte erstmals von Anderson et al. (1995) \cite{anderson} experimentell überprüft werden.
Seitdem sind in Experiment und Theorie große Fortschritte zu verbuchen.
Die Voraussetzungen, Kondensate in vielen verschiedenen Konfigurationen herzustellen,
wurden möglich.
Anhand dreier solcher Kondensate - zigarrenförmig, ringförmig und frei expandierend - wird in
dieser Arbeit die Verknüpfung zur Gravitationsphysik gezeigt.
Es können zwei Eigenschaften im BEC, die aus der Physik in gravitativen Feldern
bekannt sind, im Kondensat entdeckt werden.
Die Bewegungsgleichung einer von außen angeregten Störung kann im hydrodynamischen Limes mit einer effektiven
Metrik formuliert werden. Abhängig vom gewählten Kondensat kann die Metrik einem
Objekt aus der Allgemeinen Relativitätstheorie zugeordnet werden, z.B. einem Schwarzen
Loch und dem de-Sitter-Universum.
Außerdem kann gezeigt werden, dass in Anwesenheit dynamischer Instabilitäten Moden im Kondensat
angeregt werden. Es handelt sich dabei um Quasiteilchen, den Phononen, die paarweise mit positiver und
negativer Energie erzeugt werden.\\

In den ersten beiden Kapiteln werden die Grundlagen zur Bose-Einstein-Kondensation
und der Allgemeinen Relativitätstheorie behandelt.
Der Hauptteil beginnt mit Kapitel (\ref{BECART}), der Verifizierung des Zusammenhangs
zwischen BEC und Gravitation.
Im Anschluss daran werden drei spezielle Kondensate analysiert.
In Kapitel (\ref{kapspeziell}) wird für das zigarrenförmige und das ringförmige Kondensat
die Metrik berechnet und mit der eines Schwarzen Lochs verglichen.
In beiden Systemen wird gezeigt, dass Instabilitäten auftauchen und wie diese sich mit der
Zeit entwickeln.
In Kapitel (\ref{kapuniversum}) ist das System ein frei expandierendes Kondensat.
Es wird eine Metrik gesucht, die das
de-Sitter-Universum beschreibt.
Am Ende der Kapitel (\ref{kapspeziell}) und (\ref{kapuniversum}) werden die
Ergebnisse jeweils zusammengefasst.
Den Schlußpunkt setzt eine kurze Diskussion aller erzielten Resultate und ein Ausblick darauf, was
aktuell in der Forschung von Interesse ist.

          \mainmatter

                    \part{Einführung in die Bose-Einstein-Kondensation\label{kap1}}
                              
          \chapter{Das ideale Bose-Gas}
\begin{figure}[h]
                              \begin{center}
                              \input{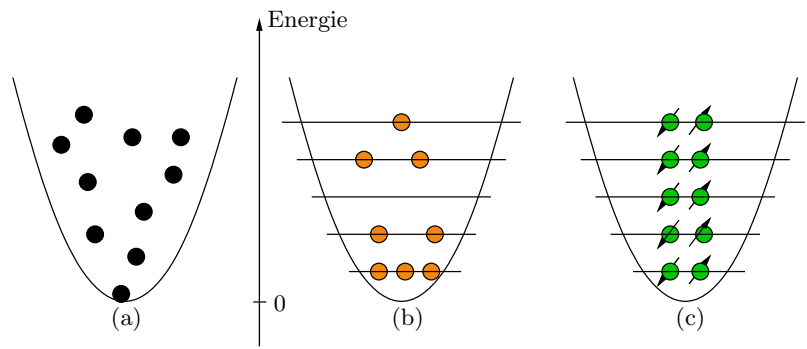}
                                        \caption[Bosonen und Fermionen im Potentialtopf]{\label{bosonenfermionen}Beim Übergang von der klassischen Beschreibung (a) eines Gases in den
                                        quantenmechanischen Formalismus treten zwei
                                        Sorten von Teilchen auf. Während die Bosonen\footnotemark[1] (b) jeden Zustand beliebig oft besetzen können, ist es
                                        zwei Fermionen\footnotemark[2] (c) nicht
                                        erlaubt, den gleichen Zustand einzunehmen. Zwei Fermionen mit der gleichen Energie müssen deshalb entgegengesetzten Spin haben.
                                        Auffallend ist, dass der Grundzustand\footnotemark[3]
                                        der Fermionen und Bosonen nicht bei $E=0$ liegt.}
                              \end{center}
                              \end{figure}

\begin{figure}[h]
                              \begin{center}
                                        \input{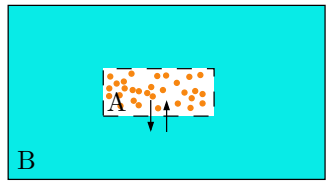}
                                        \caption[Physikalische Bedingungen für das \textit{Großkanonische Potential}]
                                        {\label{grosskanonisch}Das ideale Gas (A) befindet sich in einem viel größeren Wärmebad (B), wodurch eine konstante Temperatur erreicht wird.
                                        Ersteres ist damit im thermischen Gleichgewicht, wozu Energieaustausch vom Gas zum Wärmebad nötig ist.
                                        Die Pfeile zeigen an, dass zusätzlich Teilchenaustausch stattfindet.}
                              \end{center}
                              \end{figure}

                    Als ideal bezeichnet man ein Gas, bei welchem die Atome nicht miteinander wechselwirken. Nehmen wir zusätzlich an, die Atome sind in einem
                    äußeren Potential, so können sie nur bestimmte Energiewerte annehmen (Abb. \ref{bosonenfermionen}). \\
                              \addtocounter{footnote}{1}
                              \footnotetext[1]{Bosonen haben ganzzahligen Spin, hier werden jedoch nur Bosonen mit $s=0$ verwendet. }
                              \stepcounter{footnote}\footnotetext[2]{Fermionen haben halbzahligen Spin.}
                              \stepcounter{footnote}\footnotetext[3]{Der Grundzustand bezeichnet die niedrigst mögliche Energiekonfiguration des Systems.}
                    Von Interesse ist nun die Frage, wie die verschiedenen Energieniveaus bei gegebener Temperatur besetzt werden.
                    Diese \textit{Besetzungszahl} $n$ kann über das \textit{Großkanonische Potential} berechnet werden (Abb. \ref{grosskanonisch}).\\
                   Über den Hamilton-Operator $\hat{H}$ für das aus $N$-Teilchen
                    bestehende Gas, kann die \textit{Dichtematrix $\hat \rho$ des großkanonischen Ensembles} bestimmt werden
                              \begin{equation}  \label{dichtematrix}
                                        \hat{\rho}=\frac{1}{Z_G}exp{\left[-\beta \left( \hat{H} - \mu \hat{N} \right)\right]},
                              \end{equation}
                    wobei $Z_G$ der Normierungsfaktor ist, der als \textit{Großkanonische Zustandssumme} bezeichnet wird.
                    Im Exponenten treten noch der Teilchenzahloperator\footnote{Der Teilchenzahloperator angewendet auf einen Zustand liefert die Anzahl
                    aller Teilchen.} $\hat N$,
                    $\beta={1}/{k_B T}$ mit der Temperatur $T$ und das chemische
                    Potential $\mu$ in Erscheinung.\\
                    Mittels Gl.(\ref{n}) ist es möglich, die \textit{mittlere Besetzungszahl} $n(\epsilon_{\vec{l}})$
                              \begin{equation}  \label{n}
                                        n(\epsilon_{\vec{l}})=\frac{1}{e^{\left[\frac{1}{k_B T}(\epsilon_{\vec{l}}\,-\mu)\right]}-1}
                              \end{equation}
                    für einen Zustand mit der
                    Energie $\epsilon_{\vec{l}}$ bei einer bestimmten Temperatur $T$ zu bestimmen.
                    Oft wird in der Literatur eine Größe $z=exp({\mu}/{k_B T})$
                    eingeführt, wodurch sich für die mittlere Besetzungszahl
                              \begin{equation} \label{nz}
                                        n(\epsilon_{\vec{l}})=\frac{1}{z} e^{(\epsilon_{\vec{l}})}-1
                              \end{equation}
                    ergibt.\\
                    Als Vorbereitung für den folgenden Abschnitt sei noch darauf hingewiesen, dass die Summation der mittleren Besetzungszahl
                    über alle möglichen Zustände die Teilchenzahl $N$ ergibt:
                              \begin{equation}  \label{N}
                                        N=\sum_{\vec{l}} n_{\vec{l}}\,.
                              \end{equation}

                    Es ist Gl.(\ref{n}) zu entnehmen,
                    dass mit abfallender Temperatur energetisch tiefer gelegene Niveaus bevorzugt werden. Um zeigen zu können,
                    was passiert, wenn die Temperatur gegen Null geht, wird als Beispiel ein einfaches Potential verwendet.

                              \section{Das ideale Gas im harmonischen Potential}
                                        Die Eigenwerte $\epsilon_{\vec{l}}$ des Gases im dreidimensionalen harmonischen Potential $\frac{1}{2}m \vec\omega^2\vec{x}^2$ sind
                                                  \begin{eqnarray} \label{l}
                                                            \epsilon_{\vec{l}}=l_x\hbar \omega_x + l_y\hbar \omega_y + l_z\hbar \omega_z & & l_{x,y,z}=0,1,2,3...
                                                  \end{eqnarray}
                                        wobei die Grundzustandsenergie $\epsilon_{\vec{0}}={\hbar}/{2}=(\omega_x +\omega_y + \omega_z)$ ist.\\

                                        Das weitere Vorgehen ist, $\epsilon_{\vec{l}}$ in Gl.(\ref{N}) einzusetzen, mit dem Ziel, zu berechnen, wie sich die Teilchen
                                        bei der Annäherung an den absoluten Nullpunkt der Temperatur verhalten. Ein ausdruckstarkes Maß dafür ist das Verhältnis von
                                        Teilchen im Grundzustand $N_0$ zu den im Gas enthaltenen Teilchen $N$. Jene Teilchen, die nicht den Grundzustand bevölkern, bezeichnet
                                        man als angeregte Teilchen $N'=N-N_0$. \\
                                        Lässt man bei der Summation in der Formel für die Gesamtteilchenzahl (\ref{N}) $\vec{l}=0$ weg, ergibt sich für
                                                  \begin{equation} \label{N'}
                                                            N'=\sum_{\vec{l}\neq 0}n_{\vec{l}}=\sum_{\vec{l}\neq 0}\left[ \frac{1}{z}exp(\beta l \hbar \omega)-1 \right]^{-1}
                                                            < \sum_{\vec{l}\neq 0}\left[ exp(\beta l \hbar \omega)-1 \right]^{-1} \equiv N_{max}'
                                                  \end{equation}
                                        eine nach oben abgegrenzte Anzahl an angeregten Atomen $N_{max}'$.
                                        Diese Näherung gilt, wenn $1/z$ gegen eins geht, also für
                                        $T$ gegen Null. Zuführung von weiteren Teilchen ins Kondensat bewirkt eine direkte Anreicherung des Grundzustands.
                                         \noindent Die Grundzustandsenergie steckt im chemischen Potential $\mu$.\\

                                        Nach einigen Schritten\footnote{Eine ausführliche Herleitung findet sich in \cite{BECcastin}.} ergibt sich
                                                  \begin{equation}
                                                            \label{N0N}
                                                            \dfrac{N_0}{N}\approx 1- \left(\dfrac{T}{T_c^0}\right)^3 ,
                                                  \end{equation}
                                        mit dem als \textit{kritische Temperatur} $T_c^0$ eingeführten Wert\footnote{$\zeta(3)=1.202...$}
                                                  \begin{equation} \label{Tc}
                                                            T_c^0=\frac{\hbar \omega}{k_B}\left( \frac{N}{\zeta(3)} \right)^{1/3}. \, 
                                                  \end{equation}

                                       Unterhalb der kritischen Temperatur $T_c^0$ findet damit eine Teilchenanreicherung im Grundzustand statt. Im nächsten Abschnitt soll
                                       erklärt werden, warum es sich dabei um einen Phasenübergang handelt.

                              \section{Der Phasenübergang und sein Ordnungsparameter}
                                                  \begin{figure}[h] \begin{center}
                                                  \includegraphics*[width=15cm]{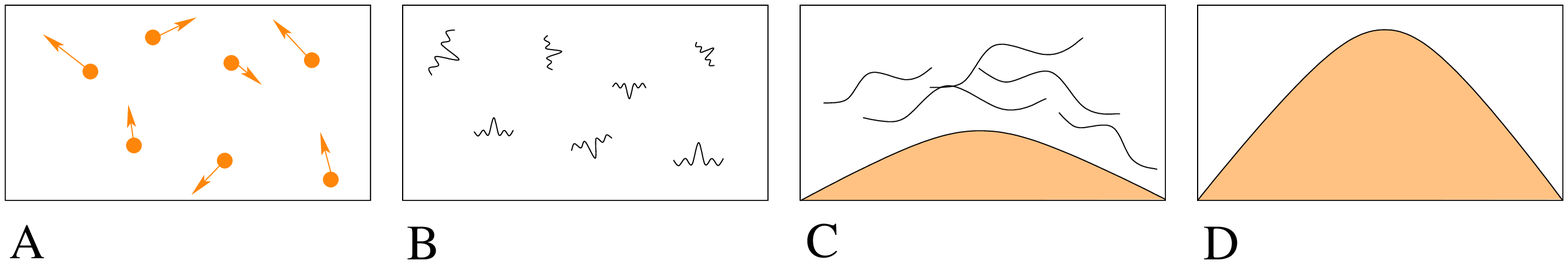}
                                                  \caption[Bildung eines Bose-Einstein-Kondensats]
                                                  {\label{phasenubergang}Ob Teilchen - oder Wellencharakter hängt von der Temperatur ab: Während für hohe Temperaturen (A) die Atome als kleine
                                                  Kugeln betrachtet werden können, nimmt das Atom für niedrige Temperaturen Wellencharakter an. Den Atomen kann
                                                  eine thermische de Broglie Welle (B) zugeordnet werden. Diese Wellenlänge nimmt mit abnehmender Temperatur zu. Bei
                                                  $T_c$ beginnen die Wellenfunktionen sich zu überlappen (C). Angekommen beim absoluten Nullpunkt der Temperatur bilden alle
                                                  Teilchen zusammen eine makroskopische Welle (D).}
                                                  \end{center} \end{figure}
                                        Jedem Atom kann eine thermische de Broglie-Wellenlänge $\lambda_T=\sqrt{{2\pi\hbar^2}/{m k_B T}}$ zugewiesen werden.
                                        Der Vergleich dieser Größe mit den mittleren Abständen $l=\frac{\sqrt{{k_B T}/{m\omega^2}}}{N^{1/3}}$ im Gas zeigt, dass sich die de Broglie-Wellen
                                        oberhalb von $T_c$ nicht überschneiden, unterhalb schon. Bei der kritischen Temperatur selbst stimmen die thermische Wellenlänge und
                                        der mittlere Abstand ungefähr überein. Dieses Anwachsen von $\lambda_T$ unterhalb $T_c$ bewirkt ein kollektives Verhalten der Atome im Gas.\\

                                        Ein Phasenübergang ist dadurch charakterisiert, dass eine Größe existiert, 
                                        die unterhalb der kritischen Temperatur endlich ist und oberhalb der kritischen Temperatur verschwindet \cite{statistischemechanik}. Diese Größe wird als Ordnungsparameter bezeichnet.
                                        Hier ist der Ordnungsparameter die makroskopische Wellenfunktion.\footnote{
                                        Der Exponent, welcher in Gl.(\ref{N0N}) auftritt und als \textit{kritischer Exponent} bezeichnet wird,
                                        weicht vom allgemeinen Fall ${N_0}/{N}\approx 1- \left({T}/{T_c^0} \right)^{3/2}$ -unendlich großer Potentialtopf - ab.}\\

                                        Generell ist aber auch die Frage zu beantworten, ob die Annahme eines idealen Gases hinreichend genau ist.
                                        Dazu verwenden wir experimentelle Daten und vergleichen sie mit den mathematischen Vorhersagen dieses Abschnitts.

                              \section{Der Vergleich mit dem Experiment \label{widerspruch}}
                                       Wird der Bruchteil der Atome im Kondensat über die Temperatur aufgetragen und mit Gl.(\ref{N0N}) verglichen,
                                       stimmt der theoretische mit dem gemessenen Wert hinreichend gut überein.
                                       Auch wenn bei der kritischen Temperatur auffällt, dass $T_{c_{exp}}^0$ unterhalb $T_{c}^0$ liegt.

                                                  \begin{figure}[h] \begin{center}
                                                  \vspace{-1.8cm}
                                                  \input{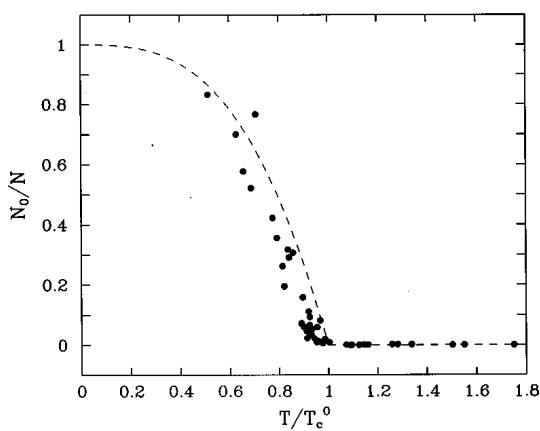}
                                                  \vspace{-2.2cm}
                                                  \caption[Bruchteil der Atome im Kondensat]
                                                  {\label{vergleichNON}Die Punkte stellen die Versuchsergebnisse von Ensher et al. \cite{Ensher} dar, während
                                                  die gestrichelte Linie den mathematischen Vorhersagen für ein ideales Gas Gl.(\ref{N0N}) entspricht.}
                                                  \end{center}
                                                  \end{figure}

                                       Dies alleine reicht jedoch nicht aus, um zu beantworten, ob die Wechselwirkung im Gas vernachlässigt werden kann.
                                       Ein probates Mittel zum Vergleich ist dagegen die Gesamt\-energie E,
                                       da die Wechselwirkung $\rm E_{int}$ nur berücksichtigt werden muss, wenn sie einen nicht verschwindenden Beitrag zur Energie
                                       $E=E_{kin} + E_{ext} + E_{int}$ leistet.\\
                                       Im Versuch von Enscher et al. \cite{Ensher} wird ein Kondensat gebildet und das äußere Potential $E_{ext}$ nach einiger Zeit abgeschaltet.
                                       Die reduzierte Energie $E=E_{kin} + E_{int}$ wird in Abhängigkeit von der Temperatur gemessen. Die Durchführung des
                                       Experiments ist jedoch mit der Schwierigkeit behaftet, dass sich ab dem Zeitpunkt der freien Expansion das Kondensat ausbreitet.
                                       Damit wachsen die Abstände zwischen den Atomen an, wodurch die Wechselwirkung abnimmt, deren Einfluss man messen
                                       möchte.

                                                  \begin{figure}[h]
                                                  \vspace{-1.5cm}\begin{center}
                                                  \input{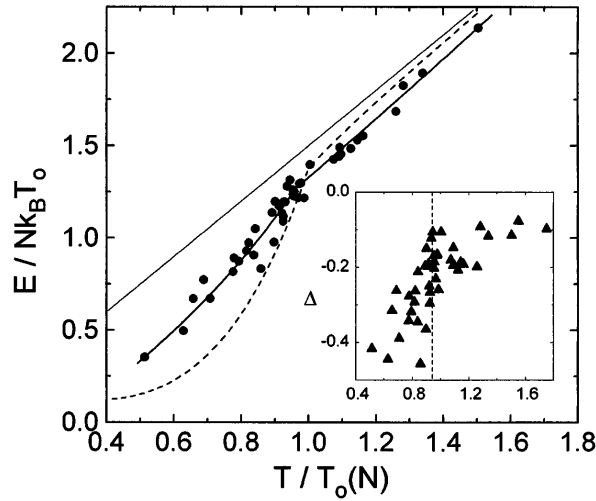}
                                                  \vspace{-2.2cm}\caption[Energieabnahme unterhalb von $T_{c}^0$]
                                                  {\label{vergleichE}Die gestrichelte Linie zeigt den theoretisch erwarteten Verlauf der Energieabnahme für ein ideales Gas.
                                                  Dieser weicht von dem gemessenen Werten deutlich ab \cite{Ensher}.}
                                                  \end{center}
                                                  \end{figure}

                                       Oberhalb von $T_{c}^0$ ist eine gute Übereinstimmung zu erkennen, das Gas kann als hinreichend ideal betrachtet werden.
                                       Die Situation ändert sich aber für Temperaturen unterhalb von $T_{c}^0$.\\
                                       Die Atome beginnen sich am untersten Energieniveau
                                       anzusammeln. Es wird also erwartet, dass sich die relative Energieabnahme
                                                  \begin{equation}
                                                            \frac{E}{Nk_BT} \propto \frac{(N-N_0)}{Nk_BT}=\left( 1-\frac{N_0}{N} \right)\frac{\hbar\omega}{k_B T}=\left(\frac{T}{T_c} \right)^3 \frac{\hbar\omega}{k_B T}
                                                  \end{equation}

                                       verhält. Betrachtet man jedoch Abb.(\ref{vergleichE}), 
                                       ist eine lineare Abnahme zu erkennen.

                                       Damit muss die Annahme eines idealen Gases aufgegeben werden.
                                       Im nächsten Kapitel werden wir uns mit den nicht zu vernachlässigenden Wechselwirkungen beschäftigen.

          \chapter{Das annähernd ideale Gas}

                    Die Widersprüche zwischen Experiment und Theorie erfordern es, die Wechselwirkung zwischen den Teilchen des Gases zu
                    berücksichtigen.
                    Es handelt sich dann um ein \textit{reales Gas}. Im allgemeinen sind die Wechselwirkungen in einem
                    N-Teilchensystem sehr kompliziert, da jedes Teilchen auf jedes andere Einfluss nimmt.\\
                    Wir bewegen uns jedoch in Temperaturbereichen von $T<1\mu K$, in welchen fast alle sich im thermischen Gleichgewicht befindlichen Stoffe in
                    der flüssigen Phase vorliegen.

                    Dies kann nur umgangen werden, indem man zur Erzeugung von Bose-Einstein-Kondensaten extrem verdünnte
                    Gase ($\rho < 10^{14}/cm^3$) verwendet\footnote{Zusätzlich muss die Kühlung so rasch erfolgen, dass ein Erstarren verhindert wird.}.\\
                    Die Wechselwirkung findet im verdünnten Gas wesentlich reduziert statt. Es stoßen immer nur zwei Teilchen gleichzeitig aneinander.
                    Damit kann die Wechselwirkung zwischen den Atomen als kleine Störung zum idealen Gas betrachtet werden.
                    Die Spezifizierung des Begriffs {\glqq verdünntes Gas\grqq} und dessen mathematische
                    Beschreibung ist Inhalt dieses Abschnitts.

                    \section{Ultrakalte verdünnte Gase}

                              Es wird ein Gas betrachtet, dessen mittlerer Teilchenabstand groß gegenüber der Wechselwirkungsreichweite\footnote{Die Wechselwirkung zwischen
                              Atomen kann proportional zu $1/r^6$ beschrieben werden, wobei $r$ der Abstand sind. Damit ist die Reichweite unendlich. Aber es ist möglich zu zeigen, dass
                              für Potentiale die stärker als $1/r^3$ mit dem Abstand abfallen, die Annahme einer endlichen Wechselwirkung zulässig ist.}
                              ist. Damit können die Teilchen nur dann wechselwirken, wenn sie sich der Reichweite entsprechend annähern.

                                                  \begin{figure} \begin{center}
                                                  \input{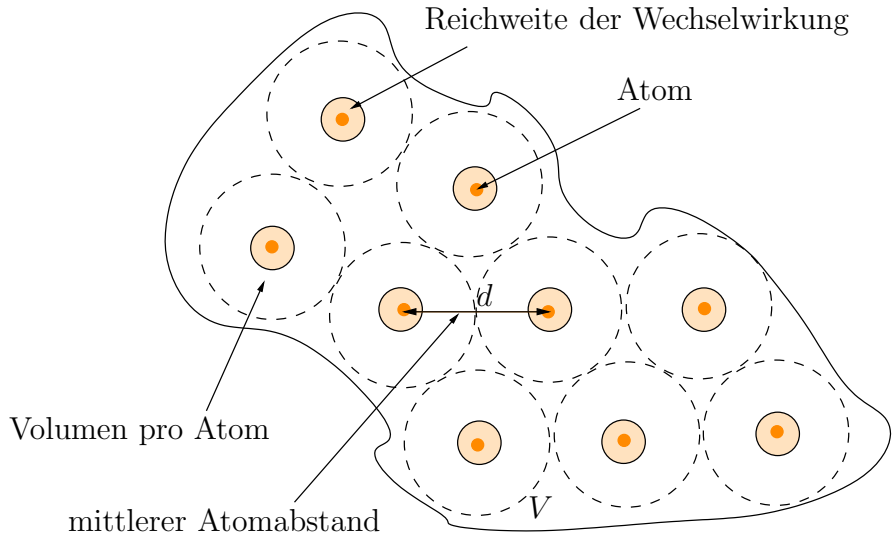}
                                                  \caption[Verdünntes Gas]
                                                  {\label{verdgas}Das Bild zeigt eine schematische Darstellung eines verdünnten Gases: Die Atomabstände liegen in einer Größenordnung von $d \sim 1000$ \AA\footnotemark,
                                                  die Wechselwirkungsreichweite der Atome kann als endlich angenommen werden und es finden hauptsächlich Zwei-Teilchen-Stöße statt.  }
                                                  \end{center}  \end{figure}

                                                  \footnotetext{W. Ketterle \textit{et al.}\cite{Ketterle1} \ verwendeten zur Herstellung eines Bose-Einstein-Kondensats (1995)
                                                  ein verdünntes Sodium-Gas, bestehend aus bis zu $5\cdot 10^{5}$ Atomen, mit einer Dichte von $10^{14}/cm^3$.
                                                  Mit $v \sim 1/d^3$ und $\rho = N_0/N$ kann damit der mittlere Abstand berechnet werden.}
                              In die Rechnungen werden also nur Zwei-Teilchen-Wechselwirkungen einbezogen.

                                        \subsection{Zwei-Teilchen-Wechselwirkung}
\begin{figure} \begin{center}
                                                  \vspace{-2.6cm}\input{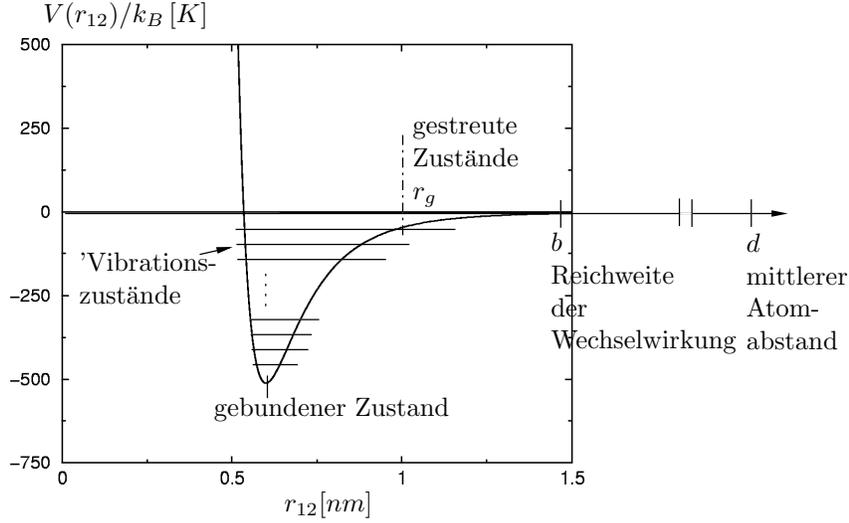}\vspace{-1.8cm}
                                                  \caption[Born-Oppenheimer-Potential]
                                                  {\label{BornOppenheimer}Der Potential-Verlauf zwischen zwei Atomen wird vom relativen Abstand $r_{12}$ bestimmt.
                                                  Das Potential kann in drei Bereiche eingeteilt werden. Für sehr kleine $r_{12}$ treten gebundene Zustände auf, ab
                                                  $r_{g}$ streuen die Atome aneinander und im Abstand $b$ voneinander dominiert der Van-der-Waals-Anteil mit $1/r^6$,
                                                  endet die Reichweite der Wechselwirkung.}
                                                  \end{center}  \end{figure}

                                                  Wir betrachten zwei Atome im Schwerpunktsystem (\ref{Schwerpunktsystem}), die im Abstand $r_{12}$ voneinander entfernt sind.
                                                  Das Wechselwirkungspotential wird durch das \textit{Born-Oppenheimer-Potential} (Abb.(\ref{BornOppenheimer}))
                                                  beschrieben.
                                                  In Abb.(\ref{BornOppenheimer}) ist dargestellt, dass für niedrige Temperaturen gebundene Zustände möglich sind.
                                                  Würde man das exakte Potential zur weiteren Berechnung verwenden, wäre der Übergang in die flüssige Phase unausweichlich.
                                                  Es muss daher ein anderes Modell gefunden werden, das keine gebundenen Zustände erlaubt und eine weitere mathematische
                                                  Erschließung ermöglicht\footnote{Der exakte Potential-Verlauf ist für bestimmte Atome sehr schwer zu bestimmen und ein kleiner Fehler
                                                  im Potential kann zu großen Abweichungen im Resultat führen.}.

                                        \subsection{Das exakte Potential}
                                                 Als Ausgangspunkt dienen die Ergebnisse der \textit{quantenmechanischen Streuprozesse},
                                                 welche in Anhang (\ref{streuergebnisse}) kurz hergeleitet werden.

                                                            $$\psi(\vec{r}_{12})=\psi_0(\vec{r}_{12})-\frac{2 \mu}{4 \pi \hbar^2}\int{d^3 \vec{r}_{12}\,'
                                                            \frac{e^{ik \vert\vec{r}_{12}-\vec{r}_{12}\,'\vert}}{\vert\vec{r}_{12}-\vec{r}_{12}\,'\vert}}V(\vec{r}_{12}\,')\psi(\vec{r}_{12}\,')$$

\begin{figure} \begin{center}
                                                            \input{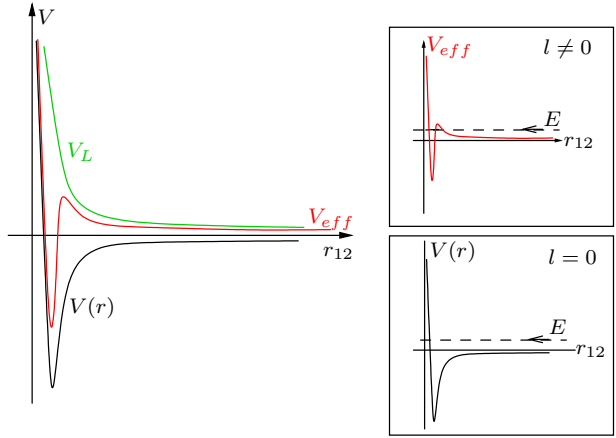}
                                                            \caption[Streuverhalten in Abhängigkeit der Drehimpulszahl $l$]
                                                            {\label{swavescatt}Im allgemeinen Fall ($l\neq 0$) muss das $V(r)$ (rot) auf ein effektives Potential $V_{eff}(r)=-V(r)+V_l(r)$ erweitert werden.
                                                            Der neue Term $V_{l}={\hbar^2 l(l+1)}/{2mr^2}$ erlaubt es nicht mehr, dass für kleine positive Energien gebundene Zustände ausgeschlossen
                                                            werden können.}
                                                            \end{center}  \end{figure}

                                                  Diese Gleichung beinhaltet auch die unerwünschten Bindungprozesse und Mehrfach-Streuprozesse. Eine Entwicklung von
                                                  $\psi(\vec{r}_{12}\,')=\psi_0(\vec{r}_{12}\,')+ \mathcal{O}(\varepsilon)$ bis zur Nullten Ordnung für schwache Potentiale schafft Abhilfe.
                                                  Elastischen Stöße können über
                                                            \begin{equation} \label{inhompsi2}
                                                                      \psi(\vec{r}_{12})=e^{i\vec{k}\vec{r}_{12}}-\frac{2 \mu}{4 \pi \hbar^2}\int{d^3\vec{r}_{12}\,'
                                                                      \frac{e^{ik \vert\vec{r}_{12}-\vec{r}_{12}\,'\vert}}{\vert\vec{r}_{12}-\vec{r}_{12}\,'\vert}}V(\vec{r}_{12}\,')e^{i\vec{k}\vec{r}_{12}\,'}
                                                            \end{equation}
                                                  berechnet werden. Abb.(\ref{swavescatt}) zeigt, dass die Streuung bei niedrigen Energien nur angewendet werden darf, wenn die Drehimpulsquantenzahl
                                                  Null ist.\footnote{Die Drehimpulsquantenzahl zeigt an, wie die Elektronenschalen besetzt sind. Bei $l=0$ darf die letzte besetzte Schale kein p-Orbital sein.}
                                                            \\

                                                  Berücksichtigen wir, dass der mittlere Abstand $\vec{r}_{12}$ im verdünnten Gas groß gegenüber - der als endlich angenommen - Reichweite $b$ ist,
                                                  gilt $\vec{r}_{12} \ll \vec{r}_{12}'$.
                                                  Folgende Näherung ermöglicht dies
                                                            \begin{equation}  \label{r-r'}
                                                                      \vert \vec{r}_{12}-\vec{r}_{12}' \vert= r_{12}-\vec{n}\vec{r}_{12}'+\mathcal{O}\left(\frac{1}{r_{12}}\right),
                                                            \end{equation}
                                                  wo $\vec{n}=\frac{\vec{r}_{12}}{r_{12}}$ die Streurichtung ist.\\
                                                  Für sehr niedrige Temperaturen liegen im thermischen Gas die auftretenden Energien in der Größenordnung von $k_B T$.
                                                  Damit muss $k$ in $E={\hbar^2 k^2}/{2\mu}$ klein sein, womit das Produkt $kb \ll 1$ ebenfalls vernachlässigt werden kann.
                                                  Unter diesen Voraussetzungen hängt der gestreute Zustand nicht mehr von der Streurichtung ab, weil die Wellenlänge groß gegenüber
                                                  den Details des Atompotentials ist.\\

                                                  Ein Einsetzen der Annahmen - kleine Energien, endliche Reichweite im stark verdünnten System - in Gl.(\ref{inhompsi2}),
                                                  liefert
                                                            \begin{equation}  \label{scatta}
                                                                      \psi(\vec{r}_{12})=e^{ikr_{12}}-a\frac{e^{ikr_{12}}}{r_{12}}.
                                                            \end{equation}
                                                  Die neu eingeführte Variable $a$ bezeichnet man als Streulänge, welche mit der Streuamplitude
                                                            \begin{equation} \label{Steuamplitude}
                                                                      f_{\vec{k}}(\vec{n})=-\frac{2\mu}{4\pi \hbar^2}\int d^3\vec{r}_{12}' e^{-ik\vec{n}\vec{r}_{12}'}V(\vec{r}_{12}')\psi_0(\vec{r}_{12}')
                                                            \end{equation}
                                                  wie folgt zusammenhängt
                                                            \begin{equation} \label{a}
                                                                      -a=\lim_{k\rightarrow 0}f_{\vec{k}}(\vec{n})=-\frac{2\mu}{4\pi \hbar^2}\int d^3\vec{r}_{12}'V(\vec{r}_{12}')\psi_0(\vec{r}_{12}').
                                                            \end{equation}

                                                  Anzumerken ist nochmals, dass $a$ nicht von der Streurichtung $\vec{n}$ abhängt, damit nur s-Wellen-Streuung betrachtet werden kann.\\

                                                  Die hier berücksichtigten Energien sind als sehr klein betrachtet worden.
                                                  Lässt man nun $k\rightarrow 0$ gehen, nimmt die Energie des zu streuenden Zustands ab.
                                                  Der Grenzfall eines Zustandes mit $E=0$
                                                            \begin{equation}  \label{scattE0}
                                                                      \psi_{E=0}=1-\frac{a}{r_{12}}
                                                            \end{equation}
                                                  kann zur numerischen Berechnung der Streuamplitude verwendet werden.\\

                                                  Das exakte Potential kann mit Gl.(\ref{scattE0}) berechnet werden. Stellvertretend dafür wird ein Modell-Potential eingeführt.
                                                  Damit kann das Problem der gebundenen Zustände umgangen werden.\\
                                                  \noindent Hinzu kommt, dass die Born-Approximation nur für schwache Potentiale gültig ist,
                                                  was für Abstände größer als die
                                                  Reichweite $b$ des Potentials erfüllt ist, aber für kleine Abstände - dort wo gebundene Zustände auftreten können - nicht.\\
                                                  \noindent Der Versuch, das starke Potential für sehr kleine Abstände durch höhere Ordnungen
                                                  in der Born-Serie zu beseitigen, würde gebundene Zustände implizieren.\\
                                                  Es wird ein Modell-Potential gesucht, das mit dem Verhalten - abgesehen von den Bindungszuständen -
                                                  übereinstimmt.

                                        \subsection{Das Modell-Potential}
\begin{figure} \begin{center}
                                                            \input{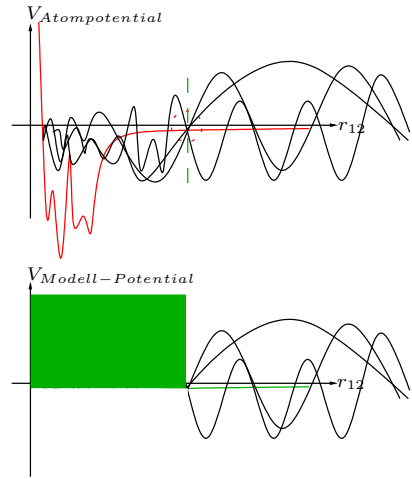}
                                                            \caption[Schematische Darstellung des Modell-Potentials]
                                                            {\label{modellpot}Dargestellt ist ein beliebiges Potential, dessen Reichweite als endlich betrachtet werden kann.
                                                            Dieses für kleine $r_{12}$ komplizierte Potential wird durch ein Kontaktpotential ersetzt, dessen Radius mit der Lage
                                                            des letzten gebundenen Zustandes zusammenhängt.}
                                                            \end{center}  \end{figure}

                                                  Die Idee für das Modell-Potential ist in Abb.(\ref{modellpot}) dargestellt.
                                                  \noindent
                                                  Es handelt sich dabei um ein \textit{Pseudo-Potential},
                                                            \begin{equation}
                                                                      \bra{\vec{r}_1\vec{r}_2}V_{int}\ket{\psi_{1,2}}\equiv g \delta(\vec{r}_1-\vec{r}_2) \left[ \frac{\partial}{\partial r_{12}}(r_{12}\psi_{12}(\vec{r}_1,\vec{r}_2)\right]_{r_{12}=0}
                                                            \end{equation}
                                                  das sich aus einem Kontaktpotential und einem Regularisierungs-Operator
                                                  zusammensetzt \cite{BECcastin}. Die \textit{Kopplungskonstante}
                                                            \begin{equation} \label{kopplungskonstante}
                                                                      g=\frac{4\pi \hbar^2}{m}a
                                                            \end{equation}
                                                  beinhaltet die einzige Bestimmungsgröße - die Streulänge - welche experimentell bestimmt wird.

                                                  Die Streulänge hängt nicht von der Reichweite des Potentials ab. Es besteht aber eine direkte Verbindung mit der Position des
                                                  letzten angeregten gebundenen Zustands.
                                                  Dessen Energieniveaus zweiatomiger Moleküle hängen von den Übergängen zwischen den Schwingungszuständen des Moleküls ab.
                                                  Dabei schwingen die beiden Atome um den metastabilen Abstand $r_0$
                                                  (Abb.(\ref{BornOppenheimer}) gegeneinander, es bildet sich ein sog. \textit{Vibrationspektrum} aus.
                                                  In der \textit{Vibarationsenergie}
                                                            \begin{equation} \label{vibenergie}
                                                                      E_v=\left( \left( \nu_D-\nu \right) K(\mu,C_6) \right)^{1/3}
                                                            \end{equation}
                                                  kommen die \textit{Vibrationsquantenzahl} $\nu$ und die beiden Konstanten $K$ und $\nu_D$ vor.
                                                  Die Vibrationsquantenzahl kann ganze positive Zahlen größer Null $\nu=0,1,2,..$ annehmen und muss kleiner gleich
                                                  $\nu_D$ sein. Ein geradzahliges $\nu_D$ entspricht damit dem letzten gebundenen Zustand, bei welchem gleichzeitig die Trennung
                                                  des Moleküls eintritt \cite{allaboutBEC}.\\
                                                  Eine exakte Berechnung der Streulänge $a$ ergibt
                                                            \begin{equation}  \label{exakta}
                                                                      a=a_{sc}\left(1-\tan(\phi)\right),
                                                            \end{equation}
                                                  mit der Konstanten
                                                            \begin{equation} \label{a_sc}
                                                                      a_{sc}=\frac{\Gamma(3/4)}{2\sqrt{2}\Gamma(5/4)}\left(\frac{2\mu C_6}{\hbar^2} \right)^{1/4}
                                                            \end{equation}

                                                  und
                                                            \begin{equation}  \label{scttphi}
                                                                      \phi=\pi(\nu_D + 1/2).
                                                            \end{equation}

\begin{figure} \begin{center}
                                                            \includegraphics*[width=10cm]{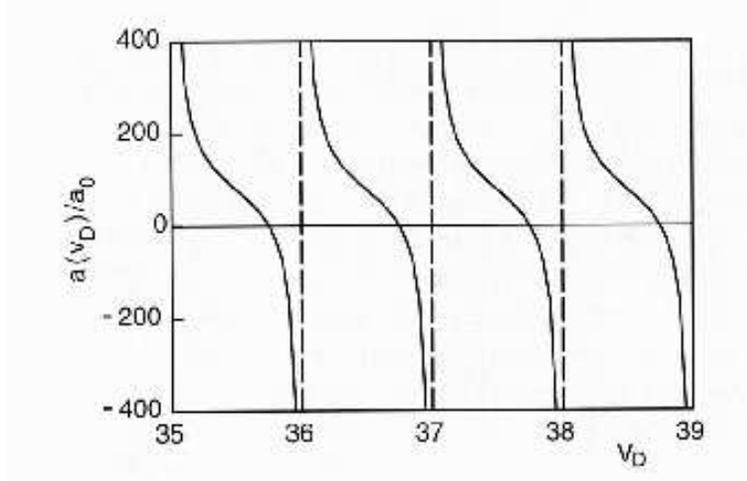}
                                                            \caption[Abhängigkeit der Streulänge von der Vibrationsquantenzahl]
                                                            {\label{scatt} Dargestellt ist die Streulänge für $Rb_2$ (im Triplett-Zustand) Potential in Abhängigkeit
                                                            von $\nu_D$. Die Divergenzen stellen sich für geradzahlige $\nu=\nu_D$ ein.
                                                            Große negative Werte von $a$ treten dann auf, wenn das Potential den nächsten gebundenen Zustand gerade vermeiden kann.
                                                            Für große positive Werte ist es genau umgekehrt. Sie liegen vor, wenn das Potential stark genug ist,
                                                            den letzten gebundenen Zustand zu unterstützen \cite{allaboutBEC}.}
                                                            \end{center}
                                                            \end{figure}

                                                  Mit $\nu_D$ und der Atommasse ist die Streulänge fixiert. Durch Manipulationen an der Tiefe des Potentials
                                                  und der Masse ist es möglich, die Streulänge zu variieren (Abb.(\ref{scatt})).

                                                  In allen weiteren Berechnungen wird als Wechselwirkungspotential dieses Modell-Potential,
                                                  ein Kontakt-Potenial mit der Streulänge $a$ als Radius, verwendet.
                                                  Damit ist es möglich, die lokal abhängigen Wechselwirkungen durch ein mittleres Feld zu ersetzten.

\chapter{Die Gross-Pitaevskii-Gleichung\label{vielteil}}

\section{Der Viel-Teilchen-Hamilton-Operator}
Die quantenmechanische Beschreibung eines Viel-Teilchen-System - in dem nur paarweise Wechselwirkung stattfindet -
erfolgt im Formalismus der zweiten Quantisierung\footnote{Zweite Quantisierung, weil analog zur ersten Quantisierung, bei
welcher die Observablen durch Operatoren ersetzt wurden, nun auch noch die Zustände durch Operatoren beschrieben werden.}.
In diesem ist es möglich, Teilchen zu erzeugen bzw. zu vernichten. Einer Anregung aus dem Grundzustand entspricht
zum Beispiel eine Vernichtung eines Teilchens im Grundzustand, kombiniert mit einer Erzeugung eines Teilchens
im angeregten Zustand.
Diese Beschreibung eignet sich zur Analyse eines Anregungsspektrums.\\
Das Verhalten eines Teilchens an einem bestimmten Ort liefert der sog.
Feldoperator $\hat\Psi^{\dag}(\vec{x})$, der ein Teilchen bei $\vec{x}$ erzeugen bzw. vernichten kann.\\
Als Ausgangspunkt verwenden wir den großkanonischen Hamilton-Operator $\hat{K}=\hat{H}-\mu \hat{N}$ im Feldformalismus
          \begin{equation} \label{hamilton2teilchen}
                    \begin{split}
                    \hat{K} = \hat{H} - \mu \hat{N}=& \int{d^3 \, \vec{x}\oppdag{\Psi}{x}\left(\kin +V_{ext}(x)\right)\opp{\Psi}{x}} \\
                    +&\frac{1}{2}\int{d^3 \vec{x} d^3 \vec{x}' \, \oppdag{\Psi}{x}\oppdag{\Psi}{x'}V(x-x')\opp{\Psi}{x'}\opp{\Psi}{x}}.
                    \end{split}
          \end{equation}
Außer der Zwei-Teilchen-Wechselwirkung wurden bislang keine weiteren Annahmen gemacht.
In den nächsten Schritten werden zusätzliche Näherungen für das ultrakalte verdünnte Gas durchgeführt.
Das chemische Potential ist in $V_{ext}$ enthalten.

\section{Der Bogoliubov-Ansatz}
Hierbei handelt es sich um ein Verfahren zur Berechnung des Anregungsspektrums für Temperaturen $T \ll T_c$.
Es wird davon ausgegangen, dass sich fast alle Atome im Grundzustand befinden, also $N-N_0 \ll N_0\approx N$.
Der Grundzustand ist die makroskopische Wellenfunktion, die sich unterhalb der kritischen Temperatur ausbildet
und durch die komplexwertige Funktion $\Phi(\vec{x})$ in der Näherung (Gl.(\ref{bog})) zur Geltung kommt.
Anregungen können dann als Fluktuationen um den Grundzustand aufgefasst werden
und sind durch den Feldoperator $\delta\hat\Psi(\vec{x})$ beschrieben.
          \begin{equation}  \label{bog}
                    \bog \,\,.
          \end{equation}
Das $\varepsilon$ hebt hervor, dass die Anregungen klein sind.\\

Bevor nun mit diesem Ansatz weiter gerechnet wird, bedarf es einer Klärung bezüglich der komplexwertigen Funktion
$\Phi(\vec{x})$.

\subsection{Spontane Symmetriebrechung}
Wie schon erwähnt, beschreibt $\Phi(\vec{x})$ den Kondensatanteil. Es wird angenommen,
dass der Feldoperator $\hat\Psi$ einen nichtverschwindenden endlichen Erwartungswert $\langle \hat\Psi \rangle=\Phi(\vec{x}) $ hat.
Der Erwartungswert der Fluktuationen sollte damit verschwinden $\langle \delta\hat\Psi \rangle=0 $.
Die Annahme eines nichtverschwindenden Erwartungswertes bringt jedoch eine Symmetriebrechung mit sich.\\
Es handelt sich um die Eichsymmetrie, welche verletzt wird. Diese fordert, dass der Hamilton-Operator
invariant gegenüber derartigen
          \begin{equation} \label{symmetrie}
                    \Psi(\vec{x})\rightarrow \Psi(\vec{x})e^{i\alpha}
          \end{equation}
Transformationen mit der Phase $\theta_0$ ist.\\
Die Forderung, dass der Erwartungswert $\bra{0}\Psi(\vec{x})\ket{0}$ des Grundzustand $\ket{0}$ nicht verschwinden darf
verletzt die Eichinvarianz und damit die Symmetrie. Konvention ist, $\alpha=0$ zu wählen. \\
Bei der Behandlung der idealen Gase wurde gezeigt, dass es sich um einen Phasenübergang handelt. Da jeder Phasenübergang
zweiter Ordnung mit gebrochener Symmetrie verbunden wird, muss es sich spontan um einen solchen handeln.

\section{Die Mean-field-Näherung\label{meanfield}}
Nachdem die Bogoliubov-Näherung diskutiert wurde, kann diese
in den Hamilton-Operator (Gl.(\ref{hamilton2teilchen})) eingesetzt werden, 
wobei zunächst für das Wechselwirkungspotential das Kontaktpotential eingesetzt wurde $V(\vec{x}-\vec{x}')=g\delta(\vec{x}-\vec{x}')$.
\noindent Dabei ergeben sich folgende Terme
          \begin{equation}\label{hbogoliubov}
                    \begin{array}{l}
                    \hat K=  \int{d^3\vec{x}\left( \Phi^*(\vec{x}) \hat{K}_0\Phi(\vec{x}) + \frac{g}{2}\vert \Phi \vert^4\right)} \\
                    +\varepsilon \int{d^3\vec{x}\left(  \Phi^*(\vec{x}) \hat{K}_0\delta\opp{\Psi}{x}  +g\Phi^*(\vec{x})\Phi^*(\vec{x})\Phi(\vec{x})\delta\opp{\Psi}{x}   + konjungiert \, komplex \right)} \\
                    +\varepsilon^2 \int{d^3\vec{x}\left( \delta\oppdag{\Psi}{x} \hat{K}_0 \delta\opp{\Psi}{x}
                    +\frac{g}{2}  \Phi^*(\vec{x})^2\delta \opp{\Psi}{x}^2+\frac{g}{2}\Phi(\vec{x})^2\delta \oppdag{\Psi}{x}^2
                    +2g\Phi(\vec{x})^2\delta\oppdag{\Psi}{x}^2  \right)}\\
                    +\varepsilon^3 g \int{d^3 \vec{x} \left( \Phi^*(\vec{x}) \delta \oppdag{\Psi}{x}\delta\opp{\Psi}{x} \delta \opp{\Psi}{x}+
                    \Phi(\vec{x}) \delta \oppdag{\Psi}{x} \delta \oppdag{\Psi}{x}\delta\opp{\Psi}{x}   \right)}  \\
                    +\varepsilon^4 \frac{g}{2} \int{d^3 \vec{x} \left( \delta \oppdag{\Psi}{x} \delta \oppdag{\Psi}{x}\delta \opp{\Psi}{x}\delta \opp{\Psi}{x}  \right)}
                    \end{array}
          \end{equation}
wobei $\hat{K}_0=\left(\kin +V_{ext}(x)-\mu\right)$ ist.

Die Terme dritter und vierter Ordnung in $\varepsilon$ können durch eine Mean-Feld-Näherung ersetzt werden.
Dazu wird das lokale Feld durch seinen Mittelwert,
also durch ein mittleres Feld ersetzt. Zum Beispiel wird aus
          $\delta\oppdag{\Psi}{x}\delta\opp{\Psi}{x} \delta\opp{\Psi}{x}\approx2\langle \delta\oppdag{\Psi}{x}\delta\opp{\Psi}{x}\rangle\delta\opp{\Psi}{x}+
          \langle\delta\opp{\Psi}{x} \delta\opp{\Psi}{x} \rangle \delta\oppdag{\Psi}{x} $.
Diese Näherung wird als Hartree-Fock-Bogoliubov-Näherung (HFB) bezeichnet \cite{Griffin}.\\
Vernachlässigt man weiterhin anomale thermische Dichten $\langle \delta\hat\Psi\delta\hat\Psi \rangle $, bekommt man die Popvo-Näherung,
die für Temperaturen bis zu $0.7\,T_c$ verwendet werden kann. Vernachlässigt man weiterhin alle thermischen Dichten
          $\langle \delta\hat\Psi\delta\hat\Psi^{\dag} \rangle $ erhält man die Gross-Pitaevskii-Gleichung, die als nächstes untersucht wird.

\subsection{Die zeitunabhängige Gross-Pitaevskii-Gleichung}
\begin{figure}  \begin{center}
          \vspace{-1.5cm}\input{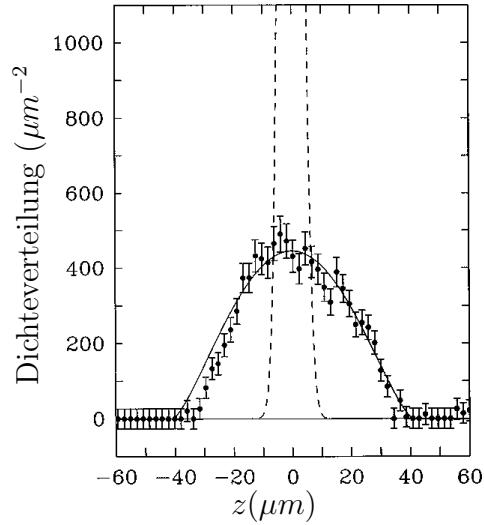}\vspace{-1.5cm}
          \caption[Dichteprofil einer Kondensatwolke]
          {\label{vergleichprofil}Die experimentellen Daten (Punkte) stimmten nicht mit den theoretischen Berechnungen überein,
          falls das Gas als ideal angenommen wird (gestrichelte Linie). Die durchgezogene Linie zeigt die
          theoretische Vorhersage der Gross-Pitaevskii-Gleichung \cite{Hau}.}
          \end{center}
          \end{figure}

In diesem Abschnitt wird der Grundzustand $\Phi_0$ des fast idealen Gases mit dem Ritzschen Variationsprinzip berechnet.
Die Forderung, dass alle Atome das niedrigste Energieniveau besetzen, wird als Nebenbedingung
          \begin{equation} \label{nb}
                    \int d^3x \vert \Phi(x)\vert^2=1
          \end{equation}
mit dem Lagrange Parameter $\lambda$ im Energiefunktional
          \begin{equation} \label{ritz1}
                    \langle \hat H_{nb} \rangle= \langle \hat H \rangle - \lambda N_0 \left( \int d^3x \vert \Phi(x)\vert^2-1\right)
          \end{equation}
berücksichtigt.\\
Als Hamilton-Operator kann Gl.(\ref{hbogoliubov}) verwendet werden, wenn man bedenkt, dass sich der großkanonische Hamilton-Operator
zu diesem reduziert, sobald alle Glieder mit $\mu$ weggelassen werden. Mit $\langle \delta\hat\Psi \rangle=0$ fallen bei der Berechnung
des Erwartungswerts alle Terme mit $\varepsilon$ weg\footnote{Die Terme dritter und vierter Ordnung fallen erst nach der Mean-Feld-Näherung
weg, weshalb Gl.(\ref{hbogoliubov}) nach der Einführung des Mittleren-Feldes zu verwenden ist.}.\\
Das sich ergebende Funktional
          \begin{eqnarray} \label{ritz2}
                    E[\Phi,\Phi^*]= \int d^3 \vec{x} \left( \Phi^*(\vec{x}) \left( \kin +V_{ext}(x) \right) \Phi(\vec{x}) + \frac{g}{2}\vert \Phi \vert^4 \right)
                    - \lambda N_0 \left( \int d^3x \vert \Phi(x)\vert^2-1\right)
          \end{eqnarray}
wird um den Grundzustand  $\Phi(\vec{x})\rightarrow \Phi(\vec{x}) + \delta \Phi(\vec{x})$ variiert.
Fordert man, dass die Funktionalableitung ${\delta}/{\delta\Phi}$ der Größe $\delta E=E[\Phi+\delta \Phi , \Phi^*+\delta\Phi^* ]$ für alle
$\delta \Phi$ verschwindet, muss folgende Gleichung erfüllt sein:
          \begin{equation} \label{GPE1}
                    \left[ \kin + V_{ext}(x) + N_0 g \vert \Phi(\vec{x}) \vert^2 - \lambda \right] \Phi(\vec{x})=0.
          \end{equation}
Dieser Ausdruck ist - mit der physikalischen Interpretation des Lagrange-Multiplikators als chemisches Potential $\lambda=\mu$ -
als zeitunabhängige Gross-Pitaevskii-Gleichung bekannt. Als Lösung ergibt sich der Grundzustand $\Phi(\vec{x})=\Phi_0(\vec{x})$.\\
Die Frage, ob diese Näherung die Realität besser beschreibt als die theoretischen Vorhersagen für das
ideale Gas, klärt Abb.(\ref{vergleichprofil}).
Sie zeigt den Vergleich zwischen idealem Gas und dem ultrakalten verdünnten Gas (Beschreibung nach Gross-Pitaevskii)
mit dem Experiment.

\subsection{Die zeitabhängige Gross-Pitaevskii-Gleichung\label{zeitGEP}}
Bei der Herleitung für den zeitabhängigen Fall wird nicht auf die Variationsrechnung zurückgegriffen
\footnote{Die Herleitung der zeitabhängigen Gross-Pitaevskii-Gleichung mittels des Variationsprinzips findet sich in \cite{BECcastin}.}.\\
Der Hamilton-Operator inklusive Zwei-Teilchen-Wechselwirkung in Feldquantisierung ist
          \begin{eqnarray}
                    \hat H=\int{dx \, \oppdag{\Psi}{x}[\kin +V_{ext}(x)]\opp{\Psi}{x}} + \nonumber   \\
                    \frac{1}{2}\int{dx dx' \, \opp{\Psi}{x}\opp{\Psi}{x'}V(x-x')\oppdag{\Psi}{x'}\oppdag{\Psi}{x}}. \label{hfort}
          \end{eqnarray}
Die Zeitentwicklung eines Feldoperators, auf den $\hat H$ wirkt, kann direkt durch die Heisenberg-Relation
          \begin{equation}  \label{Hdyn1}
                    \zeit \opp{\Psi}{x}=[\opp{\Psi}{x},\hat{H}]
          \end{equation}
angegeben werden.
Setzt man im Anschluss daran für den Feldoperator ein klassisches Feld
          \begin{equation} \label{klass}
                    \hat \Psi \rightarrow \sqrt{N_0}\Phi
          \end{equation}
für den Grundzustand ein, ergibt sich die zeitabhängige Gross-Pitaevskii-Gleichung:
          \begin{equation} \label{GPE2}
                    i\hbar\partial_t \Phi(t,\vec{x})=\left( \kin + V_{ext}(\vec{x}) + N_0 g \vert \Phi(\vec{x}) \vert^2 - \mu \right) \Phi(t,\vec{x})
          \end{equation}

\chapter{Anregungen im Kondensat\label{anregungenneu}}
In diesem Abschnitt werden kleine Anregugen\footnote{Unter dem Begriff einer kleinen Störung
versteht man eine lokale Dichtemodulation, welche den BEC-Zustand nicht vernichtet und als
kollektive Anregungen aufgefasst werden.}  im Bose-Einstein-Kondensat untersucht.
Diese k\'onnen als Schallwellen interpretiert werden, deren Ausbreitungsgeschwindigkeit als Schallgeschwindigkeit bezeichnet wird.\\
Der erste experimentielle Nachweis von Schallwellen im Bose-Einstein-Kondensat konnte erstmals \cite{exc1} in einer frei expandierenden Atomwolke erbracht werden. Dieser Versuch wird am Ende des Abschnitt kurz vorgestellt.

\section{Das Anregungsspektrum}

\begin{figure}[!h]
          \begin{center}
          \input{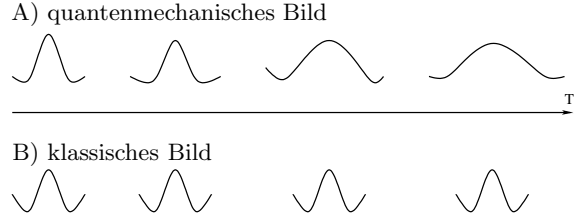}
          \caption[Lokale Dichtemodulation]
          {\label{semkl1}In der Quantenmechanik (A) kann sich keine Dichteverteilung fortbewegen, ohne deformiert zu werden.
          In der klassischen Mechanik (B) ist es dagegen möglich, die Störung als ganzes mit einem Impuls propagieren zu lassen. }
          \end{center}
          \end{figure}

Für $T = 0$ können alle auftretenden Dichten Gl.(\ref{hbogoliubov}) vernachlässigt werden.\footnote{
Es wird auf die Temperaturabhängigkeit des Spektrums hingewiesen,
so dass diese Annahme nur für Temperaturen $T \ll T_c$
gemacht werden kann. Eine Berechnung für Temperaturen bis zu $0.7\,T_c$ findet sich in \cite{exc3}.}
Nimmt man zusätzlich an, dass sich die gesamte Dichtemodulation mit einem festen Impuls $\vec{p}=\hbar \nabla \theta$
im Kondensat ausbreitet (siehe Abb.(\ref{semkl1})), ergibt eine längere Rechnung das folgende Anregungsspektrum:
          \begin{equation}\label{spek4neubitte}
                    E^{in}(\vec{p},\vec{x})=\frac{p}{2m}\left( p^2 + 4mg\rho_0(\vec{x}) \right)^{1/2}.
          \end{equation}
Eine Herleitung für Gl.(\ref{spek4neubitte}) ist in \cite{allaboutBEC} zu finden.
Für dieses Energiespektrum muss die Dichteschwankung lokal auf einen sehr kleinen Raum begrenzt sein,
was für Fluktuationen kleiner als die charakteristische Länge des Oszillators
$a_{HO}=({\hbar}/{m\omega})^{1/2}$ erfüllt ist.
Zusammen mit der  Unschärferelation ergibt $\Delta p \Delta x \geq \frac{\hbar}{2}$ für den Impuls der Welle
$p \gg {\hbar}/{a_{HO}}$.
Der Radius $R_0 \sim {2 k_B T}/{m \omega^2}$ einer thermischen Atomwolke hängt von der Temperatur ab.
Zusätzlich muss noch für den Impuls  $R_0 \gg a_{HO}$ gelten, und damit
$ T \gg {\hbar \omega}/{k_B}$.
Zu Anfang des Abschnitts wurde eine Temperatur nahezu $T \ll T_0$ vorausgesetzt.
Durch höhere Teilchenzahlen wächst $T_0$ an, wodurch beide Bedingungen erfüllt werden können.
Im  \textit{Thomas-Fermi-Limes} $Na/a_{HO}$ ist die kinetische Energie im Vergleich zur
Wechselwirkungsenergie innerhalb des Kondensats sehr klein.
Ohne den kinetischen Term $V_{ext}-\mu=g\rho_0(\vec{x})$
vereinfacht sich das Energiespektrum Gl.(\ref{spek4neubitte}) zu
          \begin{figure}[h]  \begin{center}
          \input{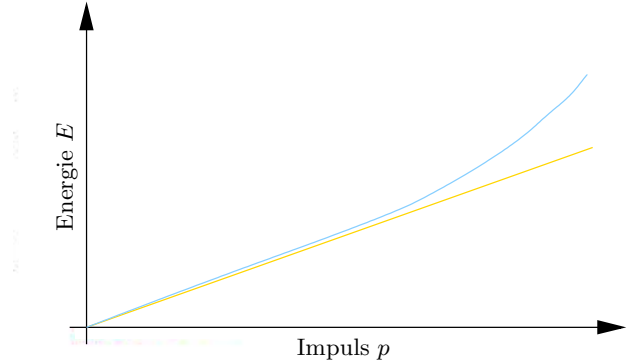}
          \caption[Anregungsspektrum]
          {\label{energie2}Die blaue Kurve stellt den Energieverlauf in Abhängigkeit vom Impuls dar.
          Die gelbe Gerade ist zur Orientierung eingezeichnet und zeigt, dass in zwei Bereiche einzuteilen ist.
          Für kleine Impulse ergibt sich ein lineare, damit ein phononenartiges Spektrum mit $E=pc$. Der quadratische Anstieg
          für größere Impulswerte lässt auf Teilchenanregung $E={p^2}/{2m}+E_0$ schließen.}
          \end{center}  \end{figure}
          \begin{equation}\label{spek4neubla}
                    E^{in}(\vec{p},\vec{x})=\frac{p}{2m}\left( p^2 + 4mg\rho_0(\vec{x}) \right)^{1/2}.
          \end{equation}
In Abb.(\ref{energie2}) ist das Energiespektrum in Abhängigkeit vom Impuls dargestellt.\\

\section{Kleine Anregungen im Kondensat}
Für kleine p-Werte lässt sich die Dispersionsrelation einfach angeben
$E=pc$
wobei
          \begin{equation} \label{schallgeschwindigkeit}
                    c=\sqrt{\frac{g \rho_0(\vec{x})}{m}}
          \end{equation}
die \textit{Schallgeschwindigkeit} ist.

\section{Experimentelle Beschreibung}
Ketterle et al. \cite{exc2} haben die Ausbreitung von Schall im Bose-Einsteinkondensat gemessen.
Im Versuch wurden $5 \times 10^6 $ Natrium-Atome abgekühlt und durch ein geeignetes Magnetfeld wurde
die Atomwolke zigarrenförmig angeordnet. Die Dichtestörung wurde im Zentrum des Kondensats durch einen
Laser erzeugt. Die Anregung propagierte dann sphärisch vom Zentrum nach außen, was direkt gemessen werden
konnte.
          \begin{figure}[t]
          \begin{center}
          \vspace*{-2cm}\input{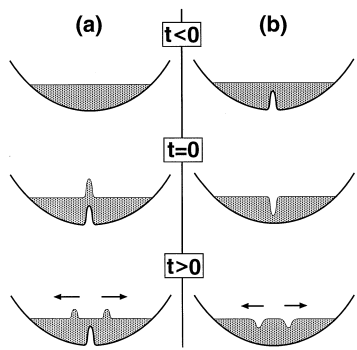}\vspace{-1cm}
          \caption[Schematische Darstellung der Schallgeschwindigkeit im Kondensat]
          {\label{schall}Ausbreitung einer kleinen Störung im Bose-Einstein-Kondensat. Zunächst wird
          das Kondensat in einer magnetischen Falle erzeugt. Zur Zeit $t=0$ wird ein Laser kurzzeitig
          an (a)- bzw. ausgeschaltet (b). Damit wird eine positive/ negative Dichtemodulation erzeugt,
          die mit der Schallgeschwindigkeit durch das Kondensat propagiert \cite{exc2}.}
          \end{center}
          \end{figure}

          \begin{figure}[h]
          \begin{center}
          \includegraphics*[width=11cm]{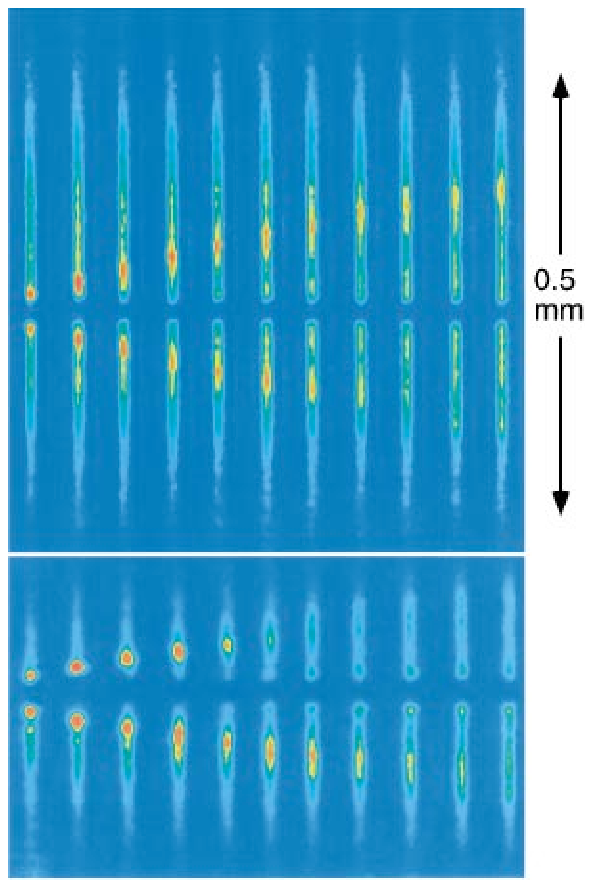}
          \caption[Messung der Schallgeschwindigkeit im Kondensat]
          {\label{schall2}Beobachtung der Schallwellen in einem Kondensat. Zur Messung wurde eine
          das Kondensat nicht zerstörende Phasen-Kontrast-Abbildung verwendet.
          Die einzelnen Plots zeigen das Kondensat im Abstand von $1.3\, \mu s$. In beiden Versuchen
          wurde das Kondensat mit einem Laserpuls - nach Anregung der Störung -
          in zwei Teile getrennt.
          Die Abbildungen zeigen
          zwei verschieden präparierte Kondensate: Die Länge der oberen Kondensatwolke ist $450\, \mu m$.
          In der unteren Sequenz ist der radikale Einschluss geringer als im Kondensat zuvor,
          was zu einem radialen Ausweiten der Dichte führt.
          Es ist deutlich zu sehen, dass die Schallgeschwindigkeit im unteren Bild langsamer ist als im oberen.
          Dort ist die Dichte, welche proportional zur Schallgeschwindigkeit ist, geringer \cite{exc2}.}
          \end{center}
          \end{figure}

                    \part{Einführung in die Allgemeine Relativitätstheorie\label{kap2}}
                              
\chapter{Eigenschaften schwarzer Löcher\label{blackholes}}

\section{Allgemeine Relativitätstheorie}

Die \textit{Allgemeine Relativitätstheorie} beschreibt gravitative Felder.
Die Dynamik ergibt sich aus der \textit{Einsteinschen Feldgleichung}
          \begin{equation} \label{einstein1}
                    G^{\alpha \beta}=-\kappa T^{\alpha \beta}.
          \end{equation}
Rechts ist der \textit{Energie-Impuls-Tensor} der Materie
          \begin{equation}
                    T^{\alpha \beta}=\frac{2}{\sqrt{-g}}\frac{\delta(\sqrt{-g}L_M)}{\delta g_{\alpha \beta}},
          \end{equation}
mit der Determinante der Metrik $g\equiv det(g_{\alpha \beta})$ und der \textit{Materie-Lagrangefunktion}
$L_M=L_M(g_{\alpha \beta}\vert \partial_{\gamma}g_{\alpha \beta}\vert \Psi^A\vert\Psi^A_{,\alpha})$ abgebildet.
$\Psi^A$ fasst alle Materievariablen zusammen.
Auf der linken Seite von Gl.(\ref{einstein1}) steht der \textit{Einstein-Tensor}
          \begin{equation} \label{einsteintensor}
                    G_{\alpha \beta}=R_{\alpha \beta}-\frac{1}{2}Rg _{\alpha \beta},
          \end{equation}
mit dem \textit{Riemannschen Krümmungstensor}
$R_{\alpha \beta}=\frac{1}{2}g^{\delta \gamma}(g_{\delta \gamma,\beta,\alpha}-g_{\delta \beta,\gamma,\alpha}+ g_{\alpha \beta,\delta,\gamma}-g_{\alpha \delta,\beta,\gamma}) $
und dem \textit{Ricci-Skalar} $R= R_{\alpha \beta}g^{\alpha \beta}$.
Die Konstante $ \kappa=\frac{8 \pi G}{c^4} $ beinhaltet die universielle \textit{Gravitationskonstante}
$G$ ist \cite{Goenner}.

Die Einstein-Gleichung ist eine Bestimmungsgleichung für die Metrik. Die Vorgehensweise ist,
zunächst die Materieverteilung zu bestimmen, in Gl.(\ref{einstein1}) einzusetzen und mit der
zweiten Einstein-Gleichung\footnote{Wobei das Semikolon die kovariante Ableitung nach $\beta$ bezeichnet.}
          \begin{equation}
                    T_{\alpha \beta;\beta}=0  
          \end{equation}
die Metrik zu ermitteln \cite{Inverno}.

Einer bestimmten Massenverteilung wird eine Metrik
          \begin{equation} \label{metrikallgemein}
                    g_{\alpha \beta}=\left( \begin{array}{cccc}
                                     g_{00} & g_{01} & g_{02} & g_{03} \\
                                     g_{01} & g_{11} & g_{12} & g_{13} \\
                                     g_{02} & g_{12} & g_{22} & g_{23} \\
                                     g_{03} & g_{13} & g_{23} & g_{33} \end{array} \right)
          \end{equation}
zugeordnet, die sich aus den Einsteinschen Feldgleichungen ergibt. Die Einträge in der
Metrik $g_{\alpha \beta}\equiv g_{\alpha \beta}(x)$ sind im Allgemeinen von der Zeit und dem Ort
abhängig\footnote{Raum und Zeit sind in der Relativitätstheorie zur Einheit verschmolzen und werden
zu einem \textit{Vierervektor} zusammengefasst. Die nullte Komponente des Vierervektors ist die Zeit,
die drei verbleibenden Einträge entsprechen den dreidimensionalen Ortskoordinaten.}.

Die Metrik ist ein symmetrischer Tensor zweiter Stufe, der es ermöglicht, den infinitesimalen Abstand $ds$
zweier benachbarter Punkte $x$ und $x+dx$ im gekrümmten Raum zu bestimmen:
          \begin{equation}\label{abstand}
                    ds^2=g_{\alpha \beta}(x)dx^{\alpha}dx^{\beta}.
          \end{equation}

Für den \textit{flachen Raum} sind die Einträge in Gl.(\ref{metrikallgemein}) unabhängig von der Raum-Zeit.
In unserem Universum kann der flache Raum durch die \textit{Minkowski-Metrik}
          \begin{equation} \label{metrikflach}
                    \eta_{\alpha \beta}=\left( \begin{array}{cccc}
                                     1 & 0 & 0 & 0 \\
                                     0 &-1 & 0 & 0 \\
                                     0 & 0 &-1 & 0 \\
                                     0 & 0 & 0 &-1 \end{array} \right)
          \end{equation}
beschrieben werden.
Das Linienelement für den flachen Raum ist
          \begin{equation}\label{abstandflach}
                    ds^2=c^2 (dx^0)^2-(dx^1)^2-(dx^2)^2-(dx^3)^2.
          \end{equation}
In dieser Arbeit sind zwei bestimmte Metriken von Interesse, welche im Folgenden vorgestellt werden.

\subsection{Die Metrik für ein schwarzes Loch}
Die exakte Lösung eines kugelsymmetrischen, statischen Himmelskörpers wurde zuerst von \textit{Schwarzschild}
berechnet.
Die Dynamik für Teilchen außerhalb einer solchen Materieverteilung ist in sphärischen Koordinaten
$(t,r,\theta,\varphi)$ gegeben durch \\
          \begin{equation} \label{schwarzschild}
                    {ds}^2= c^2 \left( 1-\frac{2m}{r} \right){dt}^2 
                    -\frac{dr^2}{\left( 1-\frac{2m}{r}\right)} - r^2 d \Omega^2,
          \end{equation}
mit dem  \textit{Raumwinkelelement} $d\Omega=sin(\theta)d\theta d\varphi$ und $m \equiv {GM}/{c^2}$, wobei $M$ die Gesamtmasse der
Materieverteilung ist.\\

In der Schwarzschild-Metrik treten zwei Besonderheiten auf.
Für $r=2m\equiv r_g$ verschwindet $g_{00}(r_{EH})=0$. $r_{EH}$ wird als \textit{Gravitationsradius} bzw.
\textit{Schwarzschildradius} bezeichnet.
Innerhalb $r<r_g$\footnote{Das Linienelement Gl.(\ref{schwarzschild})
ist nur für den Bereich $r_{EH}<r<\infty$ gültig.}
ist $g_{00}<0$ und damit der Tangentenvektor - ${1}/{c}\,{\partial_t}^2$ an die Weltlinie
$ds^2=c^2\left( 1-\frac{2m}{r} \right)dt^2$- raumartig,
für $g_{00}(r_g)=0$ lichtartig und ausserhalb $r>r_g$ ist $g_{00}>0$ zeitartig.\\
Allgemein wird eine Stelle im Raum, an welcher der Tangentenvektor sein Vorzeichen ändert als \textit{Ereignishorizont}
bezeichnet. Der Ereignishorizont ist eine zeitartige Hyperfläche, d.h. der Tangentenvektor am Ereignishorizont
ist parallel zur Zeit-Achse.
In Schwarzschild-Koordinaten wird $g_{11}=\frac{1}{\left( 1-\frac{2m}{r}\right)}$ am Ereignishorizont unendlich.
Ein sich dem Ereignishorizont radial
nähernder Körper mit der Geschwindigkeit
          \begin{equation} \nonumber
                    v^2\vert_{r_{EH}}=\lim_{r\rightarrow r_{EH}}\frac{ds^2}{dr^2}=
                    \frac{1}{\left( 1-\frac{2m}{r}\right)}\rightarrow 0
          \end{equation}
wird auf $v=0$ abgebremst. Damit ist es unmöglich, in ein schwarzes Loch einzutreten.
Eigentlich wird erwartet, dass durch die mit abnehmender Entfernung steigende gravitative Anziehungskraft
die Geschwindigkeit zu- und nicht abnimmt.
Bei der physikalischen Interpretation muss die Wahl des Koordinatensystems berücksichtigt werden.
Die Schwarzschild-Koordinaten entsprechen einem Bezugssystem außerhalb des schwarzen Lochs.
Von dort wird ein Körper - z.B. ein Raumschiff - beobachtet, wie dieses sich dem schwarzen Loch nähert.
Ein Beobachter am Ereignishorizont - mit einem anderen Bezugssystem - sieht das Raumschiff in endlicher Zeit den
Horizont passieren \cite{Goenner}. Bei $r=r_{EH}$ handelt es sich unter diesem Gesichtspunkt um keine Singularität
im eigentlichen Sinn, da sie mit einer geeigneten Koordinatentransformation zu beseitigen ist.
Am Punkt $r=0$ hingegen liegt eine echte (intrinsische) Singularität vor, die in allen Bezugssystemen zu finden ist.

\subsection{Die Metrik für das de-Sitter-Universum\label{sitter}}
Das \textit{de-Sitter-Modell} beschreibt unser Universum approximativ, indem die Gesamtmasse des Universums gegenueber dessen
Grösse verschwindend betrachtet wird. Das Universum kann daher als masselos und global flach betrachtet werde.
Es kann berechnet werden, wie sich ein solches Universum mit der Zeit verändert. Eine Herleitung
findet sich in \cite{Inverno}.
Es ergibt sich
          \begin{equation}\label{desitermetrik}
                    g_{\mu \nu}=\left(
                    \begin{array}{cccc}
                    1 & 0         & 0         & 0\\
                    0 & -\exp(2Ht) & 0         & 0\\
                    0 & 0         & -\exp(2Ht) & 0\\
                    0 & 0         & 0         & -\exp(2Ht)
                    \end{array}
                    \right)
          \end{equation}
die als \textit{de-Sitter-Metrik} bezeichnet wird.
Hier ist
          \begin{equation} \label{kosmologischekonstante}
                    H=\left( \frac{\Lambda}{3} \right)^{1/2},
          \end{equation}
mit der \textit{kosmologischen Konstanten} $\Lambda$.
Der allgemeine zeitabhängige Fall wird durch das \textit{Hubble-Gesetz}
          \begin{equation} \label{hubblegesetz}
                    H(t)=\frac{\dot{R}(t)}{R(t)},
          \end{equation}
beschrieben, wobei $H(t)$ als \textit{Hubbel-Konstante} bezeichntet wird. $R(t)$ ist ein Skalenfaktor.
Multipliziert mit einem, zum Zeitpunkt $t=0$ gemessenen Abstand, gibt er an, wie sich dieser mit
der Zeit ändert.\\
Im de-Sitter-Modell ist $\lambda>0$. Es erfährt eine exponentielle Expansion.
\footnote{Es wird noch darauf hingewiesen, dass es sich hier um das einfachste Modell für das
Universum handelt. Auch wenn die Dichte der Materie im Universum zum jetzigen Zeitpunkt
verschwindend ist, lässt sich diese Aussage nicht mehr aufrecht erhalten, wenn man lange genug in
der Zeit rückwärts wandert. Auch die Hintergrundstrahlung, die im de-Stitter-Modell berechnet werden
kann, entspricht nicht dem experimentell gemessenen Wert.}

\section{Feldquantisierung im Gravitationsfeld}
In diesem Abschnitt wird kurz erklärt, was unter dem Begriff \textit{Hawking-Strahlung} verstanden wird.
Eine ausführliche Diskussion findet sich in \cite{quantenfeld1}.\\

\begin{figure}
          \begin{center}
          \input{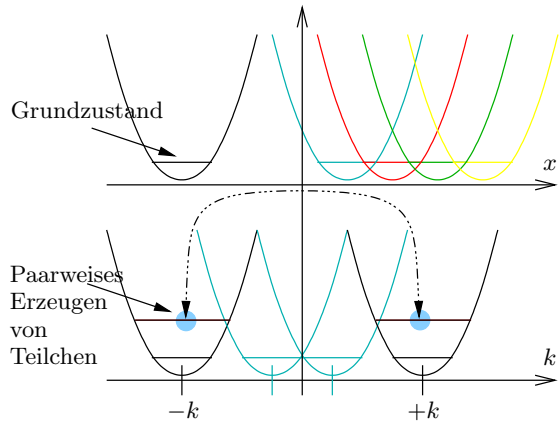}
          \caption[Gekoppelte Oszillatoren im Impulsraum]
          {\label{Fluktuationen}Die Entwicklung nach den Moden entspricht einem Wechsel vom Ortsraum (oberes Bild) in den
          Impulsraum (unteres Bild). Im Impulsraum sind die Moden mit $k$ und $-k$ miteinander gekoppelt, so dass
          nur paarweise Anregung von Moden möglich ist.}
          \end{center}
          \end{figure}

Bei der Quantisierung der Bewegungsgleichung eines massebehafteten Teilchens
im gekrümmten Raum nimmt der Hamilton-Operator die Gestalt unendlich vieler entkoppelter
Oszillatoren an. An jedem Punkt des Raumes kann man sich einen harmonischen Oszillator im Grundzustand
vorstellen. Ein Teilchen am Ort $x$ zu erzeugen, heißt den Grundzustand auf das nächstgelegene Energieniveau
anzuregen.
Ein Wechsel in den Impulsraum mittels Fourier-Transformation liefert einen
Hamilton-Operator, welcher durch die sog. \textit{Bogoliubov-Transformation} diagonalisierbar ist.
Es ergibt sich, dass nie ein einzelnes Teilchen angeregt werden kann. Es tritt Paarteilchenerzeugung
auf, wobei diese einen gleichgroß entgegengesetzten Impuls haben. Mit anderen Worten, der Gesamtimpuls des
Systems bleibt bei der Teilchenerzeugung erhalten (siehe Abb.(\ref{Fluktuationen})).
Ob Teilchen erzeugt werden oder nicht hängt vom Gravitationsfeld ab.

\subsection{Hawking-Strahlung}
Unter Hawking-Strahlung versteht man Paarteilchenerzeugung am Ereignishorizont eines schwarzen Lochs.
Es sind zwei physikalische Phänomene, die dafür verantwortlich sind.\\
Zum einen sagt die Unschärferelation $\Delta E  \cdot \Delta t\geq \hbar$ voraus,
dass überall im Raum Vakuumfluktuationen auftreten.
Darunter versteht man, dass ständig Teilchenpaare erzeugt werden, die sich nach einer gewissen Zeit
$\Delta t$ wieder vernichten.
Es sei denn, es existiert eine äussere Kraft, die stark genug ist, diese Teilchen voneinander
zu trennen. Am Ereignishorizont ist diese Bedingung erfüllt.
Ein Teilchen (mit Impuls $k$) wird als Hawking-Strahlung ins Weltall gesendet,
das andere (mit Impuls $-k$) wird in das schwarze Loch gezogen.
Dadurch verliert das schwarze Loch Energie, so dass die Teilchenerzeugung nicht ohne Aufwand betrieben werden
kann.\\

Eine wichtige Eigenschaft Schwarzer Löcher ist, dass sie dynamisch stabil
(siehe Abschnitt (\ref{forfluctiations})) sind.
Das bedeutet, dass wenn man bei den Betrachtungen komplexe Frequenzen zulässt,
die Zeitskala für das dynamische Anwachsen der dadurch produzierten Teilchen groß ist.
Schwarze Löcher werden bisher als weitgehend stabil betracht.

                    \part{Das Bose-Gas als Gravitationsmodell\label{BECART}}
                              \chapter{Die Wellengleichung für die Störung\label{wellengleichungignacio}}
Die Verbindung der Allgemeine Relativitätstheorie mit dem Bose-Einstein-Kondensat wird in diesem
und im nächsten Kapitel erläutert.\\
Im Folgenden wird für eine im Kondensat propagierende Störung die Wellengleichung
aufgestellt (siehe Kapitel (\ref{anregungenneu})) und gezeigt, dass es sich dabei um eine Wellengleichung
in einem effektiv\footnote{Der Begriff \textit{effektiv} weist darauf hin, dass es sich nicht
im eigentlichen Sinne um einen gekrümmten Raum handelt, weil dieser die Anwesenheit eines echten Gravitationsfeldes
fordern würde. Trotzdem lässt sich eine Analogie finden, da in den Gleichungen ein symmetrischer Tensor
zweiter Stufe auftaucht, der als effektive Metrik aufgefasst werden kann.
Die ganze Information des gekrümmten Raumes ist in der Metrik enthalten und damit ist es zulässig, von
einem effektiv gekrümmten Raum zu sprechen.}
gekrümmten Raum handelt.\\
Die Ergebnisse wurden bereits von Cirac et al. \cite{cirac1} publiziert. Es wird aber darauf hingewiesen,
dass hier der allgemeine Fall, dass die Kondensatgrössen $c=c(t,\vec{x})$ und $v=v(t,\vec{x})$ vom Ort und
der Zeit abhängen. In der zitierten Veröffentlichung waren die Schall - und Hintergrundgeschwindigkeit nur
vom Ort abhängig.
\footnote{In Kapitel (\ref{expandinguniverse}) wird ein zusätzlich von der Zeit abhängiges System
verwendet.}

\section{Die hydrodynamischen Gleichungen}
Ausgehend von der zeitunabhängigen Gross-Pitaevskii-Gl.(\ref{GPE2}) wird eine kleine Variation in der
Dichte $\rho=\rho_0 + \varepsilon\rho_1 $ und in der Phase $\theta=\theta_0 +  \varepsilon \theta_1 $
          \footnote{Das $\varepsilon$ zeigt die Ordnung an,
                    wir betrachten damit eine Entwicklung bis zu ersten Ordnung.}
in den Zustand $\Phi=\sqrt{\rho}e^{i\theta}$ eingesetzt
          \begin{equation} \label{GPE2mitdelta}
                    \begin{split}
                    i\hbar\partial_t \sqrt{\rho_0 + \varepsilon\rho_1}e^{i(\theta_0 +  \varepsilon \theta_1 )}
                    =&\left( \kin + V_{ext}(\vec{x}) \right) \sqrt{\rho_0 + \varepsilon\rho_1}e^{i(\theta_0 +  \varepsilon \theta_1 )}\\
                    &+  N_0 g(\rho_0 + \varepsilon\rho_1)
                    \sqrt{\rho_0 + \varepsilon\rho_1}e^{i(\theta_0 +  \varepsilon \theta_1 )},
                    \end{split}
          \end{equation}
wobei $\rho_1$ und $\theta_1$ die Störung beschreiben und das chemische Potential $\mu$ in $V_{ext}$ enthalten ist.
Im Anhang (\ref{anhanghydro}) ist Gl.(\ref{GPE2mitdelta}) ausgerechnet worden.
Es ergaben sich für die Störung:
          \begin{equation} \nonumber
                    \begin{split}
                    (\ref{hamjakobi1}):\dot{\theta_1}=&-\frac{\hbar}{m}(\nabla \theta_0 \nabla \theta_1) - \frac{N_0 g}{\hbar} \rho_1 \\
                    (\ref{kontgl1}):\dot{\rho_1}=&-\frac{\hbar}{m}\nabla(\rho_0 \nabla \theta_1 + \rho_1 \nabla \theta_0)
                    \end{split}.
          \end{equation}

\subsection{Die Wellengleichung}
Gl.(\ref{hamjakobi1}) wird nach $\rho_1$ aufgelöst
          \begin{equation}  \nonumber
                    N_0g\,\rho_1=-(v \nabla \theta_1)-\dot{\theta}_1
          \end{equation}
und anschließend nach der Zeit abgeleitet
          \begin{equation}  \nonumber
                    N_0g\,\dot{\rho}_1=-\partial_t(v \nabla \theta_1)-\ddot{\theta}_1.
          \end{equation}
Eingesetzt in Gl.(\ref{kontgl1}) liefert
          \begin{equation}  \label{wellengl1}
                    -\ddot{\theta}_1-\partial_t(v \nabla \theta_1)-\nabla(v\dot{\theta}_1)
                    +\nabla\left((c^2-v^2)\nabla\theta_1\right)=0,
          \end{equation}
eine Differentialgleichung für die Phase der Störung.\\

Dass es sich dabei um eine Wellengleichung handelt, wird sichtbar, wenn
          \begin{equation} \label{mkontra}
                    g^{\mu \nu}=\frac{-1}{c^2}\left( \begin{array}{cccc}
                    1            & v_{x}                  & v_{y}            & v_{z} \\
                    v_{x}          & -c^2+v_{x}^2         & v_{y}v_{x}     & v_{z}v_{x} \\
                    v_{y}          & v_{x}v_{y}           & -c^2+v_{y}^2   & v_{z}v_{y}  \\
                    v_{z}          & v_{x}v_{z}           & v_{y}v_{z}     & -c^2+v_{z}^2
                    \end{array} \right)
          \end{equation}
eingeführt wird.
Jetzt lässt sich Gl.(\ref{wellengl1}) schreiben als:

          \begin{equation} \label{wellengl2}
                    \partial_{\mu}(\sqrt{-g} \, g^{\mu\nu} \, \partial_{\nu}\theta_{1})=0
          \end{equation}
Formal handelt es sich um eine klassische Feldgleichung für ein masseloses Teilchen $\theta_1$ in einem
gekrümmten Raum.
Die Rolle des gekrümmten Raumes nimmt dabei das Kondensat ein, weil sämtliche Einträge in der Metrik
aus den Größen des Kondensats bestehen.
Es tritt die Schallgeschwindigkeit $c$ für die Dichtefluktuationen aus dem vorhergehenden Abschnitt
sowie die stationäre Hintergrundgeschwindigkeit $v$ auf.\\

Für ein verschwindendes $v$, wird Gl.(\ref{mkontra})
          \begin{equation} \label{mkontraflach}
                    g^{\mu \nu}=c \frac{-1}{c^2}\left( \begin{array}{cccc}
                    1            & 0                 & 0           & 0 \\
                    0          & -c^2        & 0     & 0 \\
                    0          & 0           & -c^2  & 0  \\
                    0          & 0         & 0     & -c^2
                    \end{array} \right),
          \end{equation}
zur Minkowski-Metrik, welche den flachen Raum beschreibt. Die Minkowski-Metrik für den flachen Raum, eingesetzt
in Gl.(\ref{wellengl2}), liefert die gewöhnliche Wellengleichung:
          \begin{equation} \nonumber
                    \left(c^2\partial_t^2 - \nabla^2\right)\theta_1=0
          \end{equation}

\section{Das effektive Schwarze Loch}
Mit $g_{\mu\nu}g^{\mu\nu}=\mathtt{1}$ kann die kovariante Metrik zu Gl.(\ref{mkontra}) berechnet werden.
Es ergibt sich:
          \begin{equation} \label{mkon}
                    g_{\mu \nu}=c\left( \begin{array}{cccc}
                    -(c^2-v_{0}^2)    & -v_{x}                  & -v_{y}            & -v_{z} \\
                    -v_{x}          & 1        & 0     & 0 \\
                    -v_{y}          & 0           & 1   & 0  \\
                    -v_{z}          & 0           & 0     & 1
                    \end{array} \right).
          \end{equation}
Es ist leicht zu überprüfen, ob in einem System mit dieser effektiven Metrik ein schwarzes Loch auftritt,
oder nicht. Aus Kapitel (\ref{blackholes}) ist bekannt, dass ein Ereignishorizont vorliegt, wenn an einem
Punkt $g_{00}=0$ wird. Hier ist $g_{00}=-c^2+v$, damit ist es vom System abhängig, ob ein Ereignishorizont
auftritt. Der Bereich, in dem die Hintergrundgeschwindkeit schneller fließt als
die Schallwellen sich ausbreiten können, kann mit dem Inneren eines schwarzen Lochs in Verbindung gebracht
werden. Dort können die Schallwellen nur mit der Richtung des Kondensats propagieren. Außerhalb, im
Bereich wo die Hintergrundgeschwidikgeit kleiner als die Schallgeschwindigkeit ist, kann sich der
Schall in beide Richtungen bewegen.

                              \chapter{Die Fluktuationen um den Grundzustand\label{fluktuationsaroundground}}
Die Untersuchung der Zeitentwicklung von Fluktuationen um den Grundzustand ist Bestandteil dieses
Kapitels.\\

\section{Der Bogoliubov-Ansatz bis zur ersten Ordnung}
Der Ansatz von Bogoliubov $\bog$ beschreibt ein System, das sich bis auf kleine Fluktuationen
im Grundzustand befindet. Bisher wurde die erste Ordnung nicht berücksichtigt.
Jetzt ist gerade diese von Interesse.
Dazu greifen wir die Rechnungen aus (\ref{meanfield}) auf. Anstatt der Mean-Field-Näherung
wird als nächstes $\hat{K}$ in die Heisenberg-Relation
$$ \zeit \opp{\Psi}{x}=[\opp{\Psi}{x},\hat{K}], $$
eingesetzt. Berücksichtigt man zusätzlich die Vertauschungsrelationen der Feldoperatoren für Bosoen
          \begin{eqnarray} \label{vertauschrel}
                    \left[ \hat{\psi}(\vec{x}),\hat{\psi}(\vec{x}\,')\right]=0,
                    &  \left[ \hat{\psi}^{\dag}(\vec{x}),\hat{\psi}^{\dag}(\vec{x}\,')\right]=0, &
                    \left[ \hat{\psi}(\vec{x}),\hat{\psi}^{\dag}(\vec{x}\,')\right]
                    =\delta^{(3)}(\vec{x}-\vec{x}\,')
          \end{eqnarray}
ergibt sich in erster Ordnung von $\varepsilon$ die zeitabhängige Gross-Pitaevskii-Gl.(\ref{GPE2}).
Diese verschwindet, da Fluktuationen um den Grundzustand betrachtet werden und
damit die klassische Wellenfunktion im Bogoliubov-Ansatz gerade dem Grundzustand $\Phi_s$ der
der makroskopischen Wellenfunktion entspricht.
Die Terme dritter und vierter Ordnung können vernachlässigt werden,
so dass nur Ausdrücke mit $\varepsilon^2$ zurück bleiben.
\footnote{Die Terme ohne $\varepsilon$ können als
          Konstante zusammengefasst werden, sind damit für weitere Betrachtungen uninteressant.}
Die zeitabhängige Bewegungsgleichung für die Fluktuationen ist
          \begin{equation} \label{GPme1}
                    \zeit\hat{\psi}=\left[ -\frac{\hbar^2}{2m}\nabla^2 + V_{ext} + 2Ng\rho_{s} \right] \hat{\psi} +
                    Ng\Phi_{s}^{2}\hat{\psi}^{\dag}.
          \end{equation}
Im Folgenden werden die Gleichungen auf eine Dimension $x$ beschränkt, da später nur solche Systeme betrachtet
werden.
Für weitere Berechnungen ist es zusätzlich erforderlich, die Größe $\xi\equiv \varepsilon^{-1}$
in die $x$-Koordinate mit
          \begin{equation}  \label{xepsilon}
                    x\rightarrow \frac{x}{\varepsilon}
          \end{equation}
aufzunehmen.\footnote{Der Sinn dieser Tranformation zeigt sich später, wenn die WKB-Näherung angewandt wurde.}

Mit $x$ ändert sich Gl.(\ref{GPme1}) zu
          \begin{equation} \label{GPme3}
                    \zeit\hat{\psi}=\left[ -\frac{\epsilon^2 \hbar^2}{2m}\partial_x^2 + \frac{\epsilon^2 \hbar^2}{2m}\frac{\sqrt{\rho_{s}}''}{\sqrt{\rho_{s}}}-\frac{m}{2}v_{s}^2 + Ng\rho_{s} \right] \hat{\psi} + Ng \Phi_s^2 \hat{\psi}^{\dag}
          \end{equation}
wobei hier zusätzlich das externe Potential
          \begin{equation}  \label{extV2}
                    V_{ext}(x)=\frac{\epsilon^2\hbar^2}{2m}\frac{\sqrt{\rho_s}''}{\sqrt{\rho_s}}
                    -\frac{m}{2} v_s^2- Ng \rho_s.
          \end{equation}
in die neuen Koordinaten eingesetzt wurde.

\subsection{Die Phase der Störung}
Nehmen wir an, der Zustand sei
          \begin{equation} \label{mad1}
                    \hat\Psi(t,x)=\sqrt{\rho_s(x)}e^{i\frac{m}{\hbar}\int^x v_s(x')dx'}+\varepsilon\hat\eta(t,x) e^{i\frac{m}{\hbar}\int^x v_s(x')dx'},
          \end{equation}
ausnützend, dass für eine Dimension
          \begin{eqnarray} \label{1dphase}
                    \vec{v}_s= \frac{\hbar}{m}\partial_x \theta_0 & \rightarrow
                    &\theta_0(x)=\frac{m}{\hbar} \int^x dx' v_{s}(x')
          \end{eqnarray}
gilt.
Die Phase von $\hat{\eta}$ gibt die Realtivbewegung der Störung zum Kondensat an.\\

Bevor dieser Ansatz in Gl.(\ref{GPme3}) eingesetzt werden kann, wird er durch die neuen Koordinaten (Gl.(\ref{xepsilon}))
          \begin{equation} \label{mad2}
                    \hat\Psi(t,x)=\sqrt{\rho_s(x)}e^{i \frac{m}{\hbar} \int^{x} v_s(x' )dx'}+\varepsilon\hat\eta(t,x) e^{i\epsilon\frac{m}{\hbar}\int^{x} v_s(x')dx'},
          \end{equation}
ausgedrückt.\\
Mit $\hat{\psi}=\varepsilon\hat\eta(t,x) e^{i\frac{\varepsilon m}{\hbar}\int^x v_s(x')dx'}$ erhält man somit eine Gleichung
          \begin{equation}  \label{Gleta}
                    i\hbar\dot{\hat\eta}=Ng\rho_s\left(\hat\eta+\hat\eta^{\dag} \right) -i\epsilon \left( 2v_s \hat\eta'+\hat\eta v_s' \right)
                    -\frac{\epsilon^2\hbar^2}{2m} \left( \hat\eta''-\frac{\sqrt{\rho_s}''}{\sqrt{\rho_s}} \right),
          \end{equation}
die nur noch vom Feldoperator $\hat{\eta}$ abhängt. Setzt man in Gl.(\ref{Gleta}) eine Superposition zweier
Feldoperatoren ($a\hat{\eta}_2+b\hat{\eta}_b$) ein, so muss jeder für sich diese erfüllen.
Es liegt im Ortsraum keine Koppelung zwischen den Fluktuationen vor. Es bleibt zu untersuchen,
wie die Situation im Impulsraum ist.

\subsubsection{Die Fourier-Entwicklung der Störung über einfache ebene-Wellen-Lösungen\label{fourierebenewellen}}
Liegt der Feldoperator $\hat{\psi}(\vec{x},t)$ vor,
ermöglicht die diskrete Fourier-Analyse
          \begin{equation} \label{fanalyse}
                    \hat{\psi}_{k}(t)=\frac{1}{\sqrt{V}} \int_{-L/2}^{+L/2} e^{-ikx}\hat{\psi}(\vec{x},t)
          \end{equation}
und deren Rücktransformation
          \begin{equation} \label{inversfanalyse}
                    \hat{\psi}_{x}(t)=\frac{1}{\sqrt{V}} \sum_{k=\frac{2\pi}{L}n} e^{ikx} \hat{\psi}_{k}(t)
                    \end{equation}
vom Ortsraum in den Impulsraum zu wechseln und umgekehrt, wobei $n \in \mathtt{Z}$ ist.\\

Die Operatoren $\hat{\psi}_{k}(t)$ im $k$-Raum können in
Erzeugungs - und Vernichtungsoperatoren umgeschrieben werden:
          \begin{equation}  \label{psik}
                    \hat{\psi}_{k}(t)=\bar{u}_k(x) \hat{a}_{k}(t)+\bar{v}^*_{k'}(x) \hat{a}^{\dag}_{-k}(t) .
          \end{equation}
Im Heisenberg-Bild
          \begin{equation}  \label{akt}
                    \hat{a}_{k}(t)= \hat{a}_{k}e^{-i\omega_k t}
          \end{equation}
lässt sich der Feldoperator umschreiben in
          \begin{equation} \label{allfanalyse}
                    \hat{\psi}_{x}(t)= \sum_k \left[\bar{u}_k(x) \hat{a}_{k}e^{-i\omega_k t}+
                    \bar{v}^*_{k}(x) \hat{a}^{\dag}_{k}e^{i\omega^*_{k} t}\right].
          \end{equation}
Mit dieser Entwicklung wird Gl.(\ref{GPme1}) diagonalisiert, wobei die Eigenfrequenzen $\omega_k$
der Energie eines Quasiteilchens zugeordnet werden können.
Dieses Verfahren wird als \textit{Bogoliubov-Transformation} bezeichnet \cite{BECdalfovo}.

Die Amplituden und Wellenzahlen hängen hier vom Ort ab:
          \begin{equation}  \label{uvfanalyse}
                    \begin{array} {c}
                    \bar{u}_k (x)=u(x)e^{\frac{i}{\varepsilon}\int_x dx' k(x')+i\frac{m}{\hbar} \int^x dx' v_{s}(x')} \\
                    \bar{v}_k (x)=v(x)e^{\frac{i}{\varepsilon}\int_x dx' k(x')-i\frac{m}{\hbar} \int^x dx' v_{s}(x')}
                    \end{array}.
          \end{equation}
Erzeuger und Vernichter müssen die Vertauschungsrelation
          \begin{equation}  \label{verta}
                    \left[ \hat{a}_k,\hat{a}_{k'}^{\dag}  \right]=\delta_{kk'}
          \end{equation}
erfüllen. Es ergibt sich, dass für ein Paar ($u_{k'},v_{k'}$)
immer eine dazu redundante Lösung ($v^*_{k},u^*_{k}$) mit den Frequenzen $\omega_{k'}=-\omega_{k}^*$
existiert.

Die Entwicklung nach ebenen Wellen wird im nächsten Unterabschnitt in Gl.({\ref{GPme3}}) eingesetzt.
Es ergibt sich jeweils eine Gleichung für die Erzeugung - bzw. Vernichtungsoperatoren, beide zusammen werden als
\textit{Bogoliubov-Gleichungen} bezeichnet.

\subsection{Die Bogoliubov-Gleichungen in einer Dimension}
Die Bogoliubov-Gleichung in ihrer bekanntesten Form ist\\
          \begin{equation}  \label{bGleichung1}
                    \hbar\omega_{j}
                    \left( \begin{array}{cc}  \bar{u}_{j}\\ \bar{v}_{j}
                    \end{array} \right)
                    =
                    \left( \begin{array}{cc}
                    h_{0}(x)            & mc(x)^2 e^{2i\int^x dx' v_{s}(x')} \\
                    -mc(x)^2 e^{-2i\int^x dx' v_{s}(x')}          &  -h_{0}(x)    \\
                    \end{array} \right)
                    \left( \begin{array}{cc}  \bar{u}_{j}\\ \bar{v}_{j}
                    \end{array} \right) ,
          \end{equation}
mit
          $$ h_{0}(x)=\frac{\epsilon^2 \hbar^2}{2m}\nabla^2 + V_{ext} + 2 Ng\rho_{s}$$
und
          $$  \bar{u}_j(x)=u(x)e^{\frac{im}{\epsilon\hbar}\int^x dx' v_s(x')}e^{\frac{i}{\epsilon}\int^x dx' k(x')}  $$
          $$  \bar{v}_j(x)=v(x)e^{-\frac{im}{\epsilon\hbar}\int^x dx' v_s(x')}e^{\frac{i}{\epsilon}\int^x dx' k(x')}.$$

Eingesetzt in die Bogoliubov-Gl.(\ref{bGleichung1}), ergibt sich
          \begin{equation} \label{bognewallgemein}
                    \begin{split}
                    h_{+}u + Ng\rho_{s}v-i\epsilon \left[ \left( \frac{\hbar^2}{2m}k+\hbar  v_{0}  \right)u''
                    +  \frac{\hbar}{2} \left(\frac{\hbar}{m}k''+ v_s(x)'' \right)u \right]&=0  \\
                    h_{-}v + Ng\rho_{s}u-i\epsilon \left[ \left( \frac{\hbar^2}{2m}k-\hbar  v_{0} \right)v''
                    +  \frac{\hbar}{2} \left(\frac{\hbar}{m}k''-v_s(x)'' \right)v \right]&=0
                    \end{split}
          \end{equation}
wobei
          \begin{equation} \label{h+-}
                    h_{\pm}=\frac{\hbar^2}{m}\frac{k^2}{2}+Ng\rho_{s}\pm \hbar \left(  k v_{0} + \omega_{j} \right).
          \end{equation}
eingeführt wurde.
Mit diesen Gleichungen ist es möglich, die Dispersionsrelation $\omega(k)$
und die Amplituden der erzeugten Moden zu berechnen.\\

\section{Die Zeitentwicklung der Fluktuationen\label{forfluctiations}}

Ergeben sich aus den Bogoliubov-Gleichungen komplex imaginäre oder negativ reelle Eigenfrequenzen,
wird das Kondensat instabil.

\subsection{Dynamische Instabilitäten}
Die Zeitentwicklung ist durch Gl.(\ref{akt}) gegeben.
Für komplexe Frequenzen mit negativem Imaginärteil entwickeln sich die Moden exponentiell mit der
Zeit ($exp(\omega_i t)$). Der Erwartungswert, dass Moden an einem bestimmten Ort erzeugt werden, nimmt
ebenfalls mit der Zeit exponentiell zu ($exp(2\omega_i t)$). Diese Zunahme erfolgt sehr schnell.
Die Fluktuationen können nicht mehr mit den linearisierten Gl.(\ref{GPme1}) beschrieben werden.\\

\subsection{Energetische Instabilitäten}
Für $\omega<0$ befindet sich das System nicht im Grundzustand. In einem solchen System können
eventuell Moden erzeugt werden, die Energie aus dem Kondensat befördern, wodurch der Zustand des Systems
verändert wird.

                    \part{Schwarze Löcher im Bose-Einstein-Kondensat\label{kapspeziell}}
                              
\chapter{Zigarrenförmige Kondensate mit Senke}
Es wird ein eindimensionales Kondensat untersucht, dessen Bewegung auf die $x$-Achse eingeschränkt ist.
Die Kondensatwolke nimmt eine zigarrenförmige Form mit der Länge $2D$ an und wird zusätzlich
symmetrisch zum Ursprung an zwei Stellen eingeschnürt. Dadurch ist im Bereich $2L$ der Querschnitt der
Kondensatwolke dünner als außerhalb.
Diese Konfiguration wurde im Artikel von Cirac et al. \cite{cirac1} bereits behandelt. Die Aufgabe dieses
Kapitels ist es, die dort präsentierten Ergebnisse eigenständig zu berechnen.

\section{Das Modell}

\begin{figure}[t]
          \begin{center}
          \input{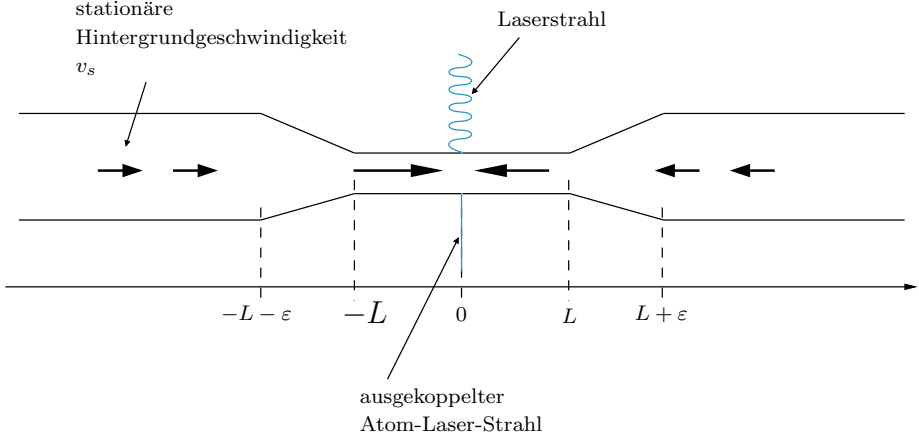}
          \caption[Zigarrenförmiges Dichteprofil]
          {\label{zigarrenform}Schematische Darstellung des Dichteprofils eines zigarrenförmigen Kondensats. Die Pfeile
          zeigen die stationäre Hintergrundgeschwindigkeit des Kondensats $v=\hbar/m \nabla\rho_0$ an.
          Für Werte $\vert x\vert>L+\epsilon$ und $\vert x\vert<L$ ist die Fließgeschwindigkeit konstant,
          wobei diese Konstante im letzteren Bereich höher ist. Dieser Sachverhalt ist durch die Länge der Pfeile angedeutet. In den
          beiden Übergangsbereichen $L< \vert x \vert <L + \epsilon$ verengt sich das Dichteprofil,
          so dass es, um die Kontinuitätsgleichung zu erfüllen, schneller fließen muss. In der Mitte bei $x=0$ treffen die beiden
          Ströme des Kondensats aufeinander, die von einem off-resonanten Laserstrahl aus dem Kondensat geleitet werden.}
          \end{center}
          \end{figure}

In Abb.(\ref{zigarrenform}) ist das Dichteprofil der Kondensatwolke skizziert.
Die Pfeile zeigen die Richtung der Hintergrundgeschwindigkeit an. Es sind zwei Ströme eingezeichnet,
die sich am Koordinatenursprung treffen. Es ist aus Gründen der Kontinuität erforderlich,
an dieser Stelle einen Laser anzubringen, der eine Senke bei ($x=0$) erzeugt.
Das Geschwindigkeitsprofil
- der Gradient der Phase - soll innerhalb $\vert x \vert <L$ um einen Faktor $\sigma>1$ größer
sein, als außerhalb $\vert x \vert> L+\epsilon $.
Für ein stationäres Problem gilt mit Gl.(\ref{kontgl0}), dass der Fluss pro Volumeneinheit und Zeit konstant
sein muss, und damit
          \begin{equation} \label{v0c0}
                    c(x)^2 v(x)=const.
          \end{equation}
Für die Schallgeschindigkeit wird
          \begin{equation} \label{csink}
                    c(x)=\left\{\begin{array}{ll}
                    c_{0},  & \vert x \vert < L \\
                    c_{0}[1+(\sigma-1)x/ \varepsilon] , & L < \vert x \vert < L+\varepsilon \\
                    \sigma c_{0}, & L+\varepsilon <  \vert x \vert
                    \end{array}  \right.
          \end{equation}
gewählt und liefert mit
          \begin{equation}  \label{vsink}
                    v(x)=-\frac{v_0 c_0^2}{c(x)^2}\frac{x}{\vert x \vert}.
          \end{equation}
die gewünschte Verteilung für die stationäre Hintergrundgeschwindigkeit.

\subsection{Die Gross-Pitaevskii-Gleichung für das zigarrenförmige Kondensat}
Das Kondensat kann durch die zeitabhängige Gross-Pitaevskii-Gleichung
          \begin{equation} \label{GPEsink}
                    \zeit \hat{\Psi}=\frac{\hbar^2}{2m}\partial{x}^2 \hat{\Psi}+(N_0g\vert\hat{\Psi}\vert^{2}
                    +V_{ext}(x)-i v_{0}\delta(x))\hat{\Psi}
          \end{equation}
beschrieben werden, wobei es erforderlich ist, den Zusatzterm $-i v_{0}\delta(x) $ einzuführen,
der die Wirkung des Lasers an der Stelle $x=0$ beschreibt.
Dort wird die Flüssigkeit mit der Geschwindigkeit $v_0$ aus dem Kondensat entfernt.

Das Potential
          \begin{equation} \label{estVsinkneu}
                    V_{ext}(\vec{x})=\frac{\hbar^2}{2m}\frac{c(x)''}{c(x)}
                    -\frac{m}{2} v^2- m c(x)^2 .
          \end{equation}
ergibt sich durch Einsetzen des Grundzustandes $\Phi_0$ in Gl.(\ref{GPEsink}).

\subsection{Stetigkeitsbedingungen am Punkt $x=0$}
Das Kondensat ist durch die Senke an der Stelle $x=0$ in zwei Teile getrennt. Die Lösungen müssen
stetig ineinander übergehen
          \begin{equation}  \label{b1}
                    \hat{\Psi}(0^{+},t)-\hat{\Psi}(0^{-},t)=0.
                    \end{equation}
Es ist $ \hat\Psi(0^{\pm},t)=\lim_{\varepsilon \rightarrow 0}\hat\Psi(\pm \varepsilon,t) $.

Eine Bedingung für die erste Ableitung ergibt sich durch Integration\footnote{Hier wurde zusätzlich für beschränkte $f(x)$  verwendet:\\
$$ \lim_{\varepsilon \rightarrow 0} \int_{0}^{x=0 \pm \varepsilon}f(x')dx'=\pm \lim_{\varepsilon \rightarrow 0} f(0)\varepsilon=0  $$
$$ \lim_{\varepsilon \rightarrow 0} \int_{0}^{x=0 \pm \varepsilon}\delta(x')f(x')dx'=\pm f(0)  $$}
von Gl.(\ref{GPEsink}):
          \begin{equation}  \label{b2}
                    \hat{\Psi}^{\prime}(0^{+},t)-\hat{\Psi}^{\prime}(0^{-},t)=-\frac{2im}{\hbar^2}\hat\Psi(0) .
          \end{equation}

\section{Die Wellengleichung\label{wellengleichung}}
Die Dynamik einer modulierten Dichtestörung wurde in Kapitel (\ref{wellengleichungignacio}) hergeleitet. In diesem Abschnitt
werden die Rechnungen auf das zigarrenförmige Kondensat angewendet.
Für ein symmetrisches System reicht es aus, das Verhalten für $x>0$ zu untersuchen.
Am Punkt $x=0$ werden die rechten und linken Lösungen stetig verbunden.

\subsection{Die Metrik für das zigarrenförmige Kondensat}
Zur Berechnung der Wellenfunktion kann $\hat{\Psi}\approx \Phi$ gesetzt werden,
was bereits in Kapitel (\ref{vielteil}) berechnet wurde.
Für eine Dimension wird Gl.(\ref{GPE2}) zu
          \begin{equation} \label{GPE21Dim}
                    i\hbar\partial_t \Phi(t,x)=\left( \frac{\hbar^2}{2m}\partial_x^2 + V_{ext}(x) + N_0 g \vert \Phi(x) \vert^2 \right) \Phi(t,x),
          \end{equation}
wobei der Term für die Senke nicht auftaucht, da nur Werte $x>0$ betrachtet werden. Das chemische Potential
$\mu$ ist in $V_{ext}$ enthalten.

Induziert man von außen eine lokale Dichteschwankung, breitet sich diese wie eine Schallwelle
in einem Superfluid aus. Die Bewegungsgleichungen für die Störung ergeben sich, wenn
in Gl.(\ref{GPEsink}) für die Dichte $\rho=\rho_0 +\varepsilon \rho_1 $ und für die Phase
$\theta=\theta_0 +\varepsilon  \theta_1 $ eingesetzt wird.
Die Rechnungen entsprechen denen aus Kapitel (\ref{wellengleichungignacio}).
Um Wiederholungen zu vermeiden, werden im Folgenden nur die Ergebnisse präsentiert.
In erster Ordnung erhält man für den Realteil
          \begin{equation} \label{dyn2a1dim}
                    \dot{\rho_0}=-\frac{\hbar}{m}\partial_x(\rho_0 \partial_x \theta_0),
          \end{equation}
und den Imaginärteil
          \begin{equation} \label{dyn2b1dim}
                    \dot{\rho_1}=-\frac{\hbar}{m}\partial_{x} (\rho_0 \partial_{x} \theta_1 + \rho_1 \partial_{x} \theta_0)
          \end{equation}
Diese beiden Gleichungen lassen sich zusammenfassen zu einer Wellengleichung
          \begin{equation} \label{wellengl2zig1D}
                    \partial_{\mu}(\sqrt{-g}g^{\mu\nu}\partial_{\nu}\theta_{1})=0,
          \end{equation}
mit der effektiven Metrik
          \begin{equation} \label{mkontrazig1D}
                    g^{\mu \nu}=\frac{-1}{c^2}\left( \begin{array}{cccc}
                    1            & v_{x}               & 0              & 0\\
                    v_{x}       & -c^2+v_{x}^2        & 0              & 0\\
                    0            & 0                    & -c^2           & 0 \\
                    0            & 0                    & 0              & -c^2
                    \end{array} \right).
                    \end{equation}

Damit ist gezeigt, dass sich die Dichteschwankungen analog zu masselosen Teilchen in Anwesenheit
eines äußeren Gravitationsfeldes ausbreiten. Als nächstes wird untersucht, welchem Gravitationsfeld
die effektive Metrik entspricht.

\subsection{Die kovariante Metrik}
Ein Umschreiben der Metrik (\ref{mkontrazig1D}) in ihre kovariante Form\footnote{Es gilt,
$g^{\mu \nu}g_{\mu \nu}=\mathsf{1}$.} ergibt
          \begin{equation} \label{mkonzig1D}
                    g_{\mu \nu}=c\left( \begin{array}{cccc}
                    -c^2+v_{x}^2           & -v_x                  & 0              & 0\\
                    -v_x                    & 1                    & 0              & 0\\
                    0                      & 0                    & 1              & 0 \\
                    0                      & 0                    & 0              & 1
                    \end{array} \right),
                    \end{equation}
mit $g_{00}=-c^2+v_{x}^2$.

Ein schwarzes Loch liegt vor (siehe Kapitel \ref{blackholes}), wenn
sich das Vorzeichen von $g_{00}$ ändert. Der Punkt mit $g_{00}=0$ wird als Ereignishorizont bezeichnet.
Damit kann überprüft werden, ob die effektive Metrik mit der eines schwarzen Lochs vergleichbar ist.

\subsection{Der Ereignishorizont}

\paragraph{$g_{00}>0$:}\hfill\parbox[t]{13.5cm}{
Nehmen wir an, für $x<L$ sei $v(x)^2>c(x)^2$, dann gilt mit Gl.(\ref{csink}) und Gl.(\ref{vsink}),
dass $v_0^2>c_0^2$, wobei $v_0={\hbar}/{m}\,\partial_x \theta_0(x)$ und $c_0={\hbar}/{m}\,\sqrt{4\pi a \rho_{0}(x)}$ sind.
Da beide Geschwindigkeiten in diesen Bereichen konstant sind, kann man mit einer Konstante $s>1$ schreiben,
dass $v_0^2=sc_0^2$.}

\paragraph{$g_{00}<0$:}\hfill\parbox[t]{13.5cm}{
Es muss möglich sein, die vorherige Bedingung und gleichzeitig $v(x)^2<c(x)^2$ für
$x>L+\epsilon$ zu erfüllen.
In diesem Bereich ist $v(x)^2={v_0^2}/{\sigma^4}$ und $c(x)^2=\sigma^2 c_0^2$. Damit kann
man erneut $s={v(x)^2}/{c(x)^2}={v_0^2}/({c_0^2\, \sigma^6})$ bestimmen.
Für $s<1$ ist dann $v(x)^2<c(x)^2$ erfüllt, was durch geeignete Wahl von $\sigma$ erreicht werden kann.}\\[5mm]
          \begin{figure}[h]
          \begin{center}
          \input{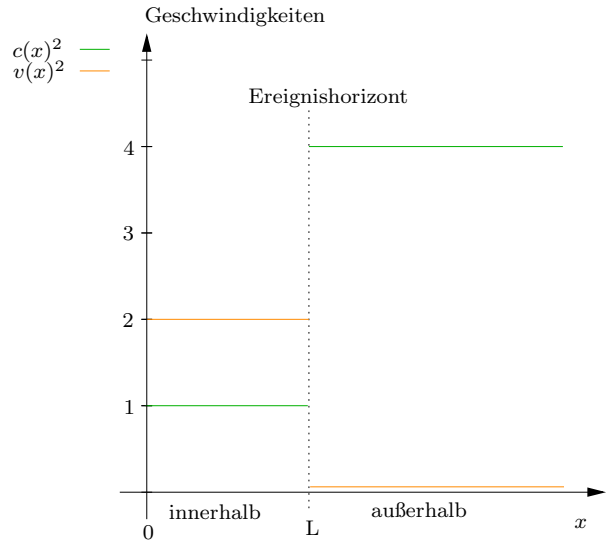}
          \caption[Geschwindigkeiten im zigarrenförmigen Kondensat mit Senke]
          {\label{cundv}Für $\epsilon \rightarrow 0$ sind hier die Quadrate der Schallgeschwindigkeit $c$ und der lokalen
          Hintergrundgeschwindigkeit $v$ in Abhängigkeit von $x$ schematisch dargestellt. Es wurden hier $c_0=1$, $s=2$
          und $\sigma=2$ gewählt. Innerhalb des schwarzen Lochs ist die Schallgeschwindigkeit kleiner, außerhalb größer als die
          Geschwindigkeit des Kondensats.}
          \end{center}
          \end{figure} \noindent

Es gibt einen Bereich, in dem die Hintergrundströmung schneller fließt (supersonic) als die Störung sich ausbreiten
kann. In diesem Bereich muss die Dichtemodulation in Richtung des Kondensats fließen.
Damit befindet sich diese im Inneren des schwarzen Lochs.\\
Außerhalb ist ein Propagieren in beide Richtungen möglich.
Dort fließt die Hintergrundströmung langsamer als die Schallgeschwindigkeit (subsonic).
Dazwischen befindet sich der Ereignishorizont.
Wenn $\epsilon$ gegen Null geht, ist der Ereignishorizont bei $x_{EH}=L$.

\section{Die Zeitentwicklung der Fluktuationen\label{zigarrenstrahlung}}
Aus Kapitel (\ref{fluktuationsaroundground}) ist bekannt, dass Fluktuationen um den Grundzustand
nach der Modenentwicklung die Bogoliubov-Gleichungen (\ref{bognewallgemein}) erfüllen.
Diese werden nun speziell auf das zigarrenförmige Kondensat angewendet.

\subsubsection{Die Lösungen der Bogoliubov-Gleichungen}
Es ist nicht notwendig, in den Bereichen $\vert x \vert< L $ und
$\vert x \vert > L + \varepsilon $
die integrale Form für die Phase $ \int^x dx' v(x') $ zu verwenden,
da dort die lokale Hintergrundgeschwindigkeit $ v $ konstant ist. Damit ist:
$$  \bar{u}_j(x)=u(x)e^{-\frac{im}{\epsilon\hbar} \vert v_0 \vert(x-L)}e^{\frac{i}{\epsilon}k (x-L)}  $$
$$  \bar{v}_j(x)=v(x)e^{\frac{im}{\epsilon\hbar}\vert v_0 \vert(x-L)}e^{\frac{i}{\epsilon}k (x-L)}. $$
Damit lauten die Bogoliubov-Gleichungen (\ref{bognewallgemein}) für den inneren und äußeren Bereich
\begin{equation}  \label{bognewsink}
\begin{split}
h_{+}u + Ng\rho_{s}v-i\epsilon \left[ \left( \frac{\hbar^2}{2m}k-\hbar \vert v \vert \right)u''
+  \frac{\hbar}{2} \left(\frac{\hbar}{m}k'' \right)u \right]&=0  \\
h_{-}v + Ng\rho_{s}u-i\epsilon \left[ \left( \frac{\hbar^2}{2m}k+\hbar \vert v \vert\right)v''
+  \frac{\hbar}{2} \left(\frac{\hbar}{m}k'' \right)v \right]&=0
\end{split}
\end{equation}
wobei
\begin{equation} \label{h+-sink}
h_{\pm}=\frac{\hbar^2}{m}\frac{k^2}{2}+Ng\rho_{s}\mp \hbar \left(  k\vert v \vert + \omega_{j} \right).
\end{equation}
ist.
Diese Gleichungen sind auf ganz $\mathbb{R}$ analytisch.
Da $c$ und $v$ sich in den drei Bereichen - siehe Abb.(\ref{randbedingungen}) -
unterschiedlich verhalten, werden für
jeden Bereich die Bogoliubov-Gleichungen separat gelöst und an den Grenzen verbunden.

\subsubsection{Die Randbedingungen}
\begin{figure}[t]
          \begin{center}
          \input{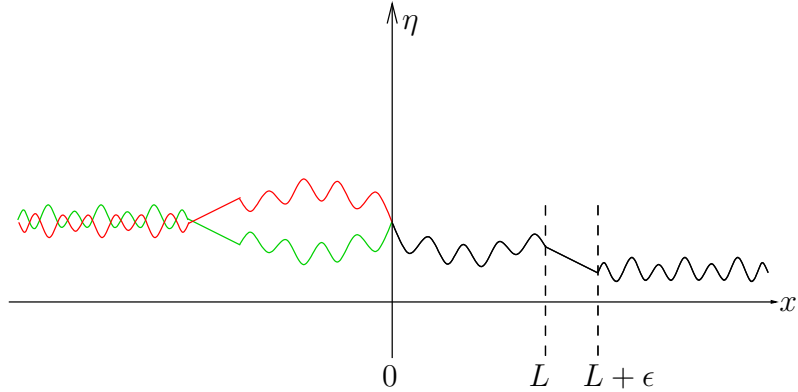}
          \caption[Randbedingungen für zigarrenförmiges Kondensat mit Senke]
          {\label{randbedingungen}Schematische Darstellung der Randbedingungen.
          Die schwarze Kurve zeigt eine Lösung für $x>0$. Im Bereich $L<x<L+\epsilon$ können die
          Wellenfunktionen linearisiert werden und die drei Bereiche müssen stetig ineinander übergehen.
          Eine Erweiterung der Lösung auf die ganze $x$-Achse ist möglich. Die Stetigkeitsbedingung an
          der Stelle $x=0$ erlaubt nur gerade - grüne Kurve -  bzw. ungerade - rote Kurve - Wellenfunktionen.}
          \end{center}
          \end{figure}

Setzt man $\hat{\psi}=\varepsilon\hat\eta(t,x) e^{i\frac{\varepsilon m}{\hbar}\int^x v(x')dx'}$ in die Stetigkeitsbedingungen
Gl.(\ref{b1}) und Gl.(\ref{b2}) ein, erhält man für $\hat{\eta}(0)$
          \begin{equation} \label{b1eta}
                    \hat{\eta}(0^+,t) -\hat{\eta}(0^-,t)=0
          \end{equation}
und
          \begin{equation} \label{b2eta}
                    \hat{\eta}'(0^+,t) -\hat{\eta}'(0^-,t)=0   ,
          \end{equation}
nur gerade und ungerade Lösungen (Abb.(\ref{randbedingungen})).
Es ist ausreichend, die rechte Seite $x>0$ zu untersuchen.\footnote{Am Schlu\ss \, sind von allen Lösungen
nur die am Punkt $x=0$ stetigen erlaubt.}\\

Im offenen Intervall $x>0$ gehen an den Stellen $x=L$ und $x=L+\varepsilon$ die Lösungen der abgetrennten
Bereiche ineinander über. Während innerhalb und außerhalb des schwarzen Lochs die Lösungen explizit
ermittelt werden können, kann durch die folgende Integralabschätzung
für ein $x \in L < x <  L + \epsilon$
          \begin{equation} \label{integalab}
                    \left\vert \int_L^x dx' v(x') \right\vert \leq \int_L^{L+\epsilon} dx \frac{v_0}{[1+(\sigma-1)x/\epsilon]^2}\leq v_0 \epsilon \ll 1,
          \end{equation}
der Zustand vereinfacht werden zu
          \begin{equation} \label{linwelle}
                    \begin{array}{c}
                    u_{\omega,k}=\alpha_{\omega,k}+\beta_{\omega,k}\frac{x}{\varepsilon}+\mathcal{O}(\varepsilon^2)\\
                    v_{\omega,k}=\gamma_{\omega,k}+\kappa_{\omega,k}\frac{x}{\varepsilon}+\mathcal{O}(\varepsilon^2)
                    \end{array},
                    \end{equation}
wodurch die zweiten Ableitungen verschwinden.\\

Das Verhalten an den Kontaktstellen wird von der Schallgeschwindigkeit bestimmt, deren zweite Ableitung
          \begin{equation}  \label{problemc}
                    \frac{c''(x)}{c(x)}=\frac{\sigma-1}{\epsilon}\left[ \delta(\vert x \vert -L)-\frac{1}{\sigma} \delta(\vert x \vert -L-\epsilon) \right]
          \end{equation}
singuläre Punkte hat.\footnote{Zur Berechnung ist die erste Ableitung der Schallgeschwindigkeit Gl.(\ref{csink}) erforderlich:
          $$\begin{array}{cc}
          c'(L^-)=0                                         & c'(L^+)=\frac{\sigma-1}{\epsilon} \\
          c'((L+\epsilon)^-)=-\frac{\sigma-1}{\epsilon}     & c'((L+\epsilon)^+)=0
          \end{array}$$}

Analog zur Bestimmung von Gl.(\ref{b1}) und Gl.(\ref{b2}) lassen sich nun die Randbedingungen an den Stellen
$x=L$ und $x=L+\varepsilon$ herleiten:
          \begin{equation} \label{binden}
          \begin{split}
          \hat{\eta}(L^+)-\hat{\eta}(L^-)=&0 \\
          \hat{\eta}'(L^+)-\hat{\eta}'(L^-)=&\frac{\sigma-1}{\varepsilon}\hat{\eta}(L) \\
          \hat{\eta}(L+\varepsilon^+)-\hat{\eta}(L+\varepsilon^-)=&0 \\
          \hat{\eta}'(L+\varepsilon^+)-\hat{\eta}'(L+\varepsilon^-)=& \frac{\sigma-1}{\sigma\varepsilon}\hat{\eta}(L+\varepsilon)
          \end{split}
          \end{equation}

In Abb.(\ref{randbedingungen}) sind alle Ergebnisse für die Randbedingungen zusammengefasst.
Die beiden ebenen Wellen - die paarweise erzeugt werden - müssen jede für sich die Gl.(\ref{binden}) erfüllen.
Es ergeben sich zwei Gleichungen für $u_k$ und $v_k$.\\
Aus den Anschlussbedingungen und der Form der Moden (Gl.(\ref{linwelle})) im Übergangsbereich ergibt sich
          \begin{equation} \label{hilfe1}
                    \frac{u_k(L+ \epsilon)}{u_k(L)}=\frac{v_k(L+ \epsilon)}{v_k(L)}=-\frac{1}{\sigma}.
          \end{equation}
Ist $\epsilon$ klein, kann
          $$ u(L+\epsilon^+)\approx u(L^+)+\epsilon u'(L^+) $$
als Taylor-Reihe bis zur ersten Ordnung entwickelt werden.
Ein Umschreiben in $u(L^+)=u_{out}(L)$ bzw. $u(L^-)=u_{in}(L)$ und Einsetzen in obige Gleichung
ergibt zusammen mit den Anschlussbedingungen
          \begin{equation}  \label{bindenuv}
                    \begin{split}
                    u_{in,k}=&-\varepsilon u_{out,k}'(L) + \frac{1}{\sigma}u_{out,k}(L)\\
                    u_{in,k}'=&\sigma u_{out,k}'\\
                    v_{in,k}=&-\varepsilon v_{out,k}'(L) + \frac{1}{\sigma}v_{out,k}(L)\\
                    v_{in,k}'=&\sigma v_{out,k}'
                    \end{split}.
          \end{equation}

\subsection{Zur Lösung der Bogoliubov-Gleichungen\label{potentialtopfnotwendig}}
Für Potentiale, die sich nur wenig mit der $x$-Koordinate ändern, kann die sog.
\textit{WKB-Näherung} (siehe Anhang \ref{WKB})
verwendet werden. In diesem Fall sind die Schall- und Hintergrundgeschwindigkeit in der
inneren und äußeren Region konstant.
Mit Gl.(\ref{extV2}) ist das externe Potential ebenfalls konstant.
Zunächst werden mit der WKB-Näherung die Lösungen von Gl.(\ref{bGleichung1}) bestimmt.
Anschließend werden mit den Randbedingungen die inneren und äußeren Wellen miteinander verbunden.

\subsubsection{Die WKB-Näherung}
Für ein sich wenig änderndes Potential lassen sich die Amplituden der ebenen Wellen als Taylor-Reihe
schreiben:
          \begin{eqnarray}
                    \label{uWKB}
                    u(x)=u_{0}+\varepsilon u_{1}+... . \\
                    \label{vWKB}
                    v(x)=v_{0}+\varepsilon v_{1}+... .
                    \end{eqnarray}
Eingesetzt in Gl.(\ref{bGleichung1}) ergibt sich in nullter Ordnung in $\varepsilon$ das Gleichungssystem
          \begin{equation} \label{Dis1}
                    \left( \begin{array}{cc}
                    h_{+}        & Ng\rho_{s} \\
                    Ng\rho_{s}   & h_{-}
                    \end{array}\right)
                    \left(    \begin{array}{cc}
                    u_{0} \\
                    v_{0}
                    \end{array}\right) =0,
          \end{equation}
dessen Lösungen die Determinante $ h_{+}h_{-}-(mc^2)^2=0 $ verschwinden lassen.
Das führt zu einer Dispersionsrelation,
          \begin{equation} \label{disprel}
                    \frac{\hbar^2}{m^2}\frac{k^4}{4}+(c^2-\vert v \vert^2)k^2
                    - 2 \vert v \vert \omega_j k - \omega_j^2=0,
          \end{equation}
deren Lösungen im Allgemeinen nicht zu finden sind, aber für spezielle $\omega$ analysiert werden können.\\

\begin{figure}
          \begin{center}\vspace{-0.8cm}
          \input{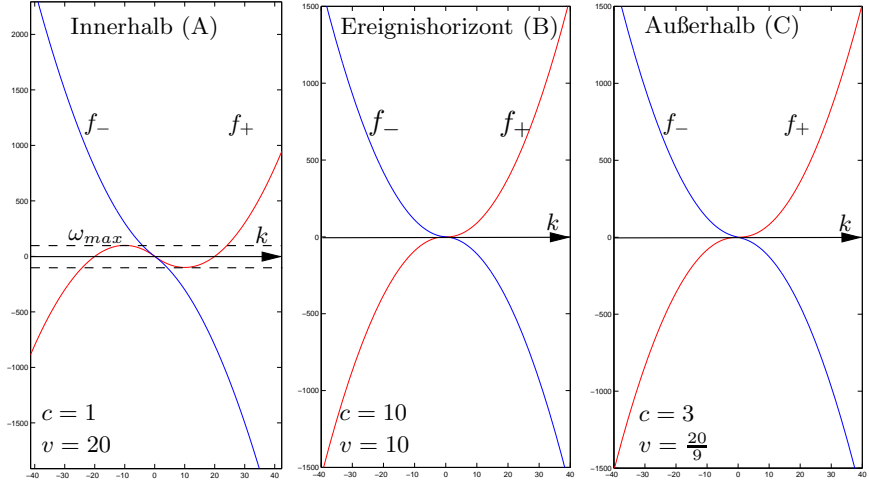}
          \caption[Reelle Frequenzen]{\label{reellefrequenzen}
          In den drei Plots ist
          die Energie $f_{\pm}$ - siehe Gl.(\ref{fpm}) - in Abhängigkeit von $k$ aufgetragen.
          Bild (A) zeigt die Situation innerhalb des schwarzen Lochs. Dort treten für Energien, die
          unterhalb $\vert E \vert<\hbar \omega_{max}$ sind, vier reelle $k$-Werte auf - oberhalb zwei.
          In (B) bzw. (C) - am Ereignishorizont bzw. außerhalb des schwarzen Lochs - ist keine maximale
          Frequenz zu finden. Zusammenfassend ist zu vermerken, dass in allen drei Bereichen mindestens
          zwei Wellen reelle Wellenzahlen $k$ mit entgegengesetzter Laufrichtung haben.}
          \end{center}
          \end{figure}

\paragraph{Reelle Frequenzen \label{realfrequ}}
Die Definition der Funktion
          \begin{equation}  \label{fpm}
                    f_{\pm}=\hbar \omega_{j}=-\hbar k \vert v \vert  \pm \hbar k\sqrt{\frac{\hbar^2}{m^2}\frac{k^2}{4}+c^2},
          \end{equation}
welche eine abgewandelte Form von Gl.(\ref{disprel}) ist, erweist sich als sinnvoll.
Die Interpretation dieser Formel wird in Abb.(\ref{reellefrequenzen}) erklärt.
Stellt man $f_{\pm}$ in Abhängigkeit von $k$ dar - wie in Abb.(\ref{reellefrequenzen})
- ist es möglich, die Lösungen
von Gl.(\ref{disprel}) graphisch zu ermitteln - vorausgesetzt, dass $\omega$ reell ist.
Die Lösungen für eine spezielle Frequenz ergeben sich, wenn eine horizontale Gerade in die Abbildung
eingezeichnet wird. Die Schnittpunkte mit den beiden Funktionen $f_+$ und $f_-$ entsprechen den Wellenzahlen
der einzelnen Moden. In den verschiedenen Bereichen kann mit diesem Verfahren die Anzahl der reellen Wellenzahlen
bestimmt werden.\\

Zur Ermittlung der nicht rein reellen Wellenzahlen wird zusätzlich verwendet, dass
Gl.(\ref{disprel}) ein komplexes Polynom vierten Grades ist und damit
genau vier Nullstellen  aufweist \cite{rudin}. Dieses Polynom besitzt nur reelle
Koeffizienten, wodurch jede komplex Konjugierte einer Nullstelle ebenfalls eine Lösung des Polynoms ist.
Die Lösungen treten in diesem Fall paarweise auf.

\subparagraph{Innerhalb des schwarzen Lochs:}$c^2<\vert v \vert^2$

\hfill\parbox[t]{15cm}{
Es existiert ein $\omega_{max}$, bei welchem genau vier reelle Nullstellen in Gl.(\ref{disprel}) auftreten.
Oberhalb dieser Frequenz liegen zwei, unterhalb vier reelle Nullstellen. Die Frequenz $\omega_{max}$ isoliert
die Bereiche verschiedener Lösungen voneinander und wird deshalb als Bifurkationspunkt bezeichnet.}
          \begin{center}
          \begin{minipage}{100mm}
          \begin{tabular}{lcl  }
          \textbf{$\omega > \omega_{max}$} & : &
          Zwei reelle und zwei komplexe $k_{c1}$ und $k_{c1}^{*}$  Lösungen.\\
          \textbf{$\omega = \omega_{max}$} & : &
          Vier reelle Lösungen, eine zweiter und zwei erster Ordnung.\\
          \textbf{$\omega < \omega_{max}$} & : &
          Vier reelle Lösungen.
          \end{tabular}
          \end{minipage}
          \end{center}

\subparagraph{Am Ereignishorizont: $c^2=\vert v \vert^2$}
          \begin{center}
          \begin{minipage}{100mm}
          \begin{tabular}{lcl  }
          \textbf{$\forall \omega$} & : &
          Zwei reelle und zwei komplexe $k_{c1}$ und $k_{c1}^{*}$ Lösungen.
          \end{tabular}
          \end{minipage}
          \end{center}

\subparagraph{Außerhalb des schwarzen Lochs: $c^2> \vert v \vert^2$}
          \begin{center}
          \begin{minipage}{100mm}
          \begin{tabular}{lcl  }
          \textbf{$\forall \omega$}  & : &
          Zwei reelle und zwei komplexe $k_{c1}$ und $k_{c1}^{*}$ Lösungen.
          \end{tabular}
          \end{minipage}
          \end{center}

\paragraph{Rein imaginäre Frequenzen:}
Einsetzen einer beliebigen, aber rein imaginären Frequenz
$\omega=i \Gamma$ in Gl.({\ref{Gleta}})
liefert
$$i\hbar \Gamma \hat{w} (x)=Ng\rho_s (\hat{w} (x) + \hat{w}(x)^{\dag})-i\varepsilon (2 v \hat{w}(x)' + \hat{w}(x)v') + \mathcal{O}(\varepsilon^2), $$
wobei für die Störung
$\hat\eta (x,t)=\hat w(x) e^{\Gamma t} $ eingesetzt wurde.\\
Anstatt gleich nach den Moden zu entwickeln, wird zunächst die WKB-Näherung durchgeführt.

In der nullten Ordnung in $\varepsilon$ erhält man
\begin{equation} \label{pureim}
i\hbar \Gamma w_0 = Ng\rho_s (w_0 + w_0^*).
\end{equation}
Diese Gleichung ist nur für $w_0=0$ erfüllt.

\subparagraph{Zusammenfassend:}
$\omega$ kann nicht rein imaginär sein!

\paragraph{Komplexe Frequenzen:}
Einsetzten einer beliebigen, aber festen komplexen Frequenz $\omega=\omega_{r}+i\omega_{i}$
in Gl.(\ref{disprel}) liefert
          $$  \frac{\hbar^2}{m^2}\frac{k^4}{4}+(c^2- v^2)k^2 - 2  v  (\omega_r+i\omega_{i})  k -\omega_{r}^2
          +\omega_{i}^2-2i \omega_{r} \omega_{i} =0 . $$
Nimmt man weiterhin an, dass $k$ reell ist, dann können
Real- und Imaginärteil dieser Gleichung voneinander separiert werden.\\
Der Imaginärteil führt zu einem Ausdruck für $\omega_{r}=-k v $, welcher zusammen mit dem reellen Anteil
zu
          $$ \omega_{i}^2=-\left( \frac{\hbar^2}{m^2}\frac{k^4}{4}+c^2 k^2 \right) $$
führt, was für reelle $k$ nicht lösbar ist.

\subparagraph{Zusammenfassend:}
Für rein komplexe Frequenzen können die Lösungen $k$ nicht rein imaginär sein.\\

Es bleibt zu klären, wie die komplexen Lösungen von $k$ aussehen.
Von ausschlaggebendem Interesse ist die Frage nach dem Vorzeichen des Imaginärteils.
Mit Abschnitt (\ref{realfrequ}) folgt, dass keine rein reellen Frequenzen  $k$ erlaubt sind. Das bedeutet aber
auch eine Unterteilung der komplexen $\omega$-Ebene in zwei Bereiche. Der eine ist die positive
komplexe Halbebene, der andere die negative.
Der Grund hierfür ist, dass alle Lösungen zu einer Frequenz das Vorzeichen
ihres Imaginärteils $k_i$ nicht wechseln können. Ansonsten würde es beim Übergang von
$+k_I$ zu $-k_I$ einen Punkt geben, in dem $k_I=0$ wäre, was nicht erlaubt ist.
Dadurch wird es möglich, kleine Frequenzen zu betrachten, deren Resultate dann
auf alle $\omega$ in der gleichen Halbebene erweitert werden können.

Gl.(\ref{disprel}) reduziert sich für $\omega=0$ zu
          \begin{equation}  \label{omega=0}
                    k^2 \left(\frac{\hbar^2}{m^2}\frac{k^2}{4}+(c^2- v^2) \right)=0.
          \end{equation}
Als Lösungen ergeben sich eine doppelte Nullstelle bei $k=0$ und zwei weitere Lösungen mit
          $k=\pm \frac{2m}{\hbar}\sqrt{ v^2-c^2}$.\\
Mit der Tatsache, dass für reelle $\omega<\omega_{max}$ alle vier Nullstellen ebenfalls reell sind,
kann man $k=k_r+i \varepsilon k_i$ setzen, da der Imaginärteil verschwinden muss,
wenn $\omega_i \rightarrow 0$.\footnote{Um Missverständnisse zu vermeiden: Es ist möglich,
in Gl.(\ref{disprel}) rein imaginäre Frequenzen einzusetzen, auch wenn diese in Gl.(\ref{Gleta}) nicht erlaubt sind. Am Ende werden
diese Lösungen ausgeschlossen. Es ist nur ihr Verhalten in der oberen komplexen $\omega$-Ebene von Interesse.}

Im nächsten Schritt werden die Annahmen - $k=k_r+i\varepsilon k_i$ und
$\omega=i\varepsilon/[sec]$ - in Gl.(\ref{disprel}) eingesetzt
          \begin{eqnarray}
                    \frac{\hbar^2}{4m^2}\left( k_r^4-\varepsilon^2 6k_r^2 k_i^2 + \varepsilon^4 k_i^4
                    +i\left( \varepsilon 4 k_r^3k_i - \varepsilon^3 4 k_r k_i^3  \right)\right) + \nonumber \\
                    \left( c^2 -  v^2 \right)\left( k_r^2 -\varepsilon^2 k_i^2 + i \varepsilon 2 k_r k_i \right)
                    +\varepsilon^2 2 k_i  v /[sec] -i \varepsilon 2 k_r  v  + \varepsilon^2 /[sec]^2 = 0 \label{disprel*}
          \end{eqnarray}
und in Real- und Imaginärteil unterteilt. Weiterhin ist noch in die verschiedenen Ordnungen von $\varepsilon$
zu unterteilen. Es wurde berücksichtigt, dass für $\omega \rightarrow i\varepsilon$ die Einheit der
Frequenz $1/[sec]$ nicht ohne Weiteres weggelassen werden darf.

\subparagraph{Reell und $\mathbf{\mathcal{O}(\varepsilon^0)}$}
          $$  \frac{\hbar^2}{4m^2}k_r^4 + \left( c^2 -  v^2 \right) k_r^2 = 0 $$
          \begin{eqnarray}
                    k_r=0 \label{kr1}\\
                    k_r=\pm \frac{2m}{\hbar}\sqrt{ v^2-c^2} \label{kr2}
          \end{eqnarray}

\textbf{$\mathbf{k_r=0}$ in Gl.(\ref{disprel*})}
$$ \frac{\hbar^2}{4m^2}\varepsilon^4 k_i^4 +
-\left( c^2 -  v^2 \right)\varepsilon^2 k_i^2 /[sec]
+\varepsilon^2 2 k_i  v /[sec]
+\varepsilon^2 /[sec]^2= 0 $$
und $\mathcal{O}(\varepsilon^2)$
          \begin{equation} \label{ki12}
                    k_i=\frac{1}{- v  \pm c}[sec]
          \end{equation}\\

\textbf{$\mathbf{k_r=\pm \frac{2m}{\hbar}\sqrt{ v^2-c^2}}$ in Gl.(\ref{disprel*})}
$$ -\varepsilon^2 6 \left(  v^2 -c^2  \right) k_i^2 + \varepsilon^4 \frac{\hbar^2}{4m^2}k_i^4
i\left( \pm \varepsilon \frac{m}{\hbar}\left( v^2 -c^2 \right)^{\frac{3}{2}}k_i \mp \varepsilon^3 \frac{m}{\hbar}
\sqrt{ v^2 -c^2}k_i \right) +$$
$$ \left( v^2 -c^2 \right) \left( -\varepsilon^2 k_i^2 \pm i\varepsilon \frac{m}{\hbar} \sqrt{ v^2 -c^2}k_i \right)
+ \varepsilon^2 2 k_i/[sec]  v  \mp i\varepsilon \frac{m}{\hbar}\sqrt{ v^2 -c^2} v  + \varepsilon^2/[sec]^2 =0 $$
und $\mathcal{O}(\varepsilon^1)$  \\

          \begin{equation} \label{ki3}
                    k_i=\frac{1}{2}\frac{ v }{\left( v^2 -c^2 \right)}[sec]
          \end{equation}
In Tabelle (\ref{allelosungen}) werden alle Ergebnisse zusammengefasst.

          \begin{table}[h] \begin{center}
          \begin{tabular}{|l|cc| }
          \hline
          $k$                                                  & Vorzeichen von $k_i$                 & Vorzeichen von $k_i$ \\
          $k=k_r+k_i$                                          & für $ v^2>c^2$   & für $ v^2<c^2$  \\ \hline\hline
          $k_1=i\frac{1}{- v^2 +c^2}[sec]$             & $-$                           & $+$ \\
          $k_2=i\frac{1}{- v^2 -c^2}[sec]$             & $-$                           & $-$ \\
          $k_3=+\frac{2m}{\hbar}\sqrt{ v^2-c^2}+i\frac{1}{2}\frac{ v }{ v^2-c^2}[sec]$ & $+$          & $+$\footnotemark\\
          $k_4=-\frac{2m}{\hbar}\sqrt{ v^2-c^2}+i\frac{1}{2}\frac{ v }{ v^2-c^2}[sec]$ & $+$          & $-$\addtocounter{footnote}{-1}\footnotemark\\[3mm]
          \hline
          \end{tabular}
          \caption[Zusammenfassung der Lösungen für komplexe Frequenzen]{Für ein beliebig gewähltes komplexes
          $\omega$ ist die Wellenzahl $k$ ebenfalls komplex. Jeweils lassen sich zwei Lösungen mit positivem bzw.
          negativem Imaginärteil $k_i$ finden.\label{allelosungen}}
          \end{center}
          \end{table}
          \footnotetext{Für $ v^2<c^2$ wird der Realteil von $k_3$ und $k_4$ imaginär, weshalb in diesem Fall die beiden
          zu verwenden sind. Daher sind $k_3=+i\frac{2m}{\hbar}\sqrt{  v^2-c^2 }$ und $k_4=-i\frac{2m}{\hbar}\sqrt{  v^2-c^2 }$
          komlex-konjugiert zueinander.}

\subparagraph{Zusammenfassend:}
Für kleine Frequenzen in der oberen Halbebene haben wir zwei Lösungen mit positivem und zwei
mit negativem Imaginärteil.
Weiterhin haben Lösungen für $\omega_I>0$ die gleiche Anzahl an positiven bzw. negativen $k_I$.
Das gilt für die obere $\omega$-Ebene und damit für die ganze Ebene.\\
Da alle Ergebnisse auf der Gültigkeit der WKB-Näherung beruhen, gilt es, diese zu überprüfen

\paragraph{Gültigkeit der WKB-Näherung:}
Zur Kontrolle der WKB-Näherung ist zu analysieren, ob die Variation der Amplituden der ebenen Wellen klein ist.
Dafür ist es notwendig,
$u_0 (x)$ und $v_0 (x)$ zu bestimmen.\\
Dazu werden
die Gl.(\ref{bognewsink}) bis zur ersten Ordnung in $\varepsilon$ betrachtet. Es ergibt sich
          \begin{equation} \label{bognewsinkersteOrdnungamplidutde}
                    \begin{split}
                    h_{+} u_1 + Ng\rho_s v_1 - i \left[ \left( \frac{\hbar^2}{2m}k(x)-\hbar  v  \right) u_0'
                    + \frac{\hbar^2}{2m}k(x)'u_0 \right]&=0 \\
                    h_{-} v_1 + Ng\rho_s u_1 - i \left[ \left( \frac{\hbar^2}{2m}k(x)+\hbar  v  \right) v_0'
                    + \frac{\hbar^2}{2m}k(x)'v_0 \right]&=0
                    \end{split}
          \end{equation}
Um die nullte Ordnung zu erfüllen, muss weiterhin die Gl.(\ref{Dis1}) erfüllt werden,
damit ${h_+ h_-}/{Ng \rho_s} =Ng \rho_s$.
Multiplizieren der ersten Zeile von Gl.(\ref{bognewsinkersteOrdnungamplidutde}) mit ${h_-}/{Ng \rho_s}$ führt zu
          $$ Ng\rho_s u_1 + h_- v_1 = i \frac{h_-}{Ng\rho_s}\left[ \left( \frac{\hbar^2}{2m}k(x)
          -\hbar  v  \right) u_0 ' + \frac{\hbar^2}{2m}k(x)'u_0 \right] .$$
Weiterführend ist es notwendig, dass aus Gl.(\ref{Dis1})
          \begin{equation} \label{hconst}
                    \frac{u_0}{v_0}=-\frac{h_-}{Ng\rho_s}=-\frac{Ng\rho_s}{h_+} \equiv h_{const}
          \end{equation}
folgt, was $ \frac{u_0'}{u_0}=(\log u_0)'= (\log v_0)'$ und ${h_-}/{Ng\rho_s}=h_{const}$ impliziert.
Einsetzen in die zweite Zeile von Gl.(\ref{bognewsinkersteOrdnungamplidutde}) liefert
          $$ h_{const}^2  \left[ \left( \frac{\hbar^2}{2m}k(x)-\hbar  v  \right)(\log u_0)'
          +\frac{\hbar^2}{2m}k(x)' \right] +
          \left[ \left( \frac{\hbar^2}{2m}k(x)+\hbar  v  \right)(\log u_0)'
          +\frac{\hbar^2}{2m}k(x)' \right] =0.$$
Umformung und Integration kann gezeigt werden, dass $u_0 (x)$ - und damit auch $v_0 (x)$ -
proprotional zu
          $$ u_0 \approx \frac{1}{d\omega/dk}$$
sind.
Am Ereignishorizont wird der Nenner unendlich groß, womit die Amplitude gegen Null geht.
Die WKB-Methode ist gültig, da $u_0$ und $v_0$ am Ereignishorizont endlich sind.

\subsubsection{Auswertung der Randbedingungen für komplexe Frequenzen}
Zur Analyse dynamischer Instabilitäten sind nur die Lösungen für komplexe Frequenzen von
Interesse (siehe (\ref{forfluctiations})).
Es treten für ein beliebiges komplexes $\omega$ immer vier komplexe $k$ auf, wobei zwei davon einen
positiven und zwei einen negativen Imaginärteil haben. Für $Im(k)<0$ wächst die Störung exponentiell
mit $x$ (siehe Abschnitt (\ref{forfluctiations}))
und entspricht einer räumlichen Instabilität, wenn
der Raum nicht begrenzt ist. Die äußeren Lösungen sind daher auf $Im(k_{out})>0$ zu beschränken.\\
Jede Linearkombination der ebenen Wellen in einem Bereich ist ebenfalls eine Lösung. Alle Lösungen müssen
die Stetigkeitsbedingungen Gl.(\ref{bindenuv}) erfüllen.

\paragraph{An der Stelle $\mathbf{x=L}$:}
Die inneren Lösungen in allgemeiner Form sind
          \begin{equation}
                    \begin{split}  \label{konectin}
                    u_{in,\alpha}(x)&=\sum_j F_{\alpha j}e^{i(k_{in,j}-v_0)(x-L)}     \\
                    v_{in,\alpha}(x)&=\sum_j F_{\alpha j}h_{in,j}e^{i(k_{in,j}+v_0)(x-L)},
                    \end{split}
                    \end{equation}
wobei $\alpha=1,2$ und $j=1,2,3,4$
ist.\footnote{Bei den Koeffizienten $u_0$ handelt es sich um Normierungsfaktoren,
die als Eins gewählt wurden. }
Gesucht ist die Koeffizientenmatrix
          \begin{equation} \label{konectmatrix1}
                    F_{\alpha j}=\left(
                    \begin{array}{cccc}
                    F_{11} & F_{21}\\
                    F_{12} & F_{22}\\
                    F_{13} & F_{23}\\
                    F_{14} & F_{24}
                    \end{array}
                    \right)
                    \end{equation}
und der Zusammenhang zwischen $u_{in}$ und $v_{in}$, der aus Gl.(\ref{hconst}) folgt:
          \begin{equation} \label{konecthin1}
                    h_{\alpha j}=\left(
                    \begin{array}{cccc}
                    h_{in,1} & h_{in,2} & h_{in,3} & h_{in,4}
                    \end{array}
                    \right).
                    \end{equation}
Entsprechend ergibt sich für den äußeren Bereich
          \begin{equation} \label{konectout}
                    \begin{split}
                    u_{out,\alpha}(x)&=\sum_{m} F^{\star}_{\alpha m}e^{i(k_{out,m}-v_0)(x-L)}     \\
                    v_{out,\alpha}(x)&=\sum_{m} F^{\star}_{\alpha m}h_{out,m}e^{i(k_{out,m}+v_0)(x-L)},
                    \end{split}
          \end{equation}
wobei $\alpha=1,2$ und $m=1,2$ ist, und
\begin{equation} \label{konectmatrix2}
          F^{\star}_{\alpha m}=\left(
          \begin{array}{cc}
          F^{\star}_{11} & F^{\star}_{21} \\
          F^{\star}_{12} & F^{\star}_{22}
          \end{array}
          \right)
          \end{equation}
und
          \begin{equation} \label{konecthin1_new}
                    h_{\alpha m}=\left(
                    \begin{array}{cc}
                    h_{out,1} & h_{out,2}
                    \end{array}
                    \right).
                    \end{equation}

Zusammen mit den Anschlussbedingungen (Gl.(\ref{bindenuv})) ergeben sich acht Gleichungen
für zwölf Unbekannte $F_{\alpha j}$ und $F^{\star}_{\alpha m}$.
Vier Unbekannte können daher frei gewählt werden:
          \begin{equation} \label{konectmatrix2special}
                    F^{\star}_{\alpha m}=\left(
                    \begin{array}{cc}
                    1 & 0 \\
                    0 & 1
                    \end{array}
                    \right).
          \end{equation}
Mit dieser Wahl ergibt sich
          \begin{equation} \label{achtgleichungen}
                    \begin{split}
                    u_{in,1}&= F_{11}e^{ik^-_{in,1}(x-L)} + F_{12}e^{ik^-_{in,2}(x-L)} +  F_{13}e^{ik^-_{in,3}(x-L)} + F_{14}e^{ik^-_{in,4}(x-L)} \\
                    u_{in,2}&= F_{21}e^{ik^-_{in,1}(x-L)} + F_{22}e^{ik^-_{in,2(x-L)}} +  F_{23}e^{ik^-_{in,3}(x-L)} + F_{24}e^{ik^-_{in,4}(x-L)}  \\
                    u_{out,1}&=      e^{ik^-_{out,1}(x-L)}                                                                                          \\
                    u_{out,2}&=      e^{ik^-_{out,2}(x-L)}                                                                                            \\
                    v_{in,1}&= F_{11}h_{in,1}e^{ik^+_{in,1}(x-L)} + F_{12}h_{in,2}e^{ik^+_{in,2}(x-L)} +  F_{13}h_{in,3}e^{ik^+_{in,3}(x-L)} + F_{14}h_{in,4}e^{ik^+_{in,4}(x-L)}\\
                    v_{in,2}&= F_{21}h_{in,1}e^{ik^+_{in,1}(x-L)} + F_{22}h_{in,2}e^{ik^+_{in,2}(x-L)} +  F_{23}h_{in,3}e^{ik^+_{in,3}(x-L)} + F_{24}h_{in,4}e^{ik^+_{in,4}(x-L)}  \\
                    v_{out,1}&=      h_{out,1}e^{ik^+_{out,1}(x-L)}\\
                    v_{out,2}&=      h_{out,2}e^{ik^+-_{out,2}(x-L)}
                    \end{split}
                    \end{equation}
                    wobei
                    \begin{equation}  \label{umschreiben}
                    \begin{split}
                    k^{\pm}_{in,j}&= k_{in,j}\pm v_0\\
                    k^{\pm}_{out,\alpha}&= k_{out,\alpha}\pm \frac{v_0}{\sigma^2}
                    \end{split}
          \end{equation}
ist.
Einsetzen von Gl.(\ref{achtgleichungen}) in Gl.(\ref{bindenuv}) und Umschreiben führt zu
          \begin{equation}
                    \left(      \begin{array}{cccc}
                    1&                  1&                  1&                  1                 \\
                    k^-_{in,1}&         k^-_{in,2}&         k^-_{in,3}&         k^-_{in,4}         \\
                    h_1&                h_2&                h_3&                h_4                \\
                    h_1 k^+_{in,1}&     h_2 k^+_{in,2}&     h_3 k^+_{in,3}&     h_4 k^+_{in,4}
                    \end{array} \right)
                    \left(  \begin{array}{c}
                    F_{\alpha 1}\\
                    F_{\alpha 2}\\
                    F_{\alpha 3}\\
                    F_{\alpha 4}\\
                    \end{array}
                    \right)=
                    \left(  \begin{array}{c}
                    -i\epsilon k^-_{out,\alpha}+\frac{1}{\sigma}\\
                    \sigma k^-_{out,\alpha}\\
                    -i\epsilon h_{out,\alpha} k^-_{out,\alpha}+\frac{1}{\sigma}h_{out,\alpha}\\
                    \sigma h_{out\alpha,}k^-_{out,\alpha}
                    \end{array}
                    \right).
                    \end{equation}
Auflösen nach $F_{\alpha j}$ ergibt
          \begin{equation} \label{konectmatrixfertig}
                    F_{\alpha j}=M^{-1}C_{\alpha},
          \end{equation}
mit
          \begin{equation}
                    M=\left(      \begin{array}{cccc}
                    1&                  1&                  1&                  1                 \\
                    k^-_{in,1}&         k^-_{in,2}&         k^-_{in,3}&         k^-_{in,4}         \\
                    h_1&                h_2&                h_3&                h_4                \\
                    h_1 k^+_{in,1}&     h_2 k^+_{in,2}&     h_3 k^+_{in,3}&     h_4 k^+_{in,4}
                    \end{array} \right)
                    \end{equation}
und
          \begin{equation}
                    C_{\alpha}=\left(  \begin{array}{c}
                    -i\epsilon k^-_{out,\alpha}+\frac{1}{\sigma}\\
                    \sigma k^-_{out,\alpha}\\
                    -i\epsilon h_{out,\alpha} k^-_{out,\alpha}+\frac{1}{\sigma}h_{out,\alpha}\\
                    \sigma h_{out\alpha,}k^-_{out,\alpha}
                    \end{array}
                    \right).
                    \end{equation}
Es bleibt noch, die Lösungen der rechten und linken Seite zu verbinden.

\paragraph{Anschlussbedingungen an der Stelle $\mathbf{x=0}$:}
Die Anschlussbedingung Gl.(\ref{b1}) an der Stelle $x=0$ lässt nur gerade $\hat{\eta}_{g}$ und ungerade
$\hat{\eta}_{u}$ zu.

\begin{description}
\item[Ungerade Lösungen:]
Für eine gerade Funktion gilt
          \begin{equation} \label{gerade}
                    \hat{\eta}_g(x,t)= -\hat{\eta}_g(-x,t).
          \end{equation}
Einsetzen in Gl.(\ref{b1}) liefert
          \begin{equation} \label{geradeb1}
                    \hat{\eta}_g(0,t)= -0.
          \end{equation}
Damit müssen $u_{in}(0,t)$ und $v_{in}(0,t)$ unabhängig voneinander an
der Stelle $x=0$ verschwinden ($u_{in}(0,t)=v_{in}(0,t)=0$).
Da innerhalb zwei Lösungen existieren, muss jede
Linearkombination der beiden dort ebenfalls Null sein.
Die beiden Gleichungen
          \begin{equation}  \label{geardehilfe}
                    \begin{split}
                    u_{in,1}(0)-v_{in,1}(0)&=0\\
                    u_{in,2}(0)-v_{in,2}(0)&=0
                    \end{split}
          \end{equation}
müssen gleichzeitig verschwinden.
Zusammengefasst ergibt sich die Anschlussbedingung für gerade Lösungen
          \begin{equation} \label{bedgerade}
                    u_{in,1}(0)v_{in,2}(0)- u_{in,2}(0)v_{in,1}(0)=0,
          \end{equation}
in welche Gl.(\ref{konectin}) eingesetzt werden kann:
          \begin{equation}  \label{gesamtelosgerade}
                    \sum_{ij} F_{1i} F_{2j}(h_{in,i}-h_{in,j})e^{-i(k_{in,i}+k_{in,j})L}=0.
          \end{equation}

\item[Gerade Lösungen:]
          Für eine ungerade Funktion gilt
                    \begin{equation} \label{ungerade}
                              \hat{\eta}_u(x,t)= \hat{\eta}_u(x,t).
                    \end{equation}
Einsetzen in Gl.(\ref{b1}) liefert
          \begin{equation} \label{ungeradeb1}
                    \hat{\eta}_{u}(0,t)+iv_0\hat{\eta}_{u}(0,t)= 0.
          \end{equation}
In Gl.(\ref{konectin}) eingesetzt ergibt sich
          \begin{equation}  \label{gesamtelosungerade}
                    \sum_{ij} F_{1i} F_{2j}(h_{in,i}-h_{in,j})k_{in,i}k_{in,j}e^{-i(k_{in,i}+k_{in,j})L}=0.
          \end{equation}
\end{description}

\subsection{Zur Lösung der Gleichungen}
In der Veröffentlichung von Cirac et al. \cite{cirac1} findet sich ein numerischer Lösungsweg, der im Folgenden kurz
skizziert wird.
Die beiden Gleichungen (\ref{gesamtelosgerade}) und (\ref{gesamtelosungerade}) hängen für ein bestimmtes System nur von $\omega$ ab.
Das System ist durch die Länge $2L$ der inneren Region, die Fließgeschwindigkeit an der Senke $v_0$,
den Unterschied der inneren und äußeren Geschwindigkeiten $\sigma$ und durch die Konstante
$U=\frac{N_0g}{m}$ bestimmt.
Es wird noch darauf hingewiesen, dass die Gesamtlänge $2D$ der Kondensatwolke und die Grösse $\epsilon$ des
Übergangsbereichs eine Rolle spielen, wobei hier nicht der exakte Wert ausschlaggebend ist. Ausreichend ist,
das Kondensat möglichst groß und die Verengung möglichst klein zu wählen.\\
Mit diesen Überlegungen verbleiben zwei Gleichungen
          \begin{equation} \nonumber
                    s(\omega; \sigma, U, v_0, L)=0.
          \end{equation}
Zur Berechnung werden zunächst die Systemgrößen gewählt und anschließend die Punktepaare $(\omega,s)$
numerisch ermittelt,
wobei beide im Allgemeinen komplexe Funktionen sind.
Trägt man den Betrag von $s$ über $\omega$ auf, repräsentieren die Nullstellen die Eigenfrequenzen.\\

Interessant ist, mehrere Plots für unterschiedliche Längen des inneren Bereichs zu vergleichen.
Es stellt sich heraus, dass die Anzahl der Instabilitäten empfindlich von $L$ abhängt.
Erst ab einem gewissen Duchmesser ${\pi\hbar}/{k_0m}-\delta$, wobei $\delta$ sehr viel kleiner ist als ${\pi}/{k}$,
treten komplexe Eigenfrequenzen auf. Jede Vergrößerung um ${\pi}/{k_0}$ erhöht die Anzahl der komplexen
Eigenfrequenzen um eins.
Bei der auftretenden Wellenzahl handelt es sich um
          \begin{equation} \label{ersteeigenfrequenz}
                    k_0=\frac{2m}{\hbar}\sqrt{v_0^2-c^2},
          \end{equation}
was physikalisch erklärbar ist:
Erst wenn komplexe Eigenfrequenzen auftreten, werden Moden angeregt.
Die Wellenzahlen dieser Moden sind in Tabelle (\ref{allelosungen}) aufgelistet.
Innerhalb des schwarzen Lochs kann es nur Lösungen geben, wenn
der Realteil ein Vielfaches von $\frac{2m}{\hbar}\sqrt{v_0^2-c^2}$ ist, da keine
rein imaginären $k$-Werte erlaubt sind (siehe Kapitel (\ref{potentialtopfnotwendig})).
Die kleine Abweichung $\delta$ entsteht, weil der Ereignishorizont keine exakt harte Wand ist.
Damit können die dynamischen Instabilitäten auf die Bindungszustände zurückgeführt werden.
Erst wenn die Wellenlänge der erzeugten Mode groß genug ist, um die Stetigkeitsbedingungen am
Rand des Potentialtopfs zu erfüllen, können sich gebundene Zustände ausbilden.
Damit gilt für die Frequenz des ersten gebundenen Zustands
          \begin{equation} \nonumber
                    \frac{L}{\lambda}=\frac{Lk_0}{2\pi}.
          \end{equation}
Obwohl mit der Länge $2L$ die Anzahl der gebundenen Zustände steigt, ergibt sich aus
der Simulation für kleinere Löcher eine höhere Instabilität.

\section{Zusammenfassung}
Am Ende des Kapitels werden alle Ergebnisse zusammengefasst.
Es wurden zwei Punkte ausgearbeitet.\\
Zum einen betrachtete man eine von außen angeregte Störung
im Kondensat.
Es resultierten dynamische Gleichungen für die Störung. Durch die Einführung einer effektiven
Metrik konnte gezeigt werden, dass sich die Phase der Störung wie Licht in der Nähe eines schwarzen Lochs
bewegt. Die Einträge in der Metrik bestehen nur aus Grössen des Kondensats. An dem Punkt, an dem die
Hintergrundgeschwindigkeit des Kondensats dem Betrag nach gleich der Schallgeschwindigkeit ist,
grenzen zwei Bereiche aneinander. In einem fliesst das Kondensat mit Überschallgeschwindigkeit, im anderen
unter Schallgeschwindigkeit.
Im ersteren kann die Störung nur in eine Richtung propagieren,
während im anderen Bereich beide Bewegungsrichtungen möglich sind.
Das zigarrenförmige Kondensat wurde so gewählt, dass dieses Verhalten im verengten Bereich auftritt.
Gelangt die Störung in den eingeschränkten Bereich, wird sie vom Kondensat unaufhaltsam mitgerissen
und am Koordinatenursprung über die Senke aus der Kondensatwolke ausgeschieden.
Dieser Bereich entspricht dem Inneren eines schwarzen Lochs.\\

Als Zweites wurde untersucht, wie sich der nichtkondensierte Anteil auf die makroskopische Wellenfunktion
auswirkt. Für Temperaturen nahe am absoluten Nullpunkt der Temperatur können die nichtkondensierten Atome
als Störung im Bose-Einstein-Kondensat beschrieben werden.
Diese regen Moden an, die als Quasiteilchen aufgefasst werden, es handelt sich um Phononen.
Da nur dann Phononen produziert werden, wenn im Inneren des schwarzen Lochs gebundene Zustände
auftreten, müssen diese dafür verantwortlich sein.
Je mehr gebundene Zustände möglich sind, desto mehr Moden mit komplexer Frequenz gibt es.
Diese verursachen dynamische Instabilitäten, d. h. die Anzahl der Phononen wächst exponentiell mit
der Zeit.\\
Ein instabiles Kondensat verharrt nicht im Grundzustand. Die Rechnungen basierten auf der
Annahme, dass es sich um kleine Störungen um den Grundzustand handelt. Damit sind die Bogoliubov-Gleichungen
nur eine gewisse Zeit gültig.\\

                              \chapter{Ringförmige Kondensate mit Wirbel}
Als nächstes Modell wird ein Bose-Einstein-Kondensat betrachtet, in dessen Zentrum ein Wirbel
ist. Es wird untersucht, ob in dieser Konfiguration ebenfalls ein schwarzes Loch auftritt und
wie sich die Fluktuationen um den Grundzustand mit der Zeit entwickeln.
Die Resultate dazu wurden bereits von Cirac et al. \cite{cirac1} veröffentlicht. Erneut ist es
die Aufgabe gewesen, die Rechungen zu verifizieren.
Neu in diesem Kapitel sind zwei Simulationen. In der Ersten wird gezeigt,
wie sich eine ebene Welle im Kondensat verhält. Das zweite Programm zeigt das Verhalten eines
Wellenpakets in der Nähe eines Ereignishorizonts.

\section{Das Modell}

\begin{figure}[t]
          \begin{center}
          \input{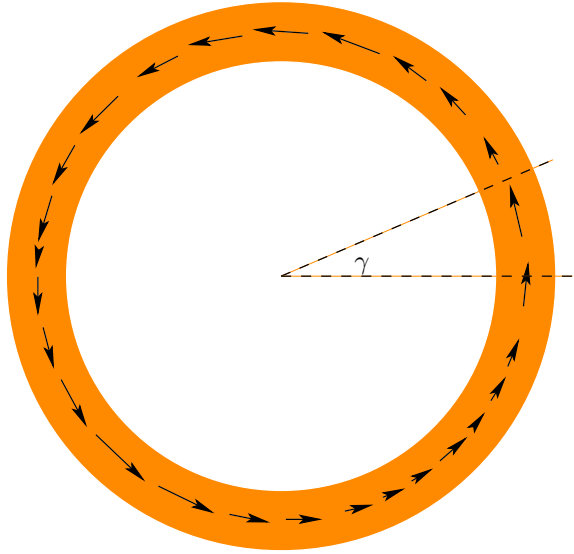}
          \caption[Dichteprofil eines Kondensats mit einem zentralen Wirbel]
          {\label{ringformig}Der Wirbel in der Mitte treibt das Kondensat - orange dargestellt - nach außen. Die
          Vektoren deuten den Verlauf der Hintergrundgeschwindigkeit an, welche vom Winkel $\gamma$ abhängt.}
          \end{center}
          \end{figure}

\begin{figure}[t]
          \begin{center}
          \input{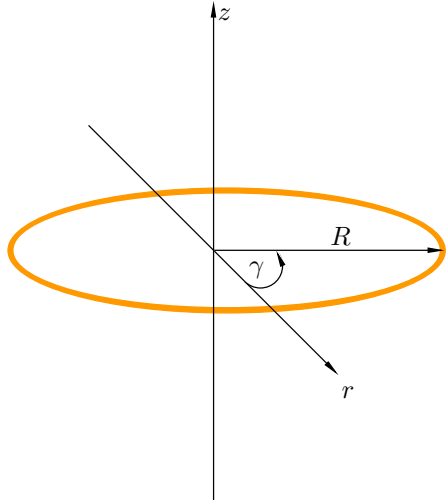}
          \caption[Koordinaten für das Kondensat mit Wirbel]
          {\label{ringformkoord}Es wird ein Kondensat (orange eingezeichnet) betrachtet, dessen
          Hintergrundgeschwindigkeit $v$ nur von $\gamma$ abhängt.
          Zur mathematischen Beschreibung werden Polarkoordinaten verwendet.}
          \end{center}
          \end{figure}

Ein Wirbel im Kondensat ist ein Punkt in der Kondensatwolke, um welchen das Kondensat rotiert.
Die Rotationsgeschwindigkeit $v\propto 1/r$ nimmt umgekehrt proportional zum Abstand zu.
Die Dichte im Zentrum des Wirbels ist Null (siehe Abb.(\ref{ringformig})).\\
Die Bewegung ist in der $r$ und
$z$ Achse durch ein externes Potential eingefroren (siehe Abb.(\ref{ringformkoord})).

Durch die Einschränkung der Bewegung auf eine Richtung kann der Zustand in Polarkoordinaten in
          \begin{equation} \label{statering}
                    \Phi(t,r,\gamma,z)=f(r,z)\tilde{\Phi}(t,\gamma)
          \end{equation}
aufgeteilt werden. Beim Wechsel von kartesischen Koordinaten zu Polarkoordinaten ändert sich der
Tangentialraum und damit der Gradient
          \begin{equation} \label{kartespolargradient}
                    \left(\begin{array}{c}
                    \partial_x \\
                    \partial_y \\
                    \partial_z
                    \end{array}\right)
                    \rightarrow
                    \left(\begin{array}{c}
                    \partial_r \\
                    \frac{1}{r}\partial_{\gamma} \\
                    \partial_z
                    \end{array}\right)
          \end{equation}
und der Kotangentialraum. Damit ist bei der Integration zu beachten, dass
          \begin{equation} \label{kartespolarintegal}
          \iiint dx dy dz \rightarrow \iiint r dr d\gamma dz.
          \end{equation}

\subsection{Die Gross-Pitaevskii-Gleichung für das ringförmige Kondensat}
          Wieder wird als Ausgangspunkt die zeitabhängige Gross-Pitaevskii-Gleichung Gl.(\ref{GPE2})
          verwendet, die in den eingeschränkten Polarkoordinaten - $dr=0$ und $dz=0$ - umgeschrieben werden kann in
                    \begin{equation} \label{GPE2polar}
                              i\hbar\partial_t \Phi(t,r,\gamma,z)=
                              \left( -\frac{\hbar^2}{2mR^2}\partial_{\gamma} + V_{ext}(\gamma) +  U \vert \Phi(t,r,\gamma,z)\vert^2
                              \right) \Phi(t,r,\gamma,z) \, ,
                    \end{equation}
          wobei das chemische Potential $\mu$ bereits in $V_{ext}$ enthalten ist.
          Einsetzen von Gl.(\ref{statering}) und der dimensionslosen Zeit $\tau=\frac{\hbar}{mR^2}$
          liefert
                    \begin{equation} \label{GPE2ring}
                              i\partial_t \tilde{\Phi}(t,\gamma)=
                              \left( -\frac{1}{2}\partial_{\gamma} + \tilde{V}_{ext}(\gamma) +
                              \frac{\tilde{U}}{N} \vert \tilde{\Phi}(t,\gamma)\vert^2
                              \right) \tilde{\Phi}(t,\gamma),
                    \end{equation}
          mit dem dimensionslosen Potential
                    \begin{equation} \label{dimlospotential}
                    \tilde{V}_{ext}\equiv V_{ext}\frac{mR^2}{\hbar^2}
                    \end{equation}
          und der neuen Wechselwirkungskonstante
                    \begin{equation}  \label{Utilde}
                              \tilde{U}=U \frac{mR^2}{\hbar^2} \int dz dr r \vert f(r,z) \vert^2,
                    \end{equation}
          wobei gleichzeitig der Grundzustand von $\tilde{\Phi}$ auf die Gesamtteilchenzahl $N$
                    \begin{equation} \label{normiert}
                              \int_0^{2 \pi} \vert \tilde{\Phi}(\gamma) \vert^2 =N
                    \end{equation}
          normiert ist.\\

          Der stationäre Zustand für Gl.(\ref{GPE2ring}) ist:
                    \begin{equation} \label{statestatring}
                              \tilde{\Phi}(\tau,\gamma)=\sqrt{\rho(\gamma)}e^{i\int d\gamma v(\gamma)}.
                    \end{equation}

\subsection{Schall- und Hintergrundgeschwindigkeit im Kondensat}
          Mit Gl.(\ref{schallgeschwindigkeit}) ist die Schallgeschwindigkeit im Kondensat durch
                    \begin{equation}
                              c(\gamma)=\sqrt{\tilde{U}\rho_{\gamma}/N}
                    \end{equation}
          gegeben und damit von der Dichte abhängig.
          Bei der Wahl der Dichte ist zu berücksichtigen, dass im Bose-Einstein-Kondensat die
          Wirbel quantisiert sind. Integration der Hintergrundgeschwindigkeit entlang einer geschlossenen Bahn
          um den Wirbel ergibt $2\pi m$, wobei $m$ eine ganze Zahl ist und als \textit{Windungszahl} bezeichnet wird.
          Die periodische Randbedingung im Kondensat kann mit der Windungszahl formuliert werden:
                    \begin{equation}  \label{windungszahl}
                              m=\frac{1}{2 \pi}\int_0^{2\pi}d\gamma v(\gamma).
                    \end{equation}
          Als Wahl erweist sich
                    \begin{equation} \label{ringdichte}
                              \rho(\gamma)=\frac{N}{2 \pi}\left( 1 + b cos(\gamma) \right)
                    \end{equation}
          passend im Sinne der Randbedingungen und für die gewünschte Analogie zum Schwarzen Loch,
          wobei $0<b<1$ ist.\\
          Zusammen mit Gl.(\ref{schallgeschwindigkeit}) und der Kontinuitätsgleichung für ein stationäres
          Kondensat $\partial_{\gamma}(c^2v)=0$ sind die Schallgeschwindigkeit
                    \begin{equation} \label{ringschall}
                              c(\gamma)=\sqrt{\frac{\tilde{U}}{2 \pi}(1+bcos(\gamma))}
                    \end{equation}
          und die Hintergrundgeschwindigkeit
                    \begin{equation} \label{ringv}
                              v(\gamma)=\frac{\tilde{U}m\sqrt{1-b^2}}{1\pi c(\gamma)^2}
                    \end{equation}
          gegeben.



\section{Die Wellengleichung}
Die dynamischen Gleichungen für eine kleine Dichteschwankung im Kondensat - injiziert durch einen Laser -
kann analog Kapitel (\ref{wellengleichung}) hergeleitet werden.

\subsection{Die Metrik für das ringförmige Kondensat}
Ausgehend von der zeitabhängigen Gross-Pitaevskii-Gl.(\ref{GPE2ring}) wird
für die Dichte $\rho=\rho_0 +\varepsilon \rho_1 $ und die Phase $\theta=\theta_0 +\varepsilon  \theta_1 $
eingesetzt.
In erster Ordnung erhält man für den Realteil
          \begin{equation} \nonumber
                    \partial_{\tau}{\rho_1}=-\frac{\hbar}{m}\partial_{\gamma}
                    \left(\rho_0 \partial_{\gamma} \theta_1 + \rho_1 \partial_{\gamma} \theta_0 \right)
          \end{equation}
und den Imaginärteil
          \begin{equation} \label{ringwellengleichung}
                    \partial_{\tau}{\theta_1}=-\frac{\hbar}{m}
                    \left(\partial_{\gamma} \theta_0 \nabla \theta_1) - \frac{\tilde{U}}{\hbar} \rho_1 \right).
          \end{equation}
Diese beiden Gleichungen lassen sich in eine Gleichung für die Phase zusammenfassen
                    \begin{equation}
                    -\ddot{\theta}_1 - \left( v\dot{\theta}_1' \right)
                    - \left( v \dot{\theta_1} \right)'
                    +  \left(\left( c^2
                    -    v^2\right) \theta_1' \right)' =0,
          \end{equation}
die mit der effektiven Metrik
          \begin{equation}    \label{ringmetrik}
                    g_{\mu \nu}=c\left( \begin{array}{cccc}
                    -c^2+v_{\gamma}^2           & v_{\gamma}                  & 0              & 0\\
                    v_{\gamma}                    & 1                    & 0              & 0\\
                    0                      & 0                    & 1              & 0 \\
                    0                      & 0                    & 0              & 1
          \end{array} \right).
          \end{equation}
zu
           \begin{equation}
                    \partial_{\mu}(\sqrt{-g}g^{\mu\nu}\partial_{\nu}\theta_{1})=0,
          \end{equation}
umgeschrieben werden kann, mit $v(\tau, \theta)\equiv v(\theta)$.

In Kapitel (\ref{wellengleichung}) wurde anhand von $g_{00}$ gezeigt, dass es sich bei Gl.(\ref{ringmetrik})
um eine Metrik für ein Schwarzes Loch handelt.

\section{Die Nullgeodäten}
          Die \textit{Nullgeodäten} beschreiben die Bahnen der masselosen Teilchen im gekrümmten Raum.
          In diesem Fall ist das Linienelement Gl.(\ref{abstand}) zeitartig ($ds^2=0$).
          Speziell betrachten wir ebene Wellen
                    \begin{equation}  \label{Ebene-Welle}
                              \tilde{\theta}=w(\gamma)e^{-i\omega t}e^{i\int d\gamma k(\gamma)}
                    \end{equation}
          die durch das Kondensat propagieren.\footnote{Zur Berechnung wurde
          $\gamma \rightarrow \frac{\gamma}{\varepsilon}$ im Exponenten eingeführt.
          Auf Gl.(\ref{ringwellengleichung}) angewendet ergibt:
          $$-\ddot{\theta}_1 + \left( v\dot{\theta}_1' \right)
                    + \varepsilon\left( v \dot{\theta_1} \right)'
                    +  \varepsilon^2\left(\left( c^2
                    -    v^2\right) \theta_1' \right)' =0.$$}
          Einsetzen in Gl.(\ref{ringwellengleichung}) liefert
                    \begin{equation} \label{ringebenewellen}
                              \begin{split}
                              0 & =fw \\
                                & +i\varepsilon \left( w \xi' +2w'\xi \right)\\
                                & +\varepsilon^2 \left( (c^2-v^2)w' \right)'\\
                              \end{split}
                    \end{equation}
          mit
                    \begin{equation} \label{ringf}
                              f=\omega^2 + 2v \omega k^2-(c^2-v^2)k^2
                    \end{equation}
          und
                    \begin{equation} \label{ringxi}
                              \xi=(c^2-v^2)k-v \omega.
                    \end{equation}
          Zur Lösung dieser Gleichung wird die WKB-Methode verwendet (siehe Anhang (\ref{WKB})).

\subsection{Die WKB-Methode für die Berechnung der Nullgeodäten}
          Bei der WKB-Methode wird die Amplitude $w(x)$ in eine Taylor-Reihe
                    \begin{equation} \label{ringWKB}
                              w(x) \approx w_{0}+\varepsilon w_{1}
                    \end{equation}
          entwickelt, wobei für ein sich wenig änderndes Potential alle Terme mit höherer Ordnung als
          $\mathcal{O}(\varepsilon)$ vernachlässigt werden.

\subsubsection{Die Dispersionsrelation}
          Für die nullte Ordnung muss die erste Zeile in Gl.(\ref{ringebenewellen}) verschwinden:
                    \begin{equation}  \label{ringf=0}
                              f=\omega^2 + 2v \omega k^2-(c^2-v^2)k^2=0.
                    \end{equation}
          Für die Dispersionsrelation ergibt sich damit
                    \begin{equation} \label{ringdisprel}
                              k(\gamma)=\frac{\omega}{-v(\gamma) \pm c(\gamma)}.
                    \end{equation}

\subsubsection{Die Amplitude}
          In erster Ordnung
                    \begin{equation} \label{ringerste}
                              fw_1 + i(\xi'w_0+2\xi w_0')=0
                    \end{equation}
          ergibt sich die Amplitude $w_0$, da $f=0$ gilt.
          Zu lösen ist
                    \begin{equation} \nonumber
                              \begin{split}
                                          &(\xi w_0^2)'=0 \\
                              \rightarrow &\xi w_0^2 = const. \\
                              \rightarrow &w_0=\frac{const.}{\sqrt{\xi}}
                              \end{split}
                    \end{equation}
          Mit Gl.(\ref{ringdisprel}) kann die Amplitude angegeben werden zu
                    \begin{equation} \label{ringamplidude}
                              w_0=\frac{1}{\sqrt{\omega \left\vert v(\gamma)-
                              \left(-v(\gamma)\pm \sqrt{c^2(\gamma)}\right)\right\vert}}.
                    \end{equation}

          Die Ebenen-Wellen im Kondensat haben die Form
                    \begin{equation} \label{ringlicht}
                              \theta_1(\tau,\gamma)=
                              \frac{e^{-i\omega t}e^{i\int{ d\gamma \frac{\omega}{-v(\gamma) \pm c(\gamma)}} }}{\sqrt{\omega \left\vert v(\gamma)-
                              \left(-v(\gamma)\pm \sqrt{c^2(\gamma)}\right)\right\vert}}
                    \end{equation}

\subsection{Einlaufende und auslaufende Koordinaten}
Mit der Metrik Gl.(\ref{ringmetrik}) für das ringförmige Kondensat kann das Linienelement
          \begin{equation} \label{ringabstand}
                    ds^2=-(c^2-v_{0\gamma})d\tau^2 -2v_{0\gamma}\gamma+\gamma^2 \, ,
          \end{equation}
(siehe Gl.(\ref{abstand})) aufgestellt werden.
Im Falle der Nullgeodäten ist $ds^2=0$ und es gilt:
          \begin{equation} \label{ringabstandnullgeo}
                 0=-(c-v_{0\gamma})-2v_{0\gamma}\dot{\gamma}+\dot{\gamma}^2.
          \end{equation}
Die Lösung der Gleichung lautet
          \begin{equation} \label{ringgammapunkt}
                    \dot{\gamma}=-v_{0\gamma}\pm c
          \end{equation}
und durch deren Integration ergeben sich die einlaufenden $\gamma_{-}^{\star}$ -
und auslaufenden $\gamma_{+}^{\star}$ Koordinaten
          \begin{equation}  \label{einauskoord}
                    \gamma_{\pm}^{\star}=\int^{\gamma}{\frac{d\gamma'}{-v(\gamma')\pm c(\gamma')}}.
          \end{equation}
Am Ereignishorizont $v(\gamma_{EH})= c(\gamma_{EH})$ divergiert $\gamma_{+}^{\star}$.
Ein Umschreiben des Linienelements in die neuen Koordinaten liefert
          \begin{equation} \label{abstandneukoordnull}
                    ds^2=-c(c^2du_+du_-),
          \end{equation}
wobei zusätzlich $u_{\pm}=\tau-\gamma_{\pm}^{\star}$ verwendet wurde.
Je näher eine auslaufende Welle dem Ereignishorizont kommt, desto langsamer vergeht die Zeit.
Es scheint für einen ruhenden Betrachter, welcher sich außerhalb des Bezugssystems der Welle befindet,
als ob die Ausbreitungsgeschwindigkeit der Welle gegen Null geht, so dass diese das Schwarze Loch nicht
verlassen kann.

\subsection{Simulation einer ebenen Welle im Kondensat}
Für $\omega=10$, $b=0.3$, $vc^2=1$ sind in Abb.(\ref{phasenring}) die Ergebnisse aus den vorherigen
Abschnitten dargestellt. Die numerischen Berechnungen wurden mit MATLAB durchgeführt. Der Quellcode zum
Programm befindet sich im Anhang (\ref{einewelle}).

          \begin{figure}
          \begin{center}
          \input{phasenring1.pstex_t}
          \caption[Ebene Welle im ringförmigen Kondensat]
          {\label{phasenring}Im obersten Bild sind die Schallgeschwindigkeit $c$ und die stationäre
          Hintergrundgeschwindigkeit $v$ dargestellt. Im mittleren Bereich fließt das Kondensat mit
          Überschallgeschwindigkeit, außerhalb darunter. Es gibt damit zwei Schnittpunkte, die zwei
          Ereignishorizonten entsprechen.\\
          In der Mitte sind die Wellenzahlen aus Gl.(\ref{ringlicht}) dargestellt. An den beiden
          Ereignishorizonten divergiert die Wellenzahl, wenn die Mode entgegengesetzt zum Kondensat
          propagiert.\\
          In der letzten Abbildung sind die einlaufenden und auslaufenden ebenen Wellen Gl.(\ref{ringlicht})
          geplotet. Während die auslaufende Welle an den beiden Ereignishorizonten divergiert, hat
          die Welle, die in das Schwarze Loch einläuft, kein divergentes Verhalten in der Wellenzahl.
          Der Ausschnitt zeigt eine Vergrößerung der Wellen am Ereignishorizont. Es ist zu sehen, dass
          die auslaufende Welle - rot dargestellt - den Ereignishorizont nicht passieren kann.}
          \end{center}
          \end{figure}

\subsection{Auswertung der Simulation}
In der Abb.(\ref{phasenring}) ist zu sehen, dass die ebene Welle, die sich mit dem Kondensat bewegt,
keine Divergenzen hat. Die gegen den Strom des Fluids laufende Welle divergiert an zwei Stellen.
Die Interpretation ist, dass beide Horizonte nur in gleicher Richtung zu überschreiten sind. Das
bedeutet, dass hier zum schwarzen Loch noch eine \textit{weiße Quelle} kommt. Zwischen den beiden
(in Stromrichtung des Kondensats) befindet sich das schwarze Loch.
Der erste Horizont (in Stromrichtung) kann nur in das schwarze Loch hinein passiert werden. Der zweite
Horizont ist nur in Richtung aus dem schwarzen Loch hinaus zu überqueren. Während beim ersten das
schwarze Loch nicht verlassen werden kann, ist es beim zweiten nicht möglich es zu betreten.

\subsection{Simulation eines Wellenpakets im Kondensat}
\begin{figure}[t]
          \begin{center}
          \input{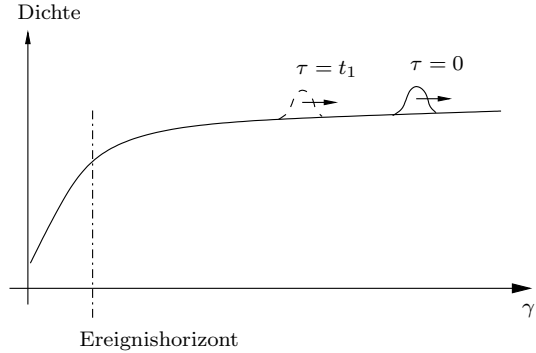}
          \caption[Ebene Welle im Kondensat]
          {\label{fourieridee}Zur Simulation wird ein Wellenpaket betrachtet, welches zum Zeitpunkt
          $\tau=0$ weit vom Ereignishorizont entfernt ist. Dazu muss die Breite des Wellenpakets klein
          sein im Vergleich zum Abstand des Ereignishorizont.}
          \end{center}
          \end{figure}

Mit einer Fourier-Analyse können die ebenen Wellen zu einem Wellenpaket überlagert werden. Damit ist es
möglich,
das Verhalten einer kleinen Dichtemodulation im Kondensat zu simulieren.
Die Fourier-Transformation in diesem Fall ist
          \begin{equation} \label{f1}
                    \Theta(\gamma,\tau)=
                    \int d\omega \theta_{\omega}(\gamma)e^{-i\gamma \tau}\bar{\theta}(\omega),
          \end{equation}
wobei
          \begin{equation} \label{f2}
                    \theta_{\omega}(\gamma)=w_0(\gamma)e^{i\gamma_{\pm}^{\star} \omega}
          \end{equation}
ist. Die Fourier-Koeffizienten $\bar{\theta}(\omega)$ werden im Folgenden bestimmt.
Zum Zeitpunkt $t=0$ wird ein Gauß'sches Wellenpaket, weit entfernt vom Ereignishorizont
betrachtet.
Ist die Breite des Wellenpakets klein gegenüber dem Abstand vom Ereignishorizont $\gamma_0$, können
$w_0(\gamma)\equiv w_0(\gamma_0)$ und $\gamma_{\pm}^{star}\equiv \gamma_{\pm}^{star}(\gamma_0) $ als
Konstanten betrachtet werden (siehe Abb.(\ref{fourieridee})).
Die Fourier-Koeffizienten im flachen Räumen sind
für ein Gauß'sches Wellenpaket
          \begin{equation} \label{f4}
                    \Theta(\gamma)=e^{-\frac{(\gamma-\gamma_0)^2}{2\sigma}}e^{ik_0\gamma}
          \end{equation}
Damit kann man die Fourier-Koeffizienten im Frequenzraum bestimmen:
           \begin{equation} \label{f3}
                    \bar{\theta}(\omega_0)=\frac{\gamma_{\pm}^{\star}(\gamma_0)}{w_0(\gamma_0)\pi}
                    \int \Theta(\gamma) e^{-i\omega_0\gamma_{\pm}^{\star}(\gamma_0) (\gamma-\gamma_0)}.
          \end{equation}
Damit ist das Wellenpaket am Ort $\gamma_0$ zur Zeit $\tau=0$ bestimmt.
Da die Fourier-Koeffizienten im Frequenz-Raum nicht vom Ort (bzw. Winkel) und der Zeit abhängen,
bleiben diese für alle Zeiten und überall im Ort gleich.\\
Zunächst kann man mit Gl.(\ref{f4}) die Fourier-Koeffizienten Gl.(\ref{f3}) an der Stelle $\gamma_0$
be\-rechen. Das Ergebnis wird dann in Gl.(\ref{f1}) eingesetzt.\\
Die Simulation wurde wieder mit MATLAB durchgeführt, und der Quelltext befindet sich im
Anhang (siehe (\ref{simulationwellenpaket})).\\
          \begin{figure}[t]
          \begin{center}
          \input{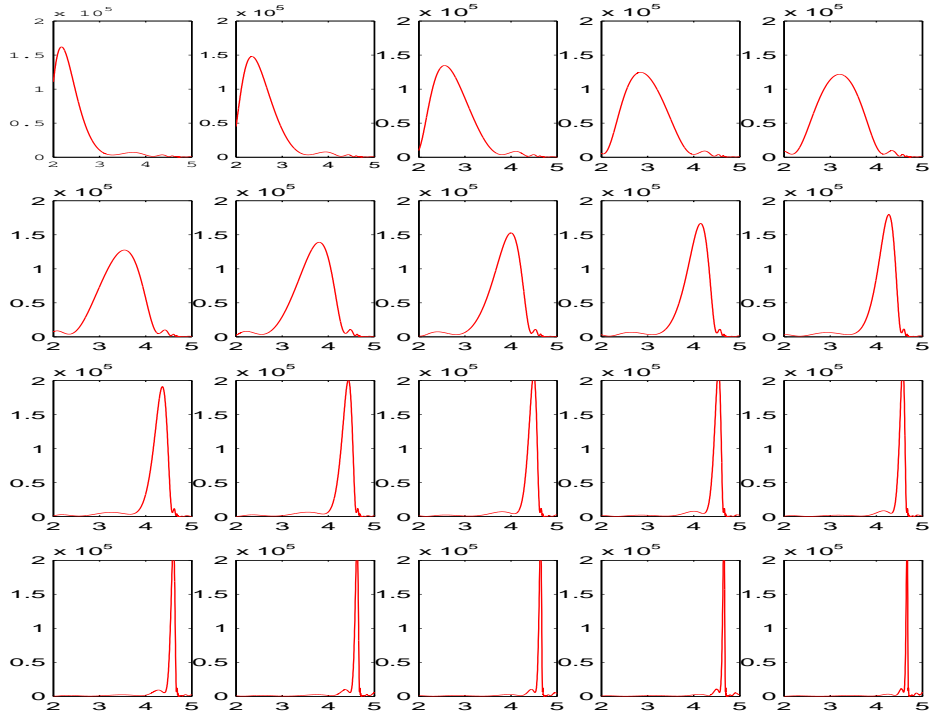}
          \caption[Wellenpaket im Kondensat]
          {\label{ringwelle}Das erste Bild links oben zeigt die Welle bei $\tau=0$, im
          zweiten Bild bei $\tau=-1$ und so weiter. Am Anfang handelt es sich um ein
          gaußförmig ausgeformtes Wellenpaket. Je länger es in der Zeit rückwärts propagiert,
          desto mehr verformt es sich. Den Ereignishorizont kann es nicht passieren.}
          \end{center}
          \end{figure}

\subsubsection{Auswertung der Simulation}
Das Ergebnis der Simulation ist in Abb.(\ref{ringwelle}) dargestellt. Es ist zu sehen,
dass es dem Wellenpaket nicht erlaubt ist, den Ereignishorizont rückwärts in der Zeit zu überschreiten.
Anders formuliert, ein Wellenpaket, das sich außerhalb eines schwarzen Lochs von diesem wegbewegt,
kann nicht innerhalb des Ereignishorizonts gestartet sein.

\section{Die Strahlung}
Die Strahlung kann mit der Feldgleichung Gl.(\ref{GPme3}) für das Kondensat berechnet werden,
welche in Kapitel (\ref{zigarrenstrahlung}) bereits hergeleitet wurde.
Für das ringförmige Kondensat - mit der dimensionslosen Zeit $\tau$ und der auf eine Dimension
eingeschränkten Bewegungsrichtung $\gamma$ - ist die Feldgleichung
          \begin{equation} \label{ringfeldgleichung}
                    i\dot{\hat{\tilde{\Phi}}}(\tau,\gamma)=-\frac{1}{2} \hat{\tilde{\Phi}}(\tau,\gamma)''
                    +\left[\frac{1}{2}\frac{c(\tau)''}{c(\tau)}-\frac{1}{2}\tilde{v}_0(\tau)^2
                    +\frac{\tilde{U}}{N}\vert\hat{\tilde{\Phi}}_0 \vert^2 \right]
                    \hat{\tilde{\Phi}}(\tau,\gamma)
                    +\frac{\tilde{U}}{N}\hat{\tilde{\Phi}}_0^2\hat{\tilde{\Phi}}^{\dag}(\tau,\gamma)
          \end{equation}
wobei $\tilde{U}$ von Gl.(\ref{Utilde}) verwendet und zusätzlich
          \begin{equation} \label{vtilde}
                    \tilde{v}_0=R\, v
          \end{equation}
eingeführt wurde.\\
Der Zustands-Operator kann mit dem Bogoliubov-Ansatz entwickelt werden:
          \begin{equation} \nonumber
                    \hat{\tilde{\Phi}}=\tilde{\Phi}_0 +
                    \hat{\tilde{\eta}}e^{\int d\gamma \tilde{v}_0(\gamma)}.
          \end{equation}
Eingesetzt in Gl.(\ref{ringfeldgleichung}) ergibt sich eine Bestimmungsgleichung für $ \hat{\tilde{\eta}}$:
          \begin{equation}  \label{ringfeldgleichungeta}
                    i\dot{\hat{\tilde{\eta}}}=
                    -\frac{1}{2} \left( \hat{\tilde{\eta}}'' - \frac{c''}{c}\hat{\tilde{\eta}} \right)
                    +i\left( \tilde{v}_0 \hat{\tilde{\eta}}' + \frac{1}{2}\tilde{v}_0'\hat{\tilde{\eta}} \right)
                    +\frac{\tilde{U}}{N}\rho_0 \left( \hat{\tilde{\eta}}+\hat{\tilde{\eta}}^{\dag} \right),
          \end{equation}

\subsection{Die Fourier-Transformierten Bogoliubov-Gleichungen}
In Kapitel (\ref{fourierebenewellen}) wurde gezeigt, dass es möglich ist, den Zustandsoperator
$\hat{\tilde{\eta}}$
          \begin{equation} \label{ringmodeexpansion}
                    \hat{\tilde{\eta}}(\tau,\gamma)=\sum_{\omega,n}\left(
                    e^{-i\omega \tau} e^{in \gamma}\hat{A}_{\omega,n}u_{\omega,n}(\gamma)
                    +
                    e^{i\omega^* \tau} e^{-in \gamma}\hat{A}_{\omega,n}^{\dag}v_{\omega,n}^*(\gamma)
                    \right)
          \end{equation}
nach ebenen Wellen zu entwickeln.
Einsetzen in die Gl.(\ref{ringfeldgleichungeta}) für $ \hat{\tilde{\eta}}$ ergeben
          \begin{equation} \label{ringfeldmitmodeexp}
          \begin{split}
                    h_- u_{\omega,n}
                    -i\left[ (n-\tilde{v}_0)u_{\omega,n}' -\frac{1}{2}(\tilde{v}_0' u_{\omega,n})  \right]
                    +\frac{1}{2}\left[ \frac{c''}{c}u_{\omega,n}-u_{\omega,n}'' \right]
                    +\tilde{c}_0^2 v_{\omega,n}=&0  \\
                    h_+ v_{\omega,n}
                    -i\left[ (n+\tilde{v}_0)v_{\omega,n}' +\frac{1}{2}(\tilde{v}_0' v_{\omega,n})  \right]
                    +\frac{1}{2}\left[ \frac{c''}{c}v_{\omega,n}-v_{\omega,n}'' \right]
                    +\tilde{c}_0^2 u_{\omega,n}=&0,
          \end{split}
          \end{equation}
wobei
          \begin{equation} \label{ringhpm}
                    h_{\pm}=\frac{n^2}{2} + \tilde{c}_0^2 \pm (n\tilde{v}_0 + \omega)
          \end{equation}
und
          \begin{equation} \label{ctilde}
                    \tilde{c}_0=\frac{\tilde{U}}{N}\rho_0
          \end{equation}
eingeführt wurden.
Ein Wechsel vom Orts- in den Impulsraum ermöglicht es, die beiden Gleichungen in der
folgenden Form zu schreiben
          \begin{equation} \label{bogfourier}
                    \omega
                    \left(\begin{array}{c}
                    u_{\omega,n}\\
                    v_{\omega,n}
                    \end{array}\right)
                    =
                    \left( \begin{array}{cc}
                    h_{np}^+ & f_{np} \\
                    -f_{np}  & h_{np}^-
                    \end{array} \right)
                              \left(\begin{array}{c}
                              u_{\omega,p}\\
                              v_{\omega,p}
                              \end{array}\right),
          \end{equation}
wobei
          \begin{equation} \label{ringfnp}
                    f_{np}=\frac{1}{2 \pi}\int_0^{2 \pi} d\gamma e^{-i(n-p)\gamma}c(\gamma)^2
          \end{equation}
und
          \begin{equation} \label{ringhnp}
                    h_{np}^{\pm}=\pm\frac{n^2}{2}\delta_{np}
                    +\frac{1}{2 \pi}\int_0^{2 \pi} d\gamma e^{-i(n-p)\gamma}
                    \left[ p\tilde{v}_0(\gamma)-\frac 12 \tilde{v}_0(\gamma)' \pm
                    \left( c(\gamma)^2 + \frac 12 \frac{\tilde{c}_0(\gamma)''}
                    {\tilde{c}_0(\gamma)} \right) \right]
          \end{equation}
sind.\footnote{Im Kapitel (\ref{zigarrenstrahlung}) wurde anstatt der Fourier-Transformation vom Orts-
in den Impulsraum die WKB-Methode angewendet. Damit war es möglich, die zweiten Ableitungen von
$u$ und $v$ zu vernachlässigen. Hier fallen die ersten und zweiten Ableitungen von $u_{\omega,n}$
und $v_{\omega,n}$ durch die Transformation weg, da
$$ u_{\omega,n}(\gamma)'=\left( \frac{1}{2 \pi}\int_{0}^{2 \pi}d\gamma u_{n,p}e^{-i(n-p)\gamma}\right)'
=u_{n,p}(-i(n-p))\delta_{np}=0 $$}\\
Diese beiden Fourier-Transformationen werden im Anhang (\ref{rechnungen}) exakt berechnet.
Es ergibt sich:
          \begin{equation}  \label{ringfnpfourier}
                    f_{np}=\frac{\tilde{U}}{2 \pi}\left( \delta_{n,p} + \frac{b}{2}\delta_{n,p+1}+ \frac{b}{2}\delta_{n,p-1}\right)
          \end{equation}
und
          \begin{equation}  \label{ringhnpfourier}
                    h_{np}^{\pm}=\frac{1}{2}(n+p)\,m^{\star}\,\sqrt{1-b^2}\,\alpha_{s=p-n}
                    \pm \left( f_{np}+\frac{4n^2-1}{8}\delta_{n,p}+\frac{1-b^2}{8} \beta_{s=n-p} \right).
          \end{equation}
Die beiden Koeffizienten $\alpha_{s=p-n}$ und $\beta_{s=n-p}$ sind im Anhang in Gl.(\ref{alpharing}) und Gl.(\ref{betaring})
zu finden. Es handelt sich um zwei unendliche Reihen.

\subsection{Zur Lösung der Gleichungen}
In der Veröffentlichung von Cirac et al. \cite{cirac1} findet sich ein numerischer Lösungsweg, der im Folgenden kurz
skizziert wird.

Bei großen Wellenlängen - kleinen $n$-Werten - treten Instabilitäten auf. Es kann gezeigt werden, dass die Einflüsse der
höheren Terme in den Fourier-Koeffizienten $\alpha_{s=p-n}$ und $\beta_{s=n-p}$, auf die Instabilitäten vernachlässigbar sind.
Es ist daher zulässig, einen Cutoff $Q$ einzuführen, der die Reihen der Fourier-Koeffizienten endlich macht.\\
Die Bogoliubov-Gl.(\ref{ringfeldmitmodeexp}) enthält nach der Einführung des Cutoffs $2(Q+1)\times 2(Q+1)$ Matrizen, welche
numerisch diagonalisiert werden können. Aus den sich ergebenden Moden sind jene erlaubt, welche zusätzlich die
Normierungsbedingung
          \begin{equation} \nonumber
                    \int d\gamma(u_{\omega^{\star},n}u_{\omega',n'}
                    -v_{\omega^{\star},n}v_{\omega',n'})=\delta_{nn'}\delta_{\omega\omega'}
          \end{equation}
erfüllen.\\
Die numerische Simulation zeigt, dass sowohl energetische wie dynamische Instabilitäten auftreten.

\section{Zusammenfassung}

Auch in diesem System ist es über die effektive Metrik möglich, eine Analogie zum schwarzen Loch herzustellen.
Eine von außen angeregte Störung hat die gleichen Bewegungsgleichungen, wie ein masseloses Teilchen im
Gravitationsfeld.
Speziell für lichtartige Teilchen, die als ebene Wellen beschrieben werden, liefert die Bewegungsgleichung
die Nullgeodäten. Es ergeben sich zwei Lösungen, die einlaufenden und auslaufenden Wellen. Erstere propagieren
in Bewegungsrichtung des Fluids, die anderen dagegen.
Die Simulation einer ebenen Welle mit einer festen Frequenz im Kondensat zeigt, dass an zwei Punkten
Divergenzen für die auslaufenden Wellen auftreten. Zwischen diesen beiden Ereignishorizonten liegt
(in Stromrichtung der Hintergrundgeschwindigkeit gesehen) das schwarze Loch. Innerhalb der Horizonte
ist es nicht möglich, entgegen dem Fluidstrom zu propagieren. Der erste Horizont kann nur in Richtung des
schwarzen Lochs überschritten werden, der zweite führt unweigerlich aus diesem hinaus. Es ist nicht möglich,
in das schwarze Loch über den zweiten Ereignishorizont zu gelangen, analog einer weissen Quelle im Universum.
Überlagert man alle Wellen zu einem Wellenpaket, kann das Verhalten der gesamten Störung
simuliert werden. Es konnte gezeigt werden, dass ein Wellenpaket außerhalb des schwarzen Lochs seinen
Ursprung nicht innerhalb des Ereignishorizonts\footnote{Hier handelt es sich um den ersten Ereignishorizont, der nur Bewegungen in Richtung des
schwarzen Lochs zulässt.}
haben kann.   \\
Für die Geodäten ist vorausgesetzt worden, dass sich die Amplituden nur wenig ändern.
Die Amplituden der ebenen Wellen divergieren jedoch am Ereignishorizont.
Im zweiten Teil werden wieder die Fluktuationen um den Grundzustand behandelt.
Hier treten die Divergenzen in den Amplituden nicht auf. Da diese Rechnungen um eine
Ordnung genauer (im Bogoliubov-Ansatz) sind, haben die Divergenzen in den Amplituden keine physikalische
Ursache.\\
          \begin{figure}[t]
          \begin{center}
          \vspace{-3cm}\input{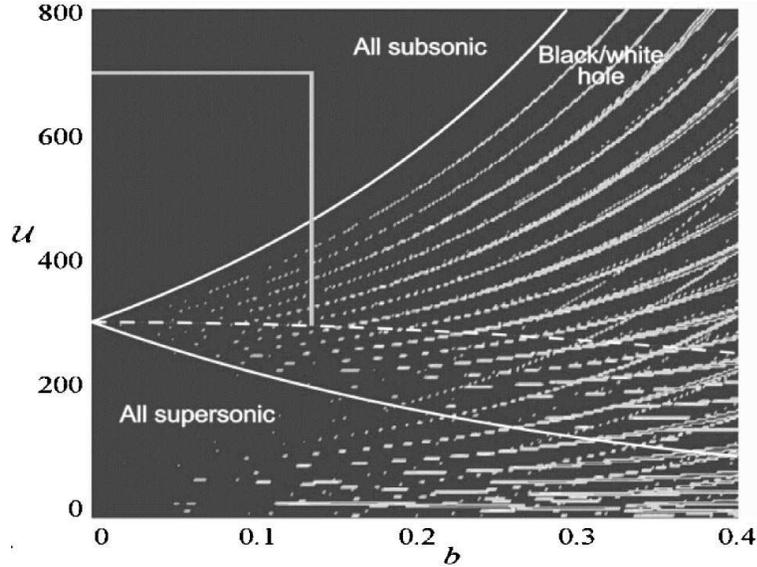} \vspace{-3cm}
          \caption[Dynamische Instabilitäten im ringförmigen Kondensat]
          {\label{ringsdyninstabil}Darstellung der dynamischen Instabilitäten.
           In Abhängigkeit von den Kondensatgrößen $\tilde{U}$ und $b$ strahlt das System (weiße Bereiche),
           oder nicht (schwarze Bereiche). Kleinere Plots mit höherer Auflösung zeigen, dass es sich bei
           den weißen Punkten tatsächlich um verbundene Bereiche handelt.
           Dem Artikel von Cirac et al. \cite{cirac1} ist zusätzlich zu entnehmen, wie eine solche Konfiguration
           zu erzeugen ist.}
          \end{center}
          \end{figure}
Die Bogoliubov-Gleichungen können durch eine Fourier-Analyse exakt berechnet werden. Als Lösungsweg der sich
ergebenden Gleichungen wurde von Cirac et al. \cite{cirac1} ein numerisches Verfahren entwickelt. Zunächst wurden
die Bogoliubov-Gleichungen vereinfacht und anschließend numerisch auf Diagonalform gebracht.
Es zeigt sich in der Simulation, dass die Fluktuationen destabilisierend auf den Kondensatzustand
wirken können.\\
Es existieren stabile und instabile Bereiche. Ob ein System stabil ist oder nicht,
hängt nur von den gewählten Kondensatvariablen ab.
Für die Instabilität kommen sowohl energetische, als auch dynamische Ursachen in Frage.
Die Zeitskala der energetischen ($\omega_{reell}<0$) Instabilitäten ist jedoch so groß, dass sie
vernachlässigbar sind. In Kapitel (\ref{forfluctiations})) wurde beschrieben, wie die Fluktuationen
sich energetisch instabil auf das Kondensat auswirken können.\\
Nicht vernachlässigbar sind hingegen die dynamischen Instabilitäten, die für komplexe Frequenzen $\omega$
auftreten. Die numerischen Berechnungen ergeben, dass
der Betrag der komplexen Eigenfrequenzen sensibel gegenüber
kleinen Änderungen der Kondensatgrößen $(\tilde{U},b,m^{\star})$ ist.
Für $m^{\star}=7$ ist die maximale Anzahl der imaginären Eigenfrequenzen
in Abb.(\ref{ringsdyninstabil}) dargestellt.

                    \part{Das Universum im Bose-Einstein-Kondensat\label{kapuniversum}}
                              \chapter{Das frei expandierende Kondensat\label{expandinguniverse}}
Im Unterschied zu den beiden vorhergehenden Kapiteln wird in diesem Abschnitt ein Kondensat
untersucht, das nicht nur vom Ort, sondern auch von der Zeit abhängt.
Am Anfang ruht das Kondensat in einem äußeren Potential, welches schlagartig abgeschaltet wird.
Das nun freie Kondensat beginnt zu expandieren.
Das Ziel dieses Abschnitts ist es, die Dynamik für eine von außen injizierte Dichtemodulation
zu berechnen.
Es wird gezeigt, dass eine Analogie zwischen dem sich frei ausbreitenden Kondensat und der
Expansion des de-Sitter-Universums besteht.\footnote{
Erw\'ahnenswert ist eine bisher unver\'offentlichte Arbeit von L. Garay \cite{garay2}, welche sich ebenfalls damit besch\'aftigt.}

\section{Das Modell}
\begin{figure}[t]
          \begin{center}
          \input{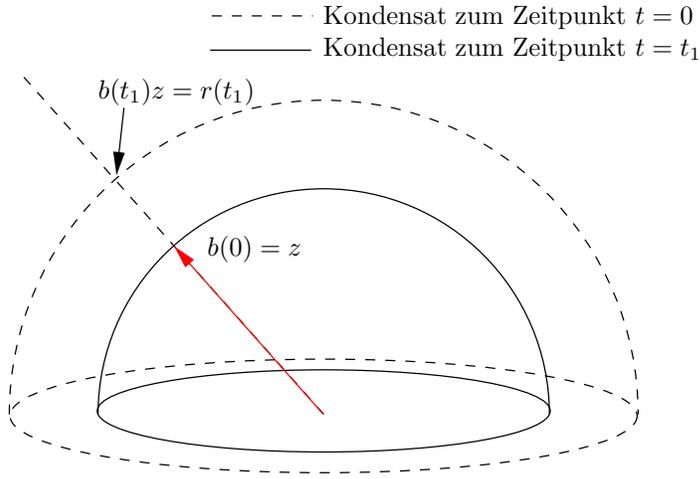}
          \caption[Frei expandierendes Kondensat]
          {\label{free}Zu Zeiten $t<0$ befindet sich das Kondensat (orange) in einem sphärischen
          Potential. Am Zeitpunkt $t=0$ verschwindet das Potential durch plötzliches Abschalten.
          Die Kondensatwolke beginnt sphärisch zu expandieren.}
          \end{center}
          \end{figure}

An das äußere Potential wird nur die Bedingung geknüpft, dass es sphärisch und dreidimensional sein muss.
Dadurch wird erreicht, dass nach dem Abschalten des Potentials
die Kondensatwolke homogen radial expandiert.
Punkte im Kondensat bewegen sich auf radialen Bahnen,
so dass
          \begin{equation} \label{neue Koordinaten}
                    r(t)=z(r,t=0) \, b(t)
          \end{equation}
gilt (siehe Abb.(\ref{free})).
Die neue Koordinate $z$ ist zeitunabhängig.
Mit der Produktregel für die Ableitung von $r(z,b)$ ergibt sich die radiale Hintergrundgeschwindigkeit
          \begin{equation} \label{vr}
                    v_r=\frac{\dot{b}}{b}r.
          \end{equation}
Das Volumen der Kondensatwolke am Zeitpunkt $t=0$ ist
          \begin{equation} \nonumber
                    V_0=\int dx \, dy \, dz \, \, z(r,t=0).
          \end{equation}
Mit der Zeit wächst das Volumen auf
          \begin{equation} \nonumber
                    V(t)=b(t)^3 V_0
          \end{equation}
an.
Das Volumen ist indirekt proportional zur Dichte. Für Gl.(\ref{schallgeschwindigkeit}) ergibt sich
          \begin{equation} \label{cschallgeschwindigkeit}
                    c(t,r)=\frac{1}{b(t)^{3/2}}c_0(r),
          \end{equation}
die radiale Schallgeschwindigkeit.
Ziel ist es, die Funktionen für $c(t,r)$ und $v(t,r)$ derart zu bestimmen, dass eine Analogie zum
de-Sitter-Universum (siehe Abschnitt (\ref{sitter})) besteht.

\section{Die Wellengleichung}
Im Falle eines frei expandierenden Kondensats ist $V_{ext}(t>0,\vec{r})=0$.\\
Induziert man mit dem Laser zusätzlich eine Störung in der Kondensatwolke, werden die Dichte
$$\rho(\vec{r},t) =\rho_{0} (\vec{r},t) + \varepsilon \rho_{1} (\vec{r},t)$$
und Phase
$$\theta(\vec{r},t) = \theta_{0} (\vec{r},t) + \varepsilon \theta_{1} (\vec{r},t),$$
minimal moduliert.\\
Die Berechnungen für die Metrik sind in den Kapiteln (\ref{wellengleichungignacio})
und (\ref{anhanghydro}) zu finden.
Es handelt sich hierbei nicht um exakte Kalkulationen. Es war erforderlich, die
Thomas-Fermi-Näherung (Gl.(\ref{fermithomas})) zu verwenden. Diese gilt aber nur,
wenn die kinetische Energie klein im Vergleich zur restlichen Energie ist:
          \begin{equation}   \nonumber
                    E_{Kin}\ll E_{pot}+E_{int}.
          \end{equation}
Am Anfang der Expansion ist
          \begin{equation}   \nonumber
                    E_{Kin}\ll E_{pot}
          \end{equation}
diese Bedingung gewährleistet.
Mit der Zeit wandelt sich potentielle in kinetsiche Energie um.
Es ist notwendig, die Rechnungen zeitlich
einzuschränken.
Für die Störung muss - um die Thomas-Fermi-Näherung anfangs verwenden zu können - zusätzlich gelten:
          \begin{equation}  \label{fermithomasstorung}
                    \begin{split}
                    \frac{\nabla^2 \sqrt{ \left( \rho_0 + \varepsilon \rho_1 \right) }}
                    {\sqrt{\rho_0 + \varepsilon \rho_1}}
                    \approx
                    \frac{\nabla^2 \sqrt{ \left( \rho_1 \right) }}
                    {\sqrt{\varepsilon \rho_1}} &\ll Ng\rho_1 \\
                    k^2  & \ll  \frac{4mNg}{\hbar^2}
                    \end{split}.
          \end{equation}
Unter diesen Voraussetzungen ergibt sich für die Metrik:
          $$             g^{\mu \nu}=c^2\left( \begin{array}{cccc}
          1            & v_x                  & v_y            & v_z \\
          v_x          & -c^2+v_{x}^2         & v_{y}v_{x}     & v_{z}v_{x} \\
          v_y          & v_{x}v_{y}           & -c^2+v_{y}^2   & v_{z}v_{y}  \\
          v_z          & v_{x}v_{z}           & v_{y}v_{z}     & -c^2+v_{z}^2
          \end{array} \right).$$
Damit kann die Wellengleichung geschrieben werden als
$$\partial_{\mu}(\sqrt{-g} \, g^{\mu\nu} \, \partial_{\nu}\theta_{1})=0,$$
was einer Wellengleichung mit Metrik und damit im gekrümmten Raum entspricht.

Der Unterschied zu den beiden vorherigen Kapiteln ist, dass es sich hier
um Geschwindigkeiten $\vec{v}(t,\vec{r})$ und $c(t,\vec{r})$ handelt, die explizit von der Zeit abhängen.
Die Herleitung in Kapitel (\ref{wellengleichungignacio}) ist jedoch allgemein gehalten und nicht auf
ein zeitunabhängiges Kondensat beschränkt.\\
Als nächstes wird die
Metrik an das frei expandierende Kondensat angepasst.

\section{Das Linienelement für ein sphärisches Kondensat}
Die Kovariante zu $g^{\mu \nu}$ ist
          $$g_{\mu \nu}=c\left( \begin{array}{cccc}
          -(c^2-v^2)             & -v_x         & -v_y              & -v_z \\
          -v_x                   & 1            & 0                 & 0 \\
          -v_y                   & 0            & 1                 & 0  \\
          -v_z                   & 0            & 0                 & 1
          \end{array}\right).$$
Das Linienelement aus Gl.(\ref{abstand}) für diese Metrik ist
          \begin{equation} \label{universumabstand}
                    dl^2=c\left[-(c^2-v^2)dt^2-2 \vec{v} d\vec{r} dt + dx^2 + dy^2 +dz^2 \right],
          \end{equation}
wobei
          \begin{equation} \nonumber
                    d\vec{r} = \left(  \begin{array}{c}
                    dx\\
                    dy\\
                    dz\\ \end{array} \right)
          \end{equation}
verwendet wurde.
Da die Kondensatwolke radial expandiert, ist es sinnvoll,
von den bisher verwendeten kartesischen in sphärische Koordinaten zu wechseln.\\
Für eine radiale Hintergrund - und Schallgeschwindigkeit
kann das Linienelement in Kugelkoordinaten angegeben werden:
          \begin{equation}
                    dl^2=c(r,t)\left[-(c(r,t)^2 - v(r,t)^2)dt^2 -2 \, v(r,t) \, dr dt + dr^2 + r^2 \, d\Omega^2 \right].
                    \label{lineelemntsph}
          \end{equation}
Um das Linienelement zu diagonalisieren, wird $r$ in den neuen Koordinaten ausgedrückt.
Dazu wird das erste
          \begin{equation}
                    dr=b(t) \,  dz + \frac{\dot{b}(t)}{b(t)} r(t) \, dt
          \end{equation}
und zweite
          \begin{equation} \nonumber
                    dr^2=b^2dz^2 + 2bv_r dz dt + v_r^2 dt^2
          \end{equation}
totale Differential berechnet.
In Gl.(\ref{universumabstand}) eingesetzt ergibt sich
          \begin{equation}
                    dl^2=c_0(r) \left[ -c_0(t)^2 \, b(t)^{-9/2} \, dt^2 + b(t)^{1/2} \, dz^2 +  b(t)^{1/2} \, z^2 \, d\Omega^2 \right].
                    \label{lineelementsphnew}
          \end{equation}

\section{Die de-Sitter-Metrik im Kondensat}
In der de-Sitter-Metrik hängt $g_{00}$ weder von der Zeit, noch vom Ort ab (siehe Gl.(\ref{desitermetrik})).
Die Zeitabhängigkeit kann durch die Transformation
          \begin{equation} \label{zeitweg}
                    \begin{split}
                    \tau=&\int \sqrt{g_{00}}dt\\
                    \rightarrow \ d\tau=&b(t)^{-9/4}dt
                    \end{split}
          \end{equation}
beseitigt werden.
Die Ortsabhängigkeit kann nicht beseitigt werden.

Für ein konkretes äußeres Potential kann geprüft werden, ob ein Bereich existiert,
in dem die Dichte annähernd konstant ist.
Zum Beispiel ist
der Grundzustand im harmonischen Potential zum Zeitpunkt $t=0$ durch
          \begin{equation}
                    \Phi_0(t=0,\vec{r})=\sqrt{\frac{\mu-\frac{1}{2}m \omega(t)^2 \vec{r}^2}{Ng}}.
                    \footnote{Der Grundzustand kann berechnet werden, indem in Gl.(\ref{GPE1}) das harmonische Potential
$ V_{ext}=\frac{1}{2} m \omega (t)^2 \vec{r}^2 $ eingesetzt wird. Mit der Thomas-Fermi-Näherung
kann die kinetische Energie vernachlässigt werden.}
          \end{equation}
gegeben.
Hierbei handelt es sich um die Kugelgleichung. In Abb.(\ref{free}) befand sich das Kondensat
in einem dreidimensionalen harmonischen Potential.
Eine Einschränkung der Beobachtung auf den Bereich um das Zentrum des Potentials
erlaubt es, die Schallgeschwindigkeit
als konstant zu betrachten.\\

Das Linienelement mit der neuen Zeitkoordinate $\tau$ auf einem beschränkten Bereich im Kondensat ist
damit:
          \begin{equation}
                    dl^2=c_0(r)\left[ -c_0(r)^2 d\tau^2 + b\left(t(\tau)\right)^{1/2} dz^2 +
                    b\left(t(\tau)\right)^{1/2} z^2 d\Omega^2 \right].
          \end{equation}
Ein Vergleich mit der de-Sitter-Metrik zeigt,
dass
          \begin{equation} \label{bedingunggefunden}
                    \sqrt{b(t)}=\exp(2H\tau)
          \end{equation}
sein muss (siehe Abschnitt (\ref{desitermetrik})).
Durch Integration von Gl.(\ref{zeitweg})
          \begin{equation} \nonumber
                    \begin{split}
                    \int_0^{\tau} d\tau' &= \int_0^{t} b(t') dt'\\
                    \int_0^{\tau} d\tau' & = \int_0^{t}\exp(4H\tau)^{-9/4} dt'\\
                    \int_0^{t} dt'&=\int_0^{\tau} \exp(H\tau)^{9} d\tau'\\
                    t & = \frac{1}{9H}\exp({9H}\tau)- \frac{1}{9H} \\
                    \rightarrow & \exp(9H\tau)=9Ht + 1
                    \end{split}
          \end{equation}
kann $b(\tau)$ in Abhängigkeit von $t$ angegeben werden (siehe \cite{garay2}):\\
          \begin{equation}
                    b(t)= (9Ht + 1)^{4/9}
          \end{equation}

In der Veröffentlichung von Castin et al. \cite{scaling} wird mit einer Skalierungs-Transformation
die Zeitentwicklung für ein frei expandierendes Kondensat berechnet. Das Kondensat befindet
sich vor der Expansion in einem dreidimensionalen harmonischen Potential.
Die Formel für $\omega$ in bewegten Koordinaten ist
          \begin{equation} \label{auspaper}
                    \ddot{b}(t)=\frac{\omega_0^2}{b(t)^4}-\omega(t)^2\ddot{b}(t).
          \end{equation}
In \cite{garay2} ist die Frequenz ebenfalls berechnet worden. Es ergibt sich:
          \begin{equation}
                    \omega(t)=\omega_0 b(t)^{-5/2}\left(1+\frac{20 \, H^2}{\omega_0^2}b(t)^{1/2} \right).
          \end{equation}

\section{Zusammenfassung}
Für das frei expandierende Kondensat kann eine effektive de-Sitter-Metrik eingeführt werden.
Das Problem der ortsabhängigen Schallgeschwindigkeit läßt sich durch Einschränkung auf einen
kleinen Raumbereich, in dem die Dichte hinreichend konstant ist, beseitigen.
Befindet sich das Kondensat vor Beginn der Expansion im harmonischen Potential, werden nur Störungen
im Zentrum des Potentials betrachtet.
Während der Ausbreitung nimmt die kinetische Energie zu. Für kleine Störungen
(siehe Gl.(\ref{fermithomasstorung})) ist die Thomas-Fermi-Näherung dennoch gültig, solange die
Energie der Störung klein gegenüber der Wechselwirkungsenergie ist.
Mit der Größe der
Kondensatwolke nehmen die atomaren Abstände im Kondensat zu, wodurch die Wechselwirkung abnimmt.
Die Gleichungen werden ungültig.

                    \part{Zusammenfassung und Ausblick\label{kapalles}}

In allen drei Kondensaten - zigarrenförmig, ringförmig und frei expandierend - konnte eine effektive
Metrik eingeführt werden.
Das zigarren - und ringförmige Kondensat sind geeignete Systeme zur Simulation eines Schwarzen
Lochs.
Eine injizierte Dichtemodulation, die gegen die Stromrichtung des Fluids propagiert,
divergiert am Ereignishorizont. Dieser Effekt kann im Experiment nachgewiesen werden.
Ohne äußere Einwirkung werden Phononen emittiert.
Diese Strahlung entsteht aus dynamischen Instabilitäten, die durch gebundene Zustände im
Inneren des Schwarzen Lochs hervorgerufen werden.
Die hier verwendete Bogoliubov-Näherung verliert jedoch nach kurzer Zeit ihre Gültigkeit.
Das Zeitfenster für die Messung der Strahlung ist daher nicht unbegrenzt.
Die Beantwortung der Frage, ob es sich bei dieser Strahlung um Hawking-Strahlung handelt, ist schwierig.
Zum einen bleibt offen, ob es sich um lokale, oder globale Strahlung handelt. Die Hawking-Strahlung
eines Schwarzen Lochs ist lokal an den Ereignishorizont gebunden. Die angeregten Moden im Kondensat
treten überall auf. Es bleibt aber zu prüfen, ob die Amplitude der Strahlung vom Ort abhängt.
Sie müsste überall, außer am Ereignishorizont, verschwindend klein sein.
Komplizierter ist das Problem der Stabilität. Schwarze Löcher werden in der Gravitationsphysik als
weitgehend stabil betrachtet. Instabil erzeugte Teilchen können als kleine Korrektur zur
eigentlichen Strahlung betrachtet werden, d. h. die Zeitskala ist groß gegenüber der Zeitskala
für stabile Hawking-Strahlung.\\

Das frei expandierende Kondensat kann als dynamisches Modell eines flachen Universums verwendet werden.
Es wurde die Metrik für eine sich radial ausbreitende Kondensatwolke bestimmt.
Unter der Bedingung, dass die Dichte in einem hinreichend großen Bereich annähernd konstant ist, und
einer Einschränkung in der Zeitdauer, ist es theoretisch möglich das de-Sitter-Universum im
Bose-Einstein-Kondensat zu simulieren. In der Arbeit von Garay \cite{garay2} wird
gezeigt, dass die Analogie zum de-Sitter-Universum erst ab drei Dimensionen möglich ist.
Die Frage, ob im frei expandierenden Kondensat Phononen erzeugt werden, bleibt offen.
In diesem Zusammenhang könnte versucht werden, dem Spektrum der Quasiteilchen eine Temperatur zuzuweisen.
Mit einem solchen Modell könnten verschiedene Szenarien für unser Universum im Labor getestet werden.
Ähnliche Anstrengungen wurden in \cite{fischer} verfolgt. Dort wurde für ein oszillierendes Kondensat
die Strahlung berechnet. Es stellte sich heraus, dass ein zweidimensionales Kondensat stabil ist,
während im dreidimensionalen Fall Phononen erzeugt werden.

          \appendix

          \part{Anhang}

\chapter{Quantenmechanische Streuprozesse}
Quantenmechanische Streuprozesse unterscheiden sich von klassischen Streuprozessen insofern, dass der Prozess nicht deterministisch
ist. Es können nur Wahrscheinlichkeiten für die einzelnen Streuwinkel angegeben werden.

\section{Reduziertes Zwei-Atom-Problem \label{Schwerpunktsystem}}
Sind zwei Atome durch ein Potential $V(\vec{r}_{12})$ verbunden, wobei $\vec{r}_{12}$ der Abstand ist,
kann das System zu einem Ein-Teilchen-Problem reduziert werden.
          \begin{figure}[h]
          \input{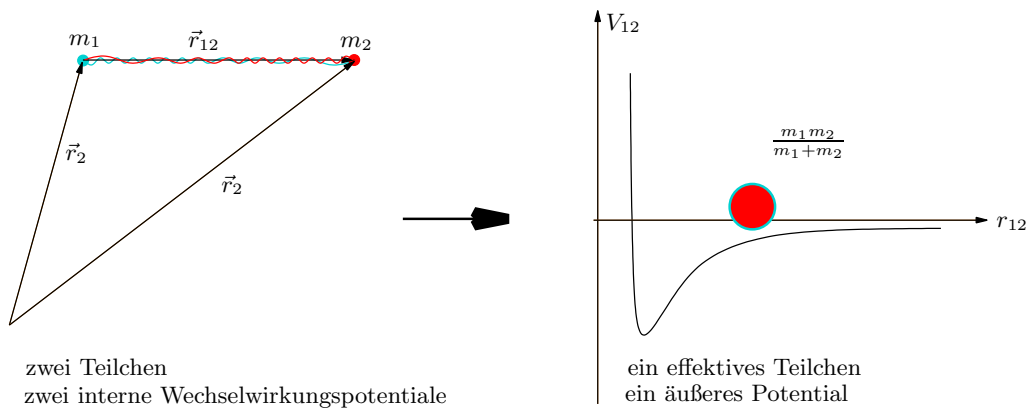}
          \caption[Reduziertes Zwei-Körper-Problem]
          {\label{externpot}Die Bewegung zweier wechselwirkender Atome zueinander kann auf die Bewegung eines Teilchens um den Schwerpunkt des Systems
          mit der \textit{reduzierten Masse}
          $\mu={m_1 m_2}/{m_1 + m_2}$, im äußeren Potential beschrieben werden.}
          \end{figure}
Die Bewegung im Schwerpunktsystem wird bestimmt durch
          \begin{equation}  \label{Hrel}
                    \hat{H}_{rel}=\frac{1}{2\mu}\nabla_{\vec{r}_{12}}^2+ V(\vec{r}_{12}),
          \end{equation}
wobei hier $m_1=m_2=m$, und damit die Masse des Schwerpunkts $\mu=m/2$,
$\vec{r}_{12}=\vec{r}_{2}-\vec{r}_{1}$, der Verbindungsvektor und der Relativimpuls
$\vec{p}_{12}=\vec{p}_{2}-\vec{p}_{1}$ als Ausgangspunkt für das
Korrespondenzprinzip\footnote{Das Korrespondenzprinzip ersetzt in der Hamilton-Funktion
Observablen durch Operatoren, hier Impuls durch Impulsoperator
$\vec{p}_{12}\rightarrow \frac{\hbar}{i}\nabla_{\vec{r}_{12}}$. }
verwendet wurde.

\section{Streuung zweier Atome \label{streuergebnisse}}
Um die Streuung zweier Atome zu beschreiben, bleiben wir in dem zuvor eingeführten Schwerpunktsystem, in welchem die Relativbewegung durch die Dynamik
eines Zustandes beschrieben werden kann.
Für ein zeitunabhängiges Potential sei der gestreute stationäre Zustand $\psi(\vec{r}_{12})$ ein Eigenzustand zum relativen
Hamilton-Operator $\hat{H}_{rel}$ mit den positiven Eigenwerten $E=\hbar^2k^2/2\mu$.
Die Schrödinger-Gleichung für das Problem lässt sich dann schreiben als
          \begin{equation}  \label{scatt1}
                    \left(\Delta+k^2 \right) \psi(\vec{r}_{12})= \frac{2\mu}{\hbar^2}V(\vec{r}_{12})\psi(\vec{r}_{12}).
          \end{equation}
Zusätzlich zu den gestreuten Zuständen können auch gebundene Zustände auftreten.
Zur Lösung dieser inhomogenen Differentialgleichung ist die
integrale Form von Gl.(\ref{scatt1})
          \begin{equation}  \label{scatt2}
                    \left(\Delta+k^2 \right) \psi(\vec{r}_{12})=\int d^3\vec{r}_{12}' \frac{2\mu}{\hbar^2}V(\vec{r}_{12}')\psi(\vec{r}_{12}')\delta(\vec{r}_{12}-\vec{r}_{12}').
          \end{equation}
aufschlussreich. Dies ist eine Differentialgleichung mit einer $\delta$-Distribution als Inhomogenität.
Die Lösungen solcher Differentialgleichungen sind bekannt, sie können mit Hilfe von sogenannten Greenschen Funktionen direkt angegeben werden.
Die $\delta-Distribution$ auf der rechten Seite von Gl.(\ref{scatt2}) wird dazu durch
          \begin{equation}  \label{green}
                    \left(\Delta+k^2 \right)\psi_G(\vec{r}_{12})=\delta(\vec{r}_{12}-\vec{r}_{12}')
          \end{equation}
ersetzt. Man bezeichnet die Lösung $\psi_G=-\frac{1}{4 \pi}\frac{e^{ik \vert\vec{r}_{12}-\vec{r}_{12}'\vert}}{\vert\vec{r}_{12}-\vec{r}_{12}'\vert}$
als Greensche Funktion zum Differentialoperator $\left(\Delta+k^2 \right)$.\\
Um die vollständige Lösung $\psi(\vec{r}_{12})$ zu erhalten, muss zuvor noch die Lösung der homogenen Differentialgleichung $\psi_0(\vec{r}_{12})$
addiert werden. Für diese muss 
          \begin{equation} \label{hompsi}
                    \left(\Delta+k^2 \right)\psi_0(\vec{r}_{12})=0
          \end{equation}
gelten. Der gestreute Zustand ist dann
          \begin{equation}  \label{inhompsi}
                    \psi(\vec{r}_{12})=\psi_0(\vec{r}_{12})-\frac{2 \mu}{4 \pi \hbar^2}\int{d^3\vec{r}_{12}'
          \frac{e^{ik \vert\vec{r}_{12}-\vec{r}_{12}'\vert}}{\vert\vec{r}_{12}-\vec{r}_{12}'\vert}}V(\vec{r}_{12}')\psi(\vec{r}_{12}').
          \end{equation}\\
Bei der homogenen Lösung $\psi_0(\vec{r}_{12})=exp(i\vec{k}\vec{r}_{12})$ handelt es sich um eine einlaufende freie Welle, wobei
die partikuläre Lösung eine auslaufende Kugelwelle darstellt.

                    \chapter{Berechnungen der hydrodynamischen Gleichungen\label{anhanghydro}}

\section{Berechnung der einzelnen Terme}
Hier werden die Terme aus Gl.(\ref{GPE2mitdelta})
         \begin{equation} \nonumber
                    \begin{split}
                    i\hbar\partial_t \sqrt{\rho_0 + \varepsilon\rho_1}e^{i(\theta_0 +  \varepsilon \theta_1 )}
                    =&\left( \kin + V_{ext}(\vec{x}) \right) \sqrt{\rho_0 + \varepsilon\rho_1}e^{i(\theta_0 +  \varepsilon \theta_1 )}\\
         &+ \left( N_0 g \left\vert \sqrt{\rho_0 + \varepsilon\rho_1}\right\vert^2
                    \right) \sqrt{\rho_0 + \varepsilon\rho_1}e^{i(\theta_0 +  \varepsilon \theta_1 )}
                    \end{split}
          \end{equation}
berechnet.

\paragraph{\fbox{$i\hbar\partial_t \sqrt{\rho_0 + \varepsilon\rho_1}e^{i(\theta_0 +  \varepsilon \theta_1 )}$}}
          \begin{equation} \nonumber
                    \begin{split}
                    =e^{i(\theta_0 +  \varepsilon \theta_1 )}\sqrt{\rho_0 + \varepsilon \rho_1} \times
                    &\left[ -\hbar \dot{\theta}_0 \right. \\
                    &\left. -\varepsilon   \hbar \dot{\theta}_1 \right]\\
          +i\frac{e^{i(\theta_0 +  \varepsilon \theta_1 )}}{\sqrt{\rho_0 + \varepsilon \rho_1}} \times
                    &\left[ +\frac{\hbar}{2} \dot{\rho}_0 \right. \\
                    &\left. +\varepsilon \frac{\hbar}{2}\dot{\rho_1} \right]\\
                    \end{split}
          \end{equation}

\paragraph{\fbox{$\kin \sqrt{\rho_0 + \varepsilon\rho_1}e^{i(\theta_0 +  \varepsilon \theta_1 )}$}}
          \begin{equation} \nonumber
                    \begin{split}
                    =e^{i(\theta_0 +  \varepsilon \theta_1 )}\sqrt{\rho_0 + \varepsilon \rho_1} \times
                    &\left[ +\frac{\hbar^2}{2m}(\nabla \theta_0)^2 \right. \\
                    &\left. +\varepsilon   \frac{\hbar^2}{m} \nabla\theta_0 \nabla\theta_1 \right.\\
                    &\left. +\varepsilon^2 \frac{\hbar^2}{2m} (\nabla \theta_1)^2 \right]\\
          +i\frac{e^{i(\theta_0 +  \varepsilon \theta_1 )}}{\sqrt{\rho_0 + \varepsilon \rho_1}} \times
                    &\left[ \frac{\hbar^2}{2m} \nabla(\rho_0 \nabla\theta_0) \right. \\
                    &\left. -\varepsilon  \frac{\hbar^2}{2m} \nabla(\rho_1 \nabla \theta_0 + \rho_0 \nabla \theta_1) \right.\\
                    &\left. -\varepsilon^2 \frac{\hbar^2}{2m} \nabla(\rho_1 \nabla \theta_1) \right]
                    \end{split}
          \end{equation}
In den Rechnungen konnte der Term
          \begin{equation} \nonumber
                    \frac{\nabla^2\sqrt{  \left( \rho_0 + \varepsilon \rho_1 \right) }}
                               {\sqrt{\rho_0 + \varepsilon \rho_1}}
          \end{equation}
mit der \textit{Thomas-Fermi-Näherung}
           \begin{equation} \label{fermithomas}
                    \frac{\nabla^2 \sqrt{ \left( \rho_0 + \varepsilon \rho_1 \right) }}
                               {\sqrt{\rho_0 + \varepsilon \rho_1}}
                    \approx  \frac{\nabla^2\sqrt{  \left( \rho_0 \right) }}
                               {\sqrt{\rho_0 }} \approx \frac{\sqrt{\rho_{max}-\rho_{\infty}}}{\sqrt{2L\rho_{0}}}
          \end{equation}
vernachlässigt werden.
          \begin{figure}
          \begin{center}
          \input{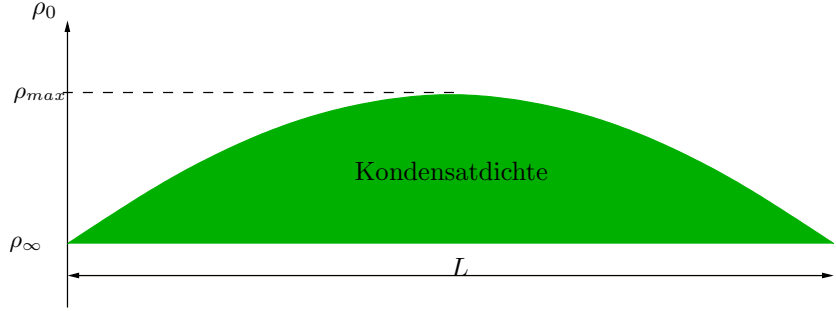}
          \caption[Thomas-Fermi-Näherung]{\label{thomasfermi}Darstellung der Größen, die in Gl.(\ref{fermithomas})
          verwendet werden.}
          \end{center}
          \end{figure}

\paragraph{\fbox{$V_{ext}+N_0 g (\rho_0+\varepsilon \rho_1) \sqrt{\rho_0 +
                 \varepsilon\rho_1}e^{i(\theta_0 +  \varepsilon \theta_1 )}$}}
          \begin{equation} \nonumber
                    \begin{split}
                    =e^{i(\theta_0 +  \varepsilon \theta_1 )}\sqrt{\rho_0 + \varepsilon \rho_1} \times
                    &\left[ V_{ext}+N_0 g\rho_0 \right.\\
                    &\left. +\varepsilon N_0g\rho_1 \right]
                    \end{split}
          \end{equation}

\section{Die Kontinuitäts- und Hamilton-Jakobi-Gleichung.}

Die Gleichungen werden einerseits nach Real - und Imaginärteil und andererseits in die verschiedenen Ordnungen aufgeteilt.
\begin{description}
\item[Realteil mit $\varepsilon^0$]
          \begin{equation} \label{hamjakobi0}
                    \dot{\theta_0}=-\frac{\hbar}{2m}(\nabla \theta_0)^2 - \frac{\hbar}{V_{ext}}-\frac{N_0g}{\hbar}
          \end{equation}

\item[Realteil mit $\varepsilon^1$]
          \begin{equation} \label{hamjakobi1}
                    \dot{\theta_1}=-\frac{\hbar}{m}(\nabla \theta_0 \nabla \theta_1) - \frac{N_0 g}{\hbar} \rho_1
          \end{equation}
\item[Imaginärteil mit $\varepsilon^0$]
          \begin{equation} \label{kontgl0}
                    \dot{\rho_0}=-\frac{\hbar}{m}\nabla(\rho_0 \nabla \theta_0)
          \end{equation}
\item[Imaginärteil mit $\varepsilon^1$]
          \begin{equation} \label{kontgl1}
                    \dot{\rho_1}=-\frac{\hbar}{m}\nabla(\rho_0 \nabla \theta_1 + \rho_1 \nabla \theta_0)
          \end{equation}
\end{description}
Gl.(\ref{hamjakobi0}) und Gl.(\ref{hamjakobi1}) entsprechen dem quantenmechanischen Pendant der Hamilton-Jakobi-Gleichung
in der klassischen Mechanik.
Die beiden letzten Gleichungen sind die Kontinuitätsgleichungen für das Kondensat Gl.(\ref{kontgl0}) und
Gl.(\ref{kontgl1}) für die Störung.

                    \chapter{WKB (Wentzel-Kramers-Brillouin)-Methode\label{WKB}}

Die ausführliche Behandlung der WKB-Methode wird motiviert durch die häufige Anwendung in dieser Arbeit.
Der Grundgedanke ist, dass ein sich wenig änderndes Potential den Erwartungswert - eines sich im Potential
bewegenden Teilchens - wenig beeinflusst.
Die Amplitude der Wellenfunktion kann in einer Taylor-Reihe entwickelt werden, wobei dessen Glieder in
höherer Ordnung verschwinden.
Dieser Zusammenhang wird im Folgenden ausführlich erörtert.

\section{Aufenthaltswahrscheinlichkeit in einem äußeren Potential}
Propagiert eine Welle in einem äußeren Potential, ändert sich die Wellenfunktion.

\subsection{Klassische Wellenfunktion in einem äußeren Potential}
\begin{figure}[h]
          \begin{center}
          \input{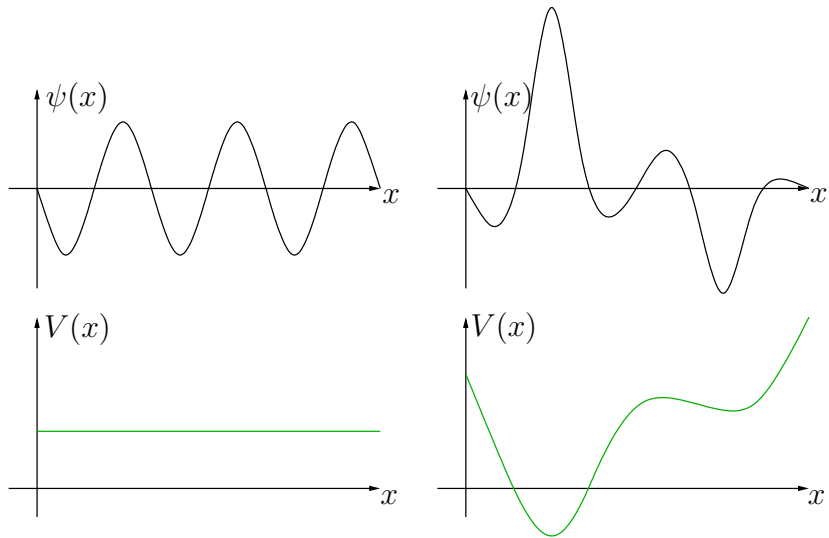}
          \caption[Klassische Wellenfunktion im Potential]
          {\label{potentialkl}Die linke Hälfte stellt eine Welle (schwarz) in einem
          konstanten äußeren Potential (grün) dar. Die Aufenthaltswahrscheinlichkeit,
          das Teilchen in einem bestimmten Intervall zu finden, hängt nicht von dessen
          Lage, sondern nur von dessen Länge ab. Für ein ortsabhängiges Potential (rechte Seite)
          ist die Aufenthaltswahrscheinlichkeit nicht mehr unabhängig von $x$.
          Nahe an einem Minimum ist es wahrscheinlicher das Teilchen zu finden, als
          bei einem Maximum.}
          \end{center}
          \end{figure}

Die klassische Wellenfunktion wird beschrieben durch
          \begin{equation}
                    \phi_{kl}(x)=we^{-i\omega t}e^{ikx}.
          \end{equation}
Im konstanten Potential ist die Aufenthaltswahrscheinlichkeit
          \begin{equation} \label{prob}
                    P(x_2-x_1)=\int_{x_1}^{x_2}dx \phi_{kl}(x)^{\star}\phi_{kl}(x)=\int_{x_1}^{x_2}dx w^2=w \Delta x
          \end{equation}
überall gleich groß (Abb.(\ref{potentialkl})).

Im Potentialtopf wird der Erwartungswert - Amplitude $w \equiv w(x)$ - ortsabhängig, während
die Wellenzahl $k$ unverändert bleibt.

\subsection{Quantenmechanischer Zustand in einem äußeren Potential}

\begin{figure}[h]
          \begin{center}
          \input{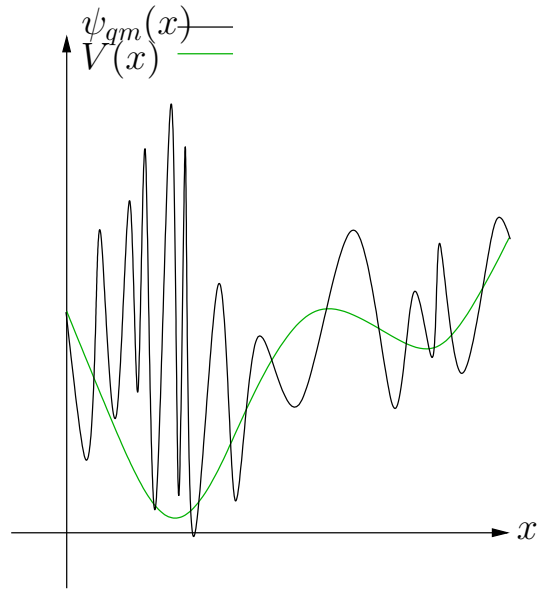}
          \caption[Quantenmechanischer Zustand im Potential]
          {\label{potentialqm}Für ein variierendes Potential (grün) ändert sich auch der Zustand (schwarz).
          Die Wellenzahl und die Amplitude sind ortsabhängig.}
          \end{center}
          \end{figure}

Die quantenmechanische Beschreibung eines Teilchens
          \begin{equation} \label{wkbzustand}
                    \phi_{qm}(x)=w(x)e^{-i\omega t}e^{i\int^{x} dx'k(x')}
          \end{equation}
gleicht der ebenen Welle, jedoch mit dem Unterschied, dass zusätzlich die Wellenzahl $k \equiv k(x)$
vom Ort abhängt, wenn es sich in einem äußeren Potential befindet (Abb.(\ref{potentialqm})).

\section{Die WKB-Näherung}
Ist die Änderung mit dem Ort schwach, kann in Gl.(\ref{wkbzustand}) die Amplitude $w(x)$ durch
          \begin{equation} \label{expandw}
                    w(x)=w_{0}+\varepsilon w_{1}+...\, ,
          \end{equation}
eine Taylor-Reihe entwickelt werden, vorausgesetzt, die Terme höherer Ordnung können vernachlässigt werden.

\section{Gültigkeit der WKB-Näherung}
Die WKB-Näherung ist gültig, wenn $ w_1 \ll w_0 $ ist. In der Talyor-Reihe sind die Terme der Ordnung $n$
proportional zur $n$ten Ableitung.
Die erste Ableitung der Amplitude nach dem Ort ist
          \begin{equation}
                    \partial_x \phi(x)=\left( \frac{w(x)'}{w(x)}+ik(x)\right) \psi(x),
          \end{equation}
damit für
          \begin{equation} \label{wkbgut}
                    \frac{w'}{w} \ll k,
          \end{equation}
d. h., wenn
sich die Amplitude nur wenig mit dem Ort ändert.

                    \chapter{Quelltext für Programme}

\section{Programm zur Simulation einer Ebenen-Welle im ringförmigen Kondensat\label{einewelle}}
\begin{verbatim}


% Festlegung der Konstanten

anzahl=10000;            % Anzahl der diskreten Schritte.
const=1;             % Es gilt: v_s*c^2=const
xanfang=0;
xende=2*pi;
omega=10;             % Frequenz
b=0.3;                 % Konstante für die Dichtefunktion. Es gilt: 0<b<1


% Hauptteil

x=linspace(xanfang,xende,anzahl);       % x-Werte
dichte_s=(1+b*cos(x));                  % Dichte
v_s=(const./dichte_s);                  % Hintergrundgeschwindigkeit

% Berechnung der Amplituden:
faktor(1,:)=1./sqrt(abs(v_s-(-v_s+sqrt(dichte_s))*omega)); % für entgegenlaufende Welle
faktor(2,:)=1./sqrt(abs(v_s-(-v_s-sqrt(dichte_s))*omega)); % für mitlaufende Welle

% Berechnung der Koordinaten:
w(1,:) =1./(-v_s+sqrt(dichte_s));       % für entgegenlaufende Welle
w(2,:) =1./(-1)*(-v_s-sqrt(dichte_s));  % für mitlaufende Welle

% Berechnung der Ereignishorizonte:
EH1=acos((const^(2/3)-1)/b)
EH2=2*pi-EH1

% Berechnung der Wellenzahl:
integrand=-omega.*w ;
deltax=(xende-xanfang) / anzahl;
for j = 1:2
    integral(j,:)=cumsum(integrand(j,:))*deltax; % Nummerische Integration
 end

% Berechnung der ebenen Wellen:
phase(1,:)= faktor(1,:).* cos(integral(1,:));
phase(2,:)= faktor(2,:).* sin(integral(2,:));
neuphase(1,:)= faktor(1,:).* real(exp(i*omega*integral(1,:)));

% Berechnung der y-Werte für die Geraden durch den Ereignishorizont
g=linspace(-2000,2000,1000);
l=linspace(-5,5,1000);

% Plotten der Funktionen:
% Schall- und Hintergrundgeschwindigkeit
subplot(3,1,1)
    plot(x,sqrt(dichte_s),'b',x,v_s,'r',EH1,l,'k',EH2,l,'k')
    axis([xanfang xende 0 2])

% Wellenzahl für beide Wellen
subplot(3,1,2)
    plot(x, integrand(1,:),'r' ,x, integrand(2,:),'b',EH1,g,'k',EH2,g,'k')
    axis([xanfang xende -2000 2000])

% Ebene Wellen
subplot(3,1,3) %
    plot(x,phase(1,:),'r',x,phase(2,:),'b',EH1,l,'k',EH2,l,'k')
    axis([xanfang xende -5 5])

% Vergrößerung eines Ausschnitts aus dem vorhergehenden Plot
    figure
    plot(x,phase(1,:),'r',x,phase(2,:),'b',EH1,l,'k',EH2,l,'k')
    axis([EH1-0.01 EH1+0.01 -2 2])
\end{verbatim}

\section{Programm zur Simulation eines Wellenpakets im ringförmigen Kondensat\label{simulationwellenpaket}}
\begin{verbatim}

% Festlegung der Konstanten

sigma=0.1;            % Standardabweichung
k0=3;               % Max. des Wellenpakets im Impulsraum (t=0)
X0=3;              % Max. des Wellenpakets im Ortsraum (t=0)

Xmin=2;
Xmax=6;
NX=1000;            % Anzahl der Intervalle im Ortsraum
Kmin=0;
Kmax=3;
NK=NX;              % Anzahl der Intervalle im Impulsraum
Tmin=0;
Tmax=-19;
NT=50;              % Anzahl der Intervalle im Zeitraum

% Hauptteil

% Diskretisierung der Räume:
n=1:NX;
Xi=Xmin+(n-1)/(NX-1)*(Xmax-Xmin);
DX=(Xmax-Xmin)/(NX-1);
Kj=Kmin+(n-1)/(NK-1)*(Kmax-Kmin);
DK=(Kmax-Kmin)/(NK-1);
t=1:NT;
Tn=Tmin+(t-1)/(NT-1)*(Tmax-Tmin);
DT=(Tmax-Tmin)/(NT-1);

% Berechnung der ebenen Wellen (siehe vorheriges Kapitel)
const=1;
vs=const/(1+b*cos(X0)^2);
rhos=(1+b*cos(X0));
A=sqrt(abs(vs-(-vs+sqrt(rhos))));
a=1/(-vs+sqrt(rhos));

dichte_s=(1+b*cos(X0));
v_s=(const./dichte_s);
EH1=acos((const^(2/3)-1)/b)
EH2=2*pi-EH1
AX=sqrt(abs(v_s-(-v_s+sqrt(dichte_s))));
aX=1./(-v_s+sqrt(dichte_s));
FX=cumsum(aX)*DX;

% Fourier-Analysen:
% Fourier-Koeffizienten im flachen Raum (siehe Gl.(9.35))
Psi=1/sqrt(sigma)*exp(-(Xi-X0).^2/(2*sigma)).*exp(i*k0*Xi);

EE=[];
MM=[];
GG=[];
for nn=1:NK
    % Fourier-Koeffizienten im "normalen" Flachen im Implusraum (siehe Gl.(9.37))
    EE=[EE;exp(-i*Kj(nn)*Xi*a)];
    % zur Vorbereitung der Fourier-Analyse im gekrümmten Raum
    % in einer Zeile sind die Frequenzen konstand, in einer Spalte die Wellenzahl
    GG=[GG;exp(i*Kj*FX(nn))./AX(nn)];
end

% Berechnung der Fourier-Koeffizienten im fast flachen Raum (siehe Gl.(9.37))
FTPsi=(exp(i*Kj*Xmin*a).').*(A*a/pi*EE*Psi.')*DX;

M=moviein(NT);  % Grafische Darstellung
for ct=1:NT
   % Fourier-Analyse in der Zeit (siehe Gl.(9.34))
   Psixt=GG*(FTPsi.*exp(-i*Kj*Tn(ct)).');
   plot(Xi,abs(Psixt).^2)
   M(ct)=getframe;
end
\end{verbatim}

                    \chapter{Lösung der Bogoliubov-Gleichungen für das ringförmige Kondensat
durch Fourier-Transformation \label{rechnungen}}

\section{Berechnung von $f_{np}$}

\paragraph{\fbox{$f_{np}=\frac{1}{2 \pi}\int_0^{2 \pi} d\gamma e^{-i(n-p)\gamma}c(\gamma)^2$}}
Die Schallgeschwindigkeit $c_{\gamma}$ ist (siehe Gl.(\ref{ringschall}))
$$ \tilde{c}_0=\sqrt{\frac{\tilde{U}}{N}\rho(\gamma)}=\sqrt{\frac{\tilde{U}}{N}\left( 1 + b cos(\gamma) \right)}. $$
        \begin{equation} \nonumber
         \begin{split}
          \frac{1}{2 \pi}\int_0^{2 \pi} d\gamma e^{-i(n-p)\gamma}\tilde{c}_0^2(\gamma)
          &= \frac{1}{2 \pi}\frac{\tilde{U}}{N}\int_0^{2 \pi} d\gamma e^{-i(n-p)\gamma}\left( 1 + b cos(\gamma) \right)\\
          &= \pm \frac{\tilde{U}}{2\pi}\int_0^{2\pi} d\gamma e^{-i(n-p)\gamma}
           \pm  \frac{\tilde{U}\,b}{2\pi} \int_0^{2 \pi} d\gamma e^{-i(n-p)\gamma} cos(\gamma)\\
          &= \pm  \frac{\tilde{U}}{(2\pi)} \delta_{n,p}
          \pm \frac{\tilde{U}}{(2\pi)^2} \int_0^{2 \pi} d\gamma \frac{1}{2}e^{-i(n-p)\gamma}e^{i\gamma}
          \pm \frac{\tilde{U}}{(2\pi)^2} \int_0^{2 \pi} d\gamma \frac{1}{2}e^{-i(n-p)\gamma}e^{-i\gamma}\\
          &= \frac{\tilde{U}}{2\pi}\left( \delta_{n,p} + \frac{b}{2}\delta_{n,p+1}+ \frac{b}{2}\delta_{n,p-1}\right)
          \end{split}
   \end{equation}

\section{Berechnung von $h_{np}^{\pm}$}

\paragraph{\fbox{$\frac{1}{2 \pi}\int_0^{2 \pi} d\gamma e^{-i(n-p)\gamma}p\tilde{v}_0(\gamma)$}}

Die Hintergrundgeschwindigkeit $v_{\gamma}$ ist Gl.(\ref{ringv})
$$ v(\gamma)=m^{\star}\frac{\sqrt{1-b^2}}{(1+bcos(\gamma)}. $$
Mit
          \begin{equation} \label{dichteinreihe}
          \begin{split}
                    \frac{1}{1+b\, cos(\gamma)}
                    =&\sum_m (-b\,cos(\gamma))^m= \\
                    \sum_m \left( -\frac{b}{2} \right)^m \left( e^{i\gamma}+e^{-i\gamma} \right)^m
                    =&\sum_m \left( -\frac{b}{2} \right)^m \sum_{k=0}^m {m \choose k}
                    e^{i\gamma(m-k)e^{-i\gamma k}}
          \end{split}
          \end{equation}
folgt
   \begin{equation} \nonumber
   \begin{split}
   \frac{1}{2 \pi}\int_0^{2 \pi} d\gamma e^{-i(n-p)\gamma}p\tilde{v}_0(\gamma)&
   =\frac{1}{2 \pi}p\, m^{\star}\,\sqrt{1-b^2}\,\sum_m \left( -\frac{b}{2} \right)^m \sum_{k=0}^m {m \choose k}
   \int_0^{2 \pi} d\gamma e^{-i(m-2k-(n-p))\gamma}\\
   &= \frac{1}{2 \pi}p\, m^{\star}\,\sqrt{1-b^2}\,\sum_m \left( -\frac{b}{2} \right)^m \sum_{k=0}^m {m \choose k}
   \left\{
          \begin{split}
              2 \pi \,\,\,\,\,\,\,\,& f"ur \,\,\,\,\,\,\,\,\,\,\,\,k=\frac{m-(n-p)}{2} \\
              0     \,\,\,\,\,\,\,\,& sonst
          \end{split}
   \right.\\
   &=p\, m^{\star} \sqrt{1-b^2}\alpha_{s=p-n},
   \end{split}
   \end{equation}
mit
          \begin{equation}  \label{alpharing}
                    \alpha_{s=p-n}=\sum_{\begin{array}{c}
                                                  m\\
                                                  m \geq \vert s \vert,\\
                                                  mit\, m+s \, gerade
                                        \end{array}}
                    \left( -\frac{b}{2} \right)^m
                    {m \choose (m+s)/2}
          \end{equation}

\paragraph{\fbox{$\frac{1}{2 \pi}\int_0^{2 \pi} d\gamma e^{-i(n-p)\gamma}\frac{1}{2} \tilde{v}_0(\gamma)'$}}

   \begin{equation} \nonumber
         \begin{split}
          \frac{1}{2 \pi}\int_0^{2 \pi} d\gamma e^{-i(n-p)\gamma}\frac{1}{2} \tilde{v}_0(\gamma)'
          &=\frac{1}{2 \pi}\left[e^{-i(n-p)\gamma}\tilde{v}_0(\gamma)  \right]_0^{\pi}
          -\frac{1}{2 \pi} \frac{1}{2} \int_0^{2 \pi} d\gamma e^{-i(n-p)\gamma}(-i(n-p))\tilde{v}_0(\gamma)\\
          &= -\frac{1}{2}i(n-p)\,m^{\star} \sqrt{1-b^2}\,\,\alpha_{s=p-n}
         \end{split}
   \end{equation}

\paragraph{\fbox{$\frac{1}{2 \pi}\int_0^{2 \pi} d\gamma e^{-i(n-p)\gamma}\frac 12 \frac{\tilde{c}_0(\gamma)''}
                    {\tilde{c}_0(\gamma)}$}}

            \begin{equation}  \label{betaring}
                    \beta_{s=p-n}=\sum_{\begin{array}{c}
                                                  m\\
                                                  m \geq \vert s \vert,\\
                                                  mit\, m+s \, gerade
                                        \end{array}}
                    \left( -\frac{b}{2} \right)^m
                    {m \choose (m+s)/2}(m+1)
          \end{equation}

          \backmatter

          \listoffigures
          \bibliography{Referenzen}

          \include{Danke}

\end{document}